\documentclass[a4paper,11pt]{article}
\usepackage{jheppub} 

\usepackage[english]{babel}

\usepackage{bbold}
\usepackage[utf8]{inputenc}
\usepackage{graphicx}
\usepackage{amssymb}
\usepackage{amsmath}
\usepackage{xcolor}
\usepackage{hyperref}
\usepackage{multirow}
\usepackage{slashed}
\usepackage{subcaption}
\usepackage{nicefrac}

\renewcommand{\vec}[1]{\boldsymbol{#1}}
\newcommand{\be}{\begin{equation}}
\newcommand{\ee}{\end{equation}}
\newcommand{\ba}{\begin{eqnarray}}
\newcommand{\ea}{\end{eqnarray}}

\newcommand{\<}{\langle}
\renewcommand{\>}{\rangle}
\newcommand{\nn}{\nonumber \\}

\newcommand{\im}{{\rm Im\,}}

\definecolor{codegray}{gray}{0.9}

\newcommand{\ahvp}{a_\mu^{\rm hvp}}
\newcommand{\ahvpnlo}{a^{\textrm{hvp}1 \gamma ^*}_\mu}

\newcommand{\sy}{\sqrt{\tilde{y}}}

\abstract{
In order to reach a precision of $0.2\%$ on the hadronic vacuum polarization (HVP) contribution to the anomalous magnetic moment of the muon, $(g-2)_\mu$, such that the full Standard-Model prediction matches in precision the direct experimental measurement, it is crucial to include the leading, O($\alpha$) electromagnetic corrections to HVP.
In this work, we determine an important contribution to the latter from a diagram comprised of two two-point quark-loops, connected by the internal photon and gluons.
This ultraviolet-finite correction is calculated from lattice QCD using a coordinate-space formalism, where photons are treated in the continuum and infinite volume. 
Our result amounts to a $(-0.89\pm 0.18)\%$ 
correction to the leading-order HVP contribution to $(g-2)_\mu$.  
To overcome the worsening statistical noise at large distances, our analysis is combined with phenomenological models featuring light pseudoscalar mesons with masses below 1 GeV.
In particular, our data show a very steep mass dependence as dictated by the charged pion loop.
Similarly, the other diagrams appearing in the O$(\alpha)$-corrections to the HVP with the internal photon connecting two separate quark loops are also ultraviolet-finite and can be computed with the same formalism. 
}

\begin{document}
\preprint{MITP-24-088}

\author[a]{Julian~Parrino,}
\emailAdd{julian.parrino@ur.de}
\affiliation[a]{PRISMA$^+$ Cluster of Excellence \& Institut f\"ur Kernphysik,
Johannes Gutenberg-Universit\"at Mainz,
D-55099 Mainz, Germany}

\author[b]{Volodymyr~Biloshytskyi,}
\affiliation[b]{Institut f\"ur Kernphysik,
Johannes Gutenberg-Universit\"at Mainz,
D-55099 Mainz, Germany}

\author[c]{En-Hung~Chao,}
\affiliation[c]{Physics Department, Columbia University, New York City, New York 10027, USA}

\author[a,d]{Harvey~B.~Meyer,}
\affiliation[d]{Helmholtz~Institut~Mainz,
Staudingerweg 18, D-55128 Mainz, Germany}

\author[b]{and Vladimir~Pascalutsa}


\title{
Computing the UV-finite electromagnetic corrections 
to the hadronic vacuum polarization in the muon $(g-2)$ from lattice QCD
}

\maketitle


\section{Introduction}

Recently, the tension on the hadronic vacuum polarization (HVP) contribution to the anomalous magnetic moment of the muon, $a_\mu$, between the determinations based on data-driven approaches~\cite{Davier:2017zfy,Keshavarzi:2018mgv,Colangelo:2018mtw,Hoferichter:2019mqg,Davier:2019can,Keshavarzi:2019abf}  -- which were used to obtain the result in the 2020 White Paper of the Muon $g-2$ Theory Initiative~\cite{Aoyama:2020ynm} -- and those from lattice Quantum Chromodynamics (lattice QCD)~\cite{Djukanovic:2024cmq,Borsanyi:2020mff,Boccaletti:2024guq, RBC:2024fic,ExtendedTwistedMassCollaborationETMC:2024xdf,Bazavov:2024eou} has become more pronounced, as the results from the latter category ameliorated over time.
This tension is further confirmed by a careful scrutiny of the so-called \textit{intermediate window observable} from many lattice QCD collaborations~\cite{Borsanyi:2020mff,Lehner:2020crt,Wang:2022lkq,Aubin:2022hgm,Ce:2022kxy,ExtendedTwistedMass:2022jpw,Chao:2022ycy,RBC:2023pvn, FermilabLattice:2024yho} with determinations at the sub-percent level.

Lattice calculations of the HVP contribution to $a_\mu$, $\ahvp$, are typically performed in the isospin-symmetric limit. 
However, in order to reach a sub-percent precision, one needs to take into account the isospin-breaking effects arising from the differences in mass and electric charge between the up and down quarks.  
These effects can be treated in a perturbative expansion in the quark-mass difference $m_u-m_d$ and the fine-structure constant $\alpha= \nicefrac{e^2}{4\pi}$ around the isospin-symmetric limit (referred to as ``isoQCD'' in the remainder of the article) respectively \cite{deDivitiis:2013xla}, as these expansion parameters are numerically small at the relevant energy scale. 
Contrary to simulating the full SU$(3)_{\rm strong}\times$U$(1)_{\rm {QED}}$ gauge group on the lattice\footnote{Note that if an accuracy beyond O($\alpha$) were sought, such simulations would still need to be corrected for lepton loops.}, including the electromagnetic (e.m.) effects with the aforementioned expansion has the advantage that it can be built on existing isoQCD gauge ensembles directly.
There are several proposals on the inclusion of the photon within this framework.
Among those where the photon is put in a finite box in accordance to the treatment of the QCD sector, the zero-mode of the photon needs to be handled with care.
A commonly used method from this category is the QED${}_{\rm L}$-prescription.
There, the zero-modes are removed by imposing that the partially-Fourier-transformed photon field vanish  on all time slices~\cite{Hayakawa:2008an}, $A_\mu(x_0, \vec{k}=0)=0 $. 
Another popular method, referred to as the QED${}_{\rm M}$-prescription, consists in performing  a series of measurements with massive photons at decreasing masses $m_\gamma$ and taking the $m_\gamma\to 0$ limit \cite{Endres:2015gda}.
Alternatively, one can also treat the QED sector in the continuum and infinite volume and combine it with hadronic matrix elements computed on the lattice, an approach also known as QED${}_{\infty}$~\cite{Green:2015mva,Feng:2018qpx}. 

In the previous paper~\cite{Biloshytskyi:2022ets}, we proposed a coordinate-space formalism to compute the e.m.\ corrections to the HVP based on QED${}_{\rm \infty}$.
The formalism handles the ultraviolet (UV) divergences due to the internal photon at large momenta with a Pauli-Villars regulator.
An appealing aspect of the formalism is the possibility of comparing the lattice results to phenomenological predictions from a Cottingham-like formula, which expresses the e.m.\ correction to HVP via the forward hadronic light-by-light (HLbL) scattering amplitude. 
A dispersive sum rule then allows one to express the latter amplitude in terms of the $\gamma^*\gamma^*\to{\rm hadrons}$ cross sections.
All the relevant counterterms can be determined unambiguously by computing the e.m.\ mass shifts of the appropriate number of hadrons (typically the average pion mass, the charged and neutral kaon mass and a baryon mass of one's choice).
These e.m. mass shifts are available from phenomenological calculations, either restricted to the elastic part of the forward Compton amplitude or going beyond that approximation, see e.g.\ \cite{Stamen:2022uqh} and \cite{Gasser:2020hzn}, respectively.

\begin{figure}
    \centering
    \includegraphics[width=0.81\textwidth]{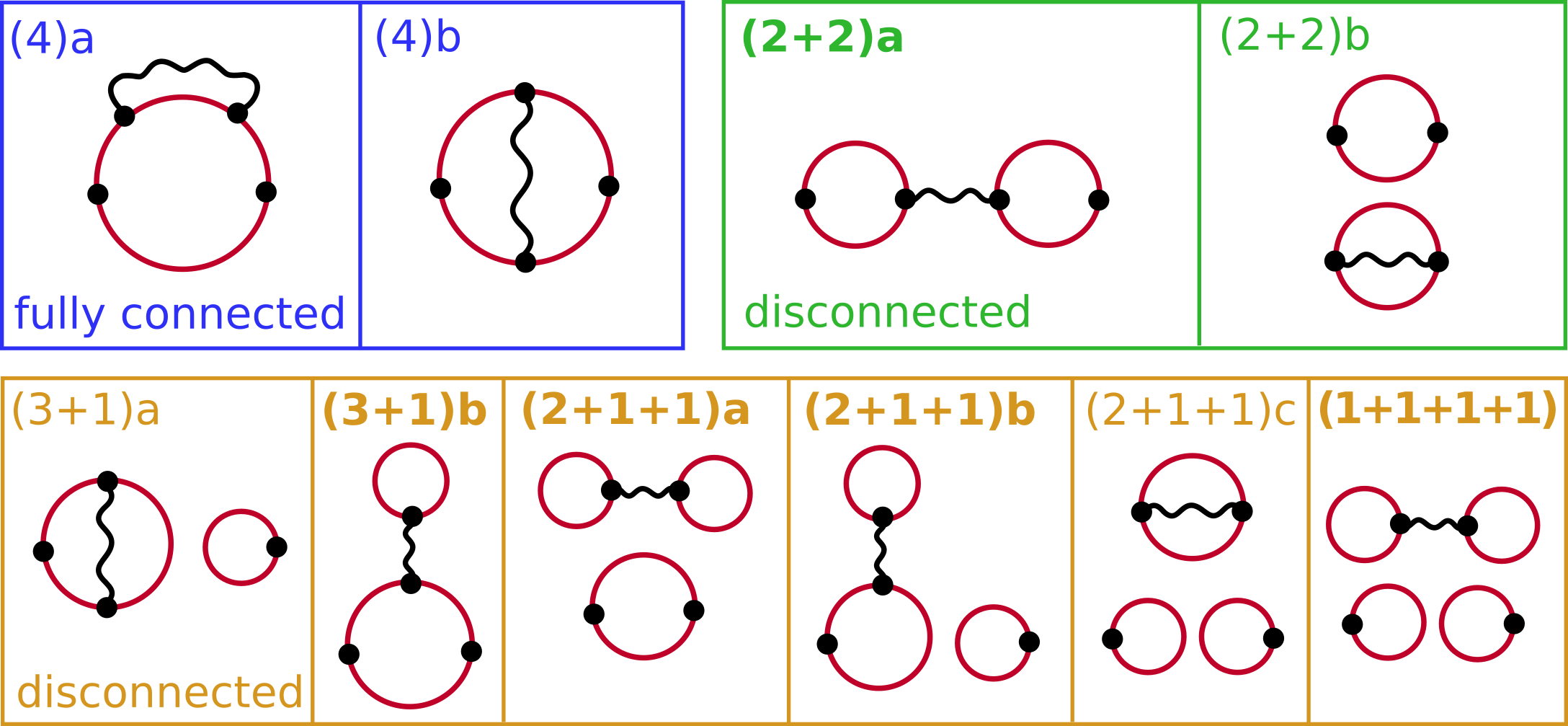}
    \caption{Wick contractions contributing to the e.m.\ corrections to the hadronic vacuum polarization as defined in Eq.~\eqref{eq:ccs_nlohvp}. The Wick contractions $(2+2)a$, $(3+1)b$, $(2+1+1)a$, $(2+1+1)b$, $(1+1+1+1)$, highlighted in boldface, are UV-finite. As usual in the lattice QCD context, gluon lines are not explicitly represented; instead, the quark propagators are computed within a non-perturbative SU(3) gauge field background, sampled according to the Euclidean path integral. 
    All diagrams are one-photon-irreducible, i.e., the disconnected quark loops are connected by gluons.
    The two vertices in each diagram that are not attached to the photon propagator are connected to the HVP kernel.}
    \label{fig:FiniteDiags}
\end{figure}

A direct comparison between results from different lattice-discretizations on each individual Wick-contraction diagram in the leading e.m. corrections to the HVP is rendered complicated by the local counterterms needed for renormalization.
However, we observe that those diagrams with disjoint quark-loops connected by the internal photon, as illustrated in figure~\ref{fig:FiniteDiags} in boldface, are in fact by themselves UV-finite when the Pauli-Villars regulator of the internal photon is removed (see section~\ref{sec:disco_def}). 
These diagrams can hence serve as useful benchmark quantities for the leading e.m. corrections to the HVP.
In this paper, we present a calculation of such a UV-finite contribution to the leading e.m. corrections to the HVP from the Wick-contraction diagrams with two two-point quark-loops, i.e. the one labelled $(2+2)a$ in figure~\ref{fig:FiniteDiags}.
From the lattice calculations of the HLbL contribution to the anomalous magnetic moment of the muon~\cite{Chao:2021tvp, Blum:2023vlm}, we expect this particular diagram to be one of the largest among the five displayed UV-finite diagrams.

Although our main result for the $(2+2)a$ contribution is obtained using lattice QCD, we use a model inspired from phenomenology in order to support this calculation. Here, we want to point out the main steps of our calculation and help the reader orient themselves across the various sections.
\begin{itemize}
    \item The basic definitions of our coordinate-space framework are established in section \ref{sec:formalism}. The master integral used to compute the $(2+2)a$ contribution is given in Eq.~\eqref{eq:2+2_master_a}. 
    \item The lattice QCD implementation of the calculation of the $(2+2)$ contributions is provided in section \ref{sect:nlohvp_setup}.
    \item In section \ref{sec:phenomenological_model} we introduce a model for the e.m. corrections to the HVP based on the $\pi^0$, $\eta$ and $\eta'$ exchange contributions as well as the charged pion and charged kaon loop. Using this model we provide an estimate of the full e.m. corrections in section \ref{sec:model_precition_full_em}. For our lattice QCD calculation this model serves two purposes: Firstly, to approximate the tail of the $(2+2)a$ contribution calculated with lattice data and secondly, to constrain the pion mass dependence of this contribution.
    \item In section \ref{sec:model_prediction_for_integrand}, we evaluate the integrand of our master integral with phenomenological models considered in section \ref{sec:phenomenological_model}. These auxiliary results in coordinate-space are used to to support our lattice calculations at large distances, where the data quality worsens.
    \item In section \ref{sec:analysis}, we discuss our method to obtain the $(2+2)a$ contribution at the physical point. A matching procedure of the model prediction to the $(2+2)a$ topology is given in section \ref{sect:hvpnlo_matching}, summarized in Eq.~\eqref{eq:full_model}.
    For each point along the chiral trajectory the contribution to Eq.~\eqref{eq:2+2_master_a} is calculated in lattice QCD, where the tail is estimated using the phenomenological model, see section \ref{sect:nlohvp_tail}.
    The result at the physical point is obtained using a model averaging procedure, which is described in section \ref{sect:nlohvp_extrapolation}.
    \item Complementary analyses of the subdominant light-quark--strange-quark and strange-quark--strange-quark components of the $(2+2)a$ diagram, as well as of the UV-divergent $(2+2)b$ diagram are given in section~\ref{sec:add-contrib}. 
    \item Our findings for the $(2+2)a$ contribution are summarized in section~\ref{sect:discu} together with concluding remarks.
\end{itemize}

\section{Formalism}
\label{sec:formalism}
In this section, we lay out the formalism to compute the electromagnetic corrections to the HVP contribution in a coordinates-space framework. In section \ref{sec:disco_def}, we give details on the coordinate-space kernel, including a photon propagator. A formal definition of the Wick contractions that we label as $(2+2)$ disconnected diagrams is provided in section \ref{sec:tpt_contributions}, followed by a discussion on the physical interpretation of those contributions in section \ref{sec:tpt_physical_interpretation}.

\subsection{Coordinate-space framework}
\label{sec:disco_def}
We start from the coordinate-space based formula for the QED corrections to the HVP from
 Ref.\ \cite{Biloshytskyi:2022ets} in Euclidean space containing one internal photon \cite{Djukanovic:2024cmq}
\ba
    \label{eq:ccs_nlohvp}
    \ahvpnlo=\lim_{\Lambda \to \infty}\Big\{-\frac{e^2}{2}\int_{x,y,z} H_{\mu\sigma}(z)\delta_{\nu \rho}\Big[G_0(y-x)\Big]_\Lambda\langle j_\mu(z) j_\nu(y) j_\rho(x) j_\sigma(0) \rangle_\mathrm{QCD}\;\\
    \nonumber
    +\,\text{counterterms} \quad \Big\},
\ea 
where 
\be 
\label{eq:tv_kernel}
H_{\mu \sigma}(z)=-\delta_{\mu \sigma} \mathcal{H}_1(|z|) + \frac{z_\mu z_\sigma}{|z|^2} \mathcal{H}_2(|z|)
\ee
is a covariant coordinate-space (CCS) kernel encoding the QED effects with $\mathcal{H}_1$ and $\mathcal{H}_2$ two scalar functions originally defined in Ref.~\cite{Meyer:2017hjv}. We use the notation $\int_{x}:= \int_{\mathbb{R}^4} d^4x$.
$[G_0(y-x)]_\Lambda$ is a UV-regularized photon propagator parameterized by a cutoff scale $\Lambda$. 
A possible choice proposed in Ref.~\cite{Biloshytskyi:2022ets} is the doubly-subtracted Pauli-Villars regularization:
\ba
\label{eq:pv}
\Big[G_0(y-x)\Big]_\Lambda=\frac{1}{4\pi^2|y-x|^2}
- 2G_{\frac{\Lambda}{\sqrt{2}}}(y-x) 
+ G_{\Lambda}(y-x),
\ea
where $G_m(x)={m K_1(m|x|)}/(4\pi^2|x|)$, with $K_1$ a modified Bessel function of the second kind, is a massive scalar propagator with mass $m$. 
The regularized photon propagator Eq.~\eqref{eq:pv} is finite at the $|y-x|\to 0$ limit. 
Finally, $\langle j_\mu(z) j_\nu(y) j_\rho(x) j_\sigma(0) \rangle_\mathrm{QCD}$ is a four-point function of electromagnetic vector currents evaluated in a QCD-only background
\ba 
j^{\rm em}_\mu (z) =\frac{2}{3}\Bar{u}(z)\gamma_\mu  u(z)  -\frac{1}{3}\Bar{d}(z)\gamma_\mu  d(z) -\frac{1}{3} \Bar{s}(z)\gamma_\mu  s(z) + \dots \quad.
\label{eq:jem}
\ea
We will only consider the contributions from the up, down and strange quarks in the QCD sector.
Making use of partial integration and exploiting the fact that the electromagnetic current is conserved $(\partial_\mu j_\mu(z)=0)$, it is possible to add a total-derivative term of type $\partial_\mu [ z_\sigma g(|x|)] $ to the CCS kernel without changing the result for $\ahvpnlo$ in the continuum and infinite volume, as long as no boundary terms contribute. 
Such a property has already be exploited in Ref.~\cite{Chao:2022ycy} to reshape the integrand to reduce statistical error coming from large separations.
In this work, instead of the original kernel Eq.~\eqref{eq:tv_kernel}, we consider two other choices: the traceless (`TL') kernel,
\begin{equation}
    \label{eq:tl_kernel}
    H_{\mu \sigma}^{\rm TL}(z)=\left(-\delta_{\mu \sigma}+ 4\frac{z_\mu z_\sigma}{|z|^2} \right) \mathcal{H}_2(|z|),
\end{equation}
and a variant proportional to $z_\mu z_\sigma$ (`XX') \cite{Ce:2018ziv},
\begin{equation}
    \label{eq:xx_kernel}
    H_{\mu \sigma}^{\rm XX}(z)=\frac{z_\mu z_\sigma}{|z|^2} \left( \mathcal{H}_2(|z|) + |z| \frac{d}{d|z|} \mathcal{H}_1(|z|) \right)\,,
\end{equation}
which are less exposed to the long-distance region, where statistical fluctuations from lattice simulations become larger~\cite{Chao:2022ycy}.

Performing Wick contractions for the four-point function in Eq.~\eqref{eq:ccs_nlohvp}, one can identify five topologies of diagrams that contribute to $a_\mu^{\textrm{hvp}1 \gamma ^*}$. 
We adopt the convention of Ref. \cite{Chao:2021tvp} to name these topologies by the number of vertices that are connected by valence quark propagators. 
In this work, we further distinguish the diagrams within the same Wick-contraction topology by how the quark loops are connected with the internal photon, leading to the categorization in figure~\ref{fig:FiniteDiags}.
Due to the relation on the electric charges of the quarks $\sum_{f=u,d,s}Q_f=0$, the diagrams containing at least one self-contracted quark loop cancel at the exact SU$(3)$-flavour symmetric point (SU$(3)_{\rm f}$), where all three quarks have equal masses. 
We therefore expect those contributions to be subdominant compared to the diagrams in the first row of figure~\ref{fig:FiniteDiags}, namely, the fully-connected and the $(2+2)$-disconnected diagrams. 
The total contribution from the $(2+2)$-disconnected diagrams can however be sizeable.
For example, in the lattice calculation of the HLbL contribution to $a_\mu$~\cite{Chao:2021tvp}, where the evaluation of a four-point correlation function of the e.m.\ current is also required, the fully-connected and the $(2+2)$-disconnected contribute at the same order of magnitude but with opposite signs.

\subsection{The $(2+2)$ contributions}
\label{sec:tpt_contributions}
In this work, we focus on the $(2+2)$-disconnected diagrams.
For the diagrams of this topology, each of the valence quark loops can either be a light (`l') or a strange (`s') quark.
For the total $(2+2)$  vector correlator, we have 
\ba 
\label{eq:nlo_vec_tpt}
\big\langle j_\mu(z) j_\nu(y) j_\rho(x) j_\sigma(0) \big\rangle_\mathrm{QCD}^{(2+2)} &=& 
\nonumber
 C^{(ll)}\Big(\widetilde{\Pi}^{(2+2)a-ll}_{\sigma\rho\nu\mu}+\widetilde{\Pi}^{(2+2)b-ll}_{\sigma\rho\nu\mu}+ \widetilde{\Pi}^{(2+2)c-ll}_{\sigma\rho\nu\mu} \Big)(x,y,z) \\ 
\nonumber
&+&C^{(ls)} \Big(\widetilde{\Pi}^{(2+2)a-ls}_{\sigma\rho\nu\mu}+\widetilde{\Pi}^{(2+2)b-ls}_{\sigma\rho\nu\mu}+ \widetilde{\Pi}^{(2+2)c-ls}_{\sigma\rho\nu\mu} \Big)(x,y,z)\\
&+& C^{(sl)}\Big(\widetilde{\Pi}^{(2+2)a-sl}_{\sigma\rho\nu\mu}+\widetilde{\Pi}^{(2+2)b-sl}_{\sigma\rho\nu\mu}+ \widetilde{\Pi}^{(2+2)c-sl}_{\sigma\rho\nu\mu} \Big)(x,y,z)\\
&+& C^{(ss)}\Big(\widetilde{\Pi}^{(2+2)a-ss}_{\sigma\rho\nu\mu}+\widetilde{\Pi}^{(2+2)b-ss}_{\sigma\rho\nu\mu}+ \widetilde{\Pi}^{(2+2)c-ss}_{\sigma\rho\nu\mu} \Big)(x,y,z). \nonumber
\ea 
with the corresponding charge factors of
\be
C^{(ll)}=\frac{25}{81}, \qquad
C^{(ls)}=C^{(sl)}=\frac{5}{81}, \qquad 
C^{(ss)}=\frac{1}{81}.
\ee
The $(2+2)$ four-point functions are composed of two two-point functions, where the different topologies for quark flavours $f$ and $f'$ are given by
\ba 
\label{eq:2+2a}
\widetilde{\Pi}^{(2+2)a-ff'}_{\sigma\rho\nu\mu}(x,y,z)=\langle \hat{\Pi}^f_{\mu \rho}(z,x)\hat{\Pi}^{f'}_{\nu \sigma}(y,0)\rangle_{U},
\ea 
\ba 
\label{eq:2+2b}
\widetilde{\Pi}^{(2+2)b-ff'}_{\sigma\rho\nu\mu}(x,y,z)= \langle \hat{\Pi}^f_{\mu \sigma}(z,0)\hat{\Pi}^{f'}_{\nu\rho}(y,x)\rangle_{U}, 
\ea 
\ba 
\label{eq:2+2c}
\widetilde{\Pi}^{(2+2)c-ff'}_{\sigma\rho\nu\mu}(x,y,z)=\langle \hat{\Pi}^f_{\mu \nu}(z,y)\hat{\Pi}^{f'}_{\rho \sigma}(x,0)\rangle_{U}.
\ea 
Here, $\langle \dots \rangle_U$ denotes the average over gauge configurations,
and $\hat\Pi^f_{\mu\nu}$ represents the two-point correlation function for local vector currents,
\ba 
\label{eq:2ptf}
\Pi^f_{\mu \nu}(x,y) = -\mathrm{Re}\Big(\mathrm{Tr}\Big[D^{-1}_f(y,x)\gamma_\mu D^{-1}_f(x,y) \gamma_\nu  \Big] \Big),
\ea 
with the vacuum expectation value (VEV) subtracted to retain the one-photon irreducible contribution:
\ba 
\label{eq:vev_subtr}
\hat{\Pi}^f_{\mu \nu}(x,y) = \Pi^f_{\mu \nu}(x,y)-\langle \Pi^f_{\mu \nu}(x,y) \rangle_{U}
\ea
and $D_f^{-1}$ is the quark propagator of flavor $f$.
To further simplify the calculation, we realize Eq.~\eqref{eq:2+2a} and Eq.~\eqref{eq:2+2c} -- both belong to the $(2+2)a$ class as defined in figure~\ref{fig:FiniteDiags} -- are related by swapping the r\^oles of $x$ and $y$.
Inserting Eqs.~\eqref{eq:nlo_vec_tpt} into Eq.~\eqref{eq:ccs_nlohvp} and using O(4)-invariance of scalar functions in the continuum and infinite-volume limit, we can rearrange the twelve-dimensional integral in a more economical way.
Concretely, the contributions from the $(2+2)a$ and $(2+2)b$ diagrams to $\ahvpnlo$ of a given quark-flavor content take the form 
\ba 
\nonumber
a_\mu^{(2+2)a-ff'}&=&-8\pi^3 \alpha C^{(ff')}\int_0^\infty d|x| |x|^3 \Big[\langle I^{(2,f)}_{\rho \sigma}(x)I^{(3,f')}_{\sigma \rho}(x)\rangle_{U} -\langle I^{(2,f)}_{\rho \sigma}(x)\rangle_{U}\langle I^{(3,f')}_{\sigma \rho}(x)\rangle_{U} \Big]\\
&\equiv& \int_0^\infty d|x| f^{(2+2)a-ff'}(|x|),
\label{eq:2+2_master_a}\\
a_\mu^{(2+2)b-ff'}&=&-4\pi^3 \alpha C^{(ff')} \int_0^\infty d|x| |x|^3 \Big[\langle I^{(1,f)}(x)I^{(4,f')}\rangle_{U} -\langle I^{(1,f)}(x)\rangle_{U}\langle I^{(4,f')}\rangle_{U} \Big]\nonumber\\ 
&\equiv& \int_0^\infty d|x| f^{(2+2)b-ff'}(|x|),\label{eq:2+2_master_b}
\ea 
where the four integrals $I^{(1,f)}(x)$, $I^{(2,f)}_{\nu \sigma}(x)$, $I^{(3,f)}_{\sigma \rho}(x)$ and $I^{(4,f)}$ are defined as follows \cite{Chao:2023lxw}:
\begin{subequations}
\ba 
I^{(1,f)}(x) &=& \delta_{\nu\rho}\int_y \Big[G_0(x-y)\Big]_\Lambda \Pi^f_{\rho \nu}(x,y), \label{eq:I1}
\\
\quad I^{(2,f)}_{\nu \sigma}(x) &=& \int_y \Big[G_0(x-y)\Big]_\Lambda \Pi^f_{\nu \sigma}(y,0), \label{eq:I2}
\\
I^{(3,f)}_{\sigma \rho}(x) &=& \int_z H_{\mu \sigma}(z) \Pi^f_{\mu \rho}(z,x),\label{eq:I3}
\\
 I^{(4,f)} &=& \int_z  H_{\mu \sigma}(z) \Pi^f_{\mu \sigma}(z,0).\label{eq:I4}
\ea
\label{eq:Ii}
\end{subequations} 
The quantities $f(|x|)$ defined in the second line of Eq.~\eqref{eq:2+2_master_a} and Eq.~\eqref{eq:2+2_master_b} will be referred to as \textit{integrands}.
The total contribution from the (2+2)-topology is then given by the linear combination 
\ba
a_\mu^{(2+2)} 
&=& C^{(ll)} (a_\mu^{(2+2)a-ll} + a_\mu^{(2+2)b-ll})+C^{(sl)} (a_\mu^{(2+2)a-sl} + a_\mu^{(2+2)b-sl})
\\ && + C^{(ls)} (a_\mu^{(2+2)a-ls} + a_\mu^{(2+2)b-ls})
+ C^{(ss)} (a_\mu^{(2+2)a-ss} + a_\mu^{(2+2)b-ss}).
\nonumber
 \ea
The (ls) and (sl) components are equal in the continuum and infinite volume.

\subsection{Physical interpretation of the $(2+2)$ contributions}
\label{sec:tpt_physical_interpretation}
As a matter of fact, the $(2+2)a$ and $(2+2)b$ diagrams have different sensitivities to short-distance physics, as they can be understood as different types of QED corrections to the HVP.
On the one hand, the $(2+2)a$ contribution can be interpreted as a QED correction to the leading-order disconnected contribution to the HVP. 
The latter is doubly SU$(3)$-flavour suppressed and contributes approximately $-2\%$ of the leading-order HVP contribution \cite{Borsanyi:2020mff}. 
This suppression is lifted by the photon propagator, leading to a relatively large effect of  $a_\mu^{(2+2)a} $ compared to the leading-order disconnected HVP contribution.
On the other hand, we note that the VEV of Eq.~\eqref{eq:I4} is precisely the leading order HVP contribution to $a_\mu$. 
Consequently, the $(2+2)b$ contribution Eq.~\eqref{eq:2+2_master_b}, which involves the VEV-subtracted version of Eq.~\eqref{eq:I4}, can be seen as the effect of sea quarks exchanging a photon.
A stark difference between the two diagrams is that the $(2+2)a$ contribution remains UV-finite when the UV-regulator on the internal photon in Eq.~\eqref{eq:ccs_nlohvp} is removed, which will be referred to as the $\Lambda\to\infty$ limit in the remainder of the paper.
More generally, all quark-disconnected diagrams with the internal photon connecting two disjoint quark-loops, namely, the $(2+2)a$, $(3+1)b$, $(2+1+1)a$, $(2+1+1)b$ and  $(1+1+1+1)$ diagrams in figure~\ref{fig:FiniteDiags}, also have the same feature. 
This property can be derived with partially-quenched QCD, where the correlators involving only a specific set of Wick-contraction diagrams can be constructed by supplementing pairs of quark and ghost of extra (``quenched'') flavors to the original QCD.
These extra degrees of freedom only affect the valence sector and cancel in the sea sector.
Let us illustrate this for the $(2+2)a$ diagram, whose contribution to the e.m. correction to the HVP can be obtained by substituting the photon-weighted correlator in Eq.~\eqref{eq:ccs_nlohvp} by~\cite{Chao:2020kwq}:
\ba 
\delta_{\mu \nu} G_0(x-y)  \langle  \overline{u}(0)\gamma_\sigma d(0) \overline{d}(y)\gamma_\mu u(y) \overline{v} (x)\gamma_\nu r(x) \overline{r}(z) \gamma_\lambda v(z) \rangle\,,
\ea 
where $v$ and $r$ are two additional quenched quark flavors.
Upon Wick contraction, the above expression gives rise exactly to the contraction appearing in Eq.~\eqref{eq:2+2_master_a}.
Now, we study the $x\to y$ limit of this quantity, which corresponds to the UV region of the internal photon propagator.
Let us first consider the case where $x$ and $y$ are both kept at a finite distance from the origin.
The operator product expansion (OPE) of the operator $\overline{d}(y)\gamma_\mu u(y) \overline{v} (x)\gamma_\nu r(x)$ must include four quark flavours, hence the lowest-dimension effective operator it can produce has a mass dimension of six.
In particular, we have
\be 
\overline{d}(y)\gamma_\mu u(y) \overline{v} (x)\gamma_\nu r(x)  \quad \overset{x\to y}{\sim} \quad \Big(\overline{d}(x)\gamma_\mu u(x) \overline{v} (x)\gamma_\nu r(x)\Big) c(|x|)+ \dots,
\ee 
where $c(|x|)$ is the Wilson coefficient. From a dimensional analysis, this Wilson coefficient can only be either a constant or exhibit a $\log(|x|)$ dependence, neither of which is sufficient to induce a divergence when integrating over $x$.
The case where $x\to y \to 0$ is more subtle. 
By rescaling $x$ and $y$ as
\ba 
x\to \ell x, \qquad  y\to \ell y,
\ea 
the relevant limit corresponds to $\ell \to 0$.
In the OPE of the product of six quark fields $ \overline{u}(0)\gamma_\sigma d(0) \overline{d}(y)\gamma_\mu u(y) \overline{v} (x)\gamma_\nu r(x) $, two quark flavours can be reduced to two propagators resulting in a $\ell^{-6}$ behaviour as $\ell \to 0$. 
From the photon propagator we get another factor $\ell^{-2}$. 
Taking the integral measure with $\ell^8$ one would still expect a logarithmic divergence in the case where the two valence quark loops are not connected by gluons. However, in the relevant case,  
the valence quark loops are connected by (at least two) gluon lines, 
as guaranteed by the VEV-subtraction Eq.~\eqref{eq:vev_subtr}.
Hence, an additional factor of the strong coupling squared, $\alpha^2_s(\mu/\ell)$, comes into play. This factor behaves asymptotically as $\alpha^2_s(\mu/\ell) \sim 1/\log^2(\mu/\ell)$, which is sufficient to make the result UV-finite.
By similar arguments, one can also generalize this result to the other diagrams mentioned earlier.

In this work, we will mainly focus on the $(2+2)a$ diagram as it alone is a well-defined, scheme-independent quantity.
On the contrary, we do not attempt a full calculation for the $(2+2)b$ contribution, as this contribution is UV-divergent when the regulator on the internal photon is removed and one needs to work with a specific QED+QCD renormalization scheme -- possibly combining with other UV-divergent diagrams under the same regularization -- to obtain a physical result;
instead, in section~\ref{sec:tptb}, we evaluate the $(2+2)b$ diagram at finite photon-regulator cutoffs and compare our lattice results to model calculations at a qualitative level.

\section{Lattice setup}
\label{sect:nlohvp_setup}
\begin{table}[]
\footnotesize
\centering
\caption{The parameters of the ensembles used in the analysis, generated by the CLS consortium. The lattice spacing $a$, the pion and kaon masses are taken from Ref.~\cite{Kuberski:2024bcj} based on the determination in Refs.~\cite{Strassberger:2021tsu,RQCD:2022xux}. The asterisk $*$ indicates that an ensemble is used to check for finite-size effects and does not enter the chiral extrapolation for obtaining the final result. On each ensemble we use \#confs gauge configurations and \#sources indicates the number of source points for propagator inversion, see Sec. \ref{sect:nlohvp_setup}.}
\label{table:nlohvp_ensemble}
\begin{tabular}{|c|c|c|c|c|c|c|c|c|c|}
\hline
Id   & $\beta$ & $N_L^3 \times N_T$& $a$ [fm] & $m_\pi$ [MeV] & $m_K$ [MeV] & $m_\pi L$ & $L$ [fm] & \#sources  &\#confs\\ \hline
H101 & 3.4 & $32^3 \times 96$ & 0.0849(9)  & 424(5) & 424(5) & 5.8 &  2.7 & 32  &182\\
N101 &     & $48^3 \times 128$ &   & 282(4) & 468(5) & 5.8 &  4.1 & 48 &200\\
H105$^*$ &     & $32^3 \times 96$ &   & 283(4) & 470(5) & 3.9 &  2.7 & 64  &180\\
C101 &    & $48^3 \times 96$             &    & 222(3)            & 478(5)          & 4.6      & 4.1          & 96    &200\\
S100$^*$&    & $32^3 \times 128$             &    & 222(3)            & 478(5)          & 2.9      & 2.7          & 128&100\\ \hline
B450 & 3.46    & $32^3 \times 96$             & 0.0751(8)    &  422(5)            & 422(5)          & 5.1       & 2.4         & 32  &200\\
N451 &     & $48^3 \times 128$             &    & 291(4)            & 468(5)          & 5.3       & 3.6          & 48                                                        &200\\ 
D450 &     & $64^3 \times 128$             &     & 219(3)            & 483(5)          & 5.3       & 4.8          & 64                                                         &200\\ \hline
H200$^*$ & 3.55    & $32^3 \times 96$             & 0.0635(6)    & 423(5)            & 423(5)          & 4.4       & 2.0          & 32                                                         &200\\ 
N202 &    & $48^3 \times 128$             &   & 418(5)            & 418(5)          & 6.5       & 3.0          & 96                                                         &200\\ 
N203 &     & $48^3 \times 128$             &    & 349(4)            & 447(5)          & 5.4       & 3.0          & 48                                                         &200\\
 E250& & $96^3 \times 192$             & & 132(2)& 495(6)& 4.1& 6.1& 192  &200\\ \hline
 N300 & 3.7    & $48^3 \times 128$             & 0.0491(5)    & 425(5)            & 425(5)          & 5.1       & 2.4         &48                                                         &200\\
E300 & & $96^3 \times 192$             &     & 177(2)            & 497(6)          & 4.2       & 4.7          & 192                                                       &200\\ \hline
\end{tabular}
\end{table}

The lattice calculations of this work are performed on gauge ensembles generated by the Coordinated Lattice Simulations (CLS) consortium.
The ensembles are generated with non-perturbatively-O$(a)$-improved Wilson-Clover fermion action with $N_{\rm f} = 2 + 1 $ dynamical flavors and tree-level-O$(a^2)$-improved L\"uscher-Weisz gauge action. 
The parameters for the ensembles included in this work are given in table~\ref{table:nlohvp_ensemble}.
The four vector currents appearing in Eq.~\eqref{eq:ccs_nlohvp} are treated differently on the lattice for the present calculation: we use the conserved discretization 
\begin{eqnarray}
    j_\mu^{\textrm{(C)}}(x) &=& {\textstyle\frac{1}{2}} \left(j_\mu^{\textrm{(N)}}(x) + j_\mu^{\textrm{(N)}}(x-a\hat\mu) \right),
  \\
  j_\mu^{\textrm{(N)}}(x) &=& \frac{1}{2}\Big[\bar{\psi}(x+a\hat{\mu}) (1+\gamma_\mu)U_\mu^\dagger(x) {\cal Q}{\psi}(x)
	 -\bar{\psi}(x)(1-\gamma_\mu)U_\mu(x){\cal Q}{\psi}(x+a\hat{\mu})\Big]\,,
\label{eq:conserved_vector_current}
\end{eqnarray}
for those located at $y$ and $z$, where $\hat{\mu}$ is a unit vector in the $\mu$-th direction, and the local discretization,
\begin{equation} 
  j_\mu^{\textrm{(L)}}(x) =  
  \bar{\psi}(x)\gamma_\mu {\cal Q} {\psi}(x)\,,
  \label{eq:local_vector_current}
\end{equation}
for the ones at the origin (`0') and $x$.
Here, $\mathcal{Q}$ is the e.m.\ charge matrix leading to Eq.~\eqref{eq:jem} in the continuum limit.
The local vector current needs to be renormalized (`R') multiplicatively from its O($a$)-improved version (`impr.')~\cite{Gerardin:2018kpy},
\ba\label{eq:renorm}
j^{({\rm L}),\textrm{R}}_\mu(x) = \hat{Z}_{\rm V}^{\textrm{(L)}}\, j^{{\textrm{(L)-impr.}}}_\mu(x).
\ea
In this work, we will use the coefficient $\hat{Z}_{\rm V}$ computed in Ref.~\cite{Gerardin:2018kpy} and neglect the additive O($a$)-improvement.

\subsection{Implementation of the two-point functions}
To get the quantities Eq.~\eqref{eq:2+2_master_a} and Eq.~\eqref{eq:2+2_master_b} for the $(2+2)a$ and $(2+2)b$ diagram respectively, we first evaluate Eqs.~\eqref{eq:Ii}
on the lattice at fixed $|x|$, and then calculate the last one-dimensional $|x|$-integral as part of our analysis program. 
We perform multiple measurements with different choices for the origin and $x$ leading to the same norm $|x|$.
The two-point functions $\Pi^f_{\mu \nu}$ with different discretized currents are constructed with point-source propagators.
Similar to Ref.~\cite{Chao:2023lxw}, we place the point sources along the diagonal $(n,n,n,N_T/2)$ for $n=\{0,1,\dots,N_L\}$ for each ensemble to efficiently increase the number of samples for each $|x|$.
On the ensembles E250, E300, C101, N202, H105, we additionally compute point-source propagators on a second spatial diagonal along the direction $(-n,n,n,N_T/2)$ to further increase the statistics. For the ensemble S100, given the small number of available gauge configurations, we also generated data with origins along the $(n,-n,n,N_T/2)$ and $(n,n,-n,N_T/2)$ direction. 
The total numbers of point sources used per configuration for each ensemble are given under \#sources in table~\ref{table:nlohvp_ensemble}.
These point-source propagators can be recycled for the measurement with the origin and $x$ swapped, giving therefore in total $2 \times$(\#sources) measurements for each value of $|x|$.

For all the lattice results presented in this work, the errors are estimated using the jackknife procedure.
The VEV-subtractions in Eq.~\eqref{eq:2+2_master_a} and Eq.~\eqref{eq:2+2_master_b} are performed on each jackknife subset to account for the correlation properly.

For the main focus of this work, the $(2+2)a$ contribution, since we know that it is UV finite, we can drop the Pauli-Villars regulator in the photon-propagator, effectively taking $\Lambda \to \infty$ directly in Eq.~\eqref{eq:pv}. 
However, in order to obtain a definite result, one needs to treat the value of Eq.~\eqref{eq:pv} in the $x\to y$ limit.
To this end, we use a modified version of the propagator in the $\Lambda\to \infty$ limit, which reads
\be
G_0^{[\lambda]}(x-y) = \begin{cases}\displaystyle
    \frac{1}{4\pi^2|\lambda a|^2} , \quad |x-y| = 0\\
    \displaystyle \frac{1}{4\pi^2|x-y|^2}, \quad \textrm{else} 
\end{cases}
\ee 
where $a$ is the lattice spacing. 
We performed calculations for $\lambda=\{0.25, 0.5, 0.75\}$ on the ensemble N203 and have found no significant change in the result within the achieved statistical precision. 
We expect a suppression by the integration measure at the $|x-y|\to 0$ limit in order for the $(2+2)a$ to be UV-finite.
For the rest of the paper, $\lambda=0.5$ will be used without further mention.

\subsection{Finite-size effects and comparison between different CCS kernels}
\label{sect:nlohvp_fv}
\begin{figure}[h]
\begin{subfigure}{0.48\textwidth}
    \centering
    \includegraphics[width=0.99\textwidth]{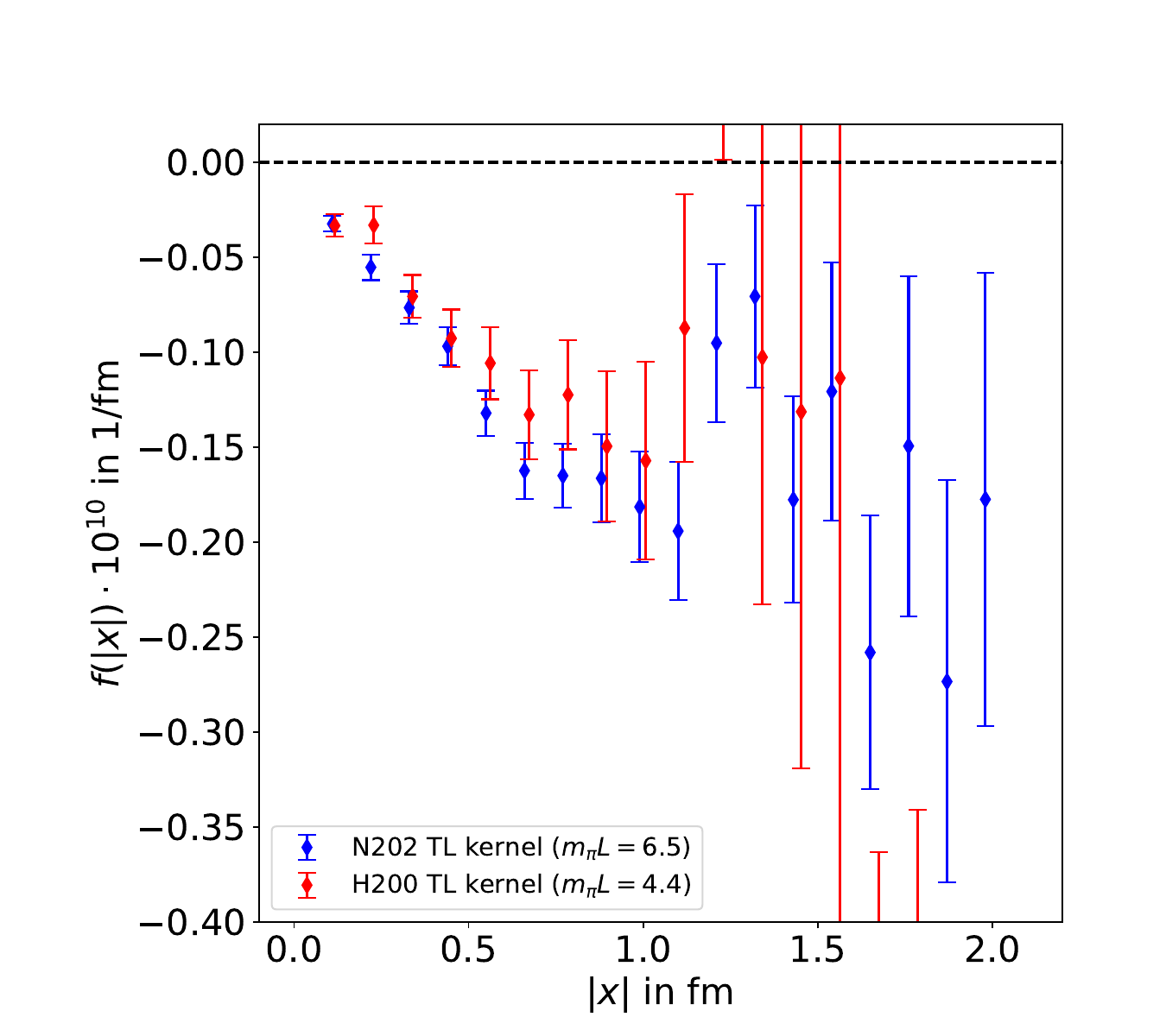}
    \caption{`TL' kernel, $m_\pi= 418(5)$ MeV}
\end{subfigure}
\begin{subfigure}{0.48\textwidth}
    \centering
    \hspace{-1.0cm}
    \includegraphics[width=0.99\textwidth]{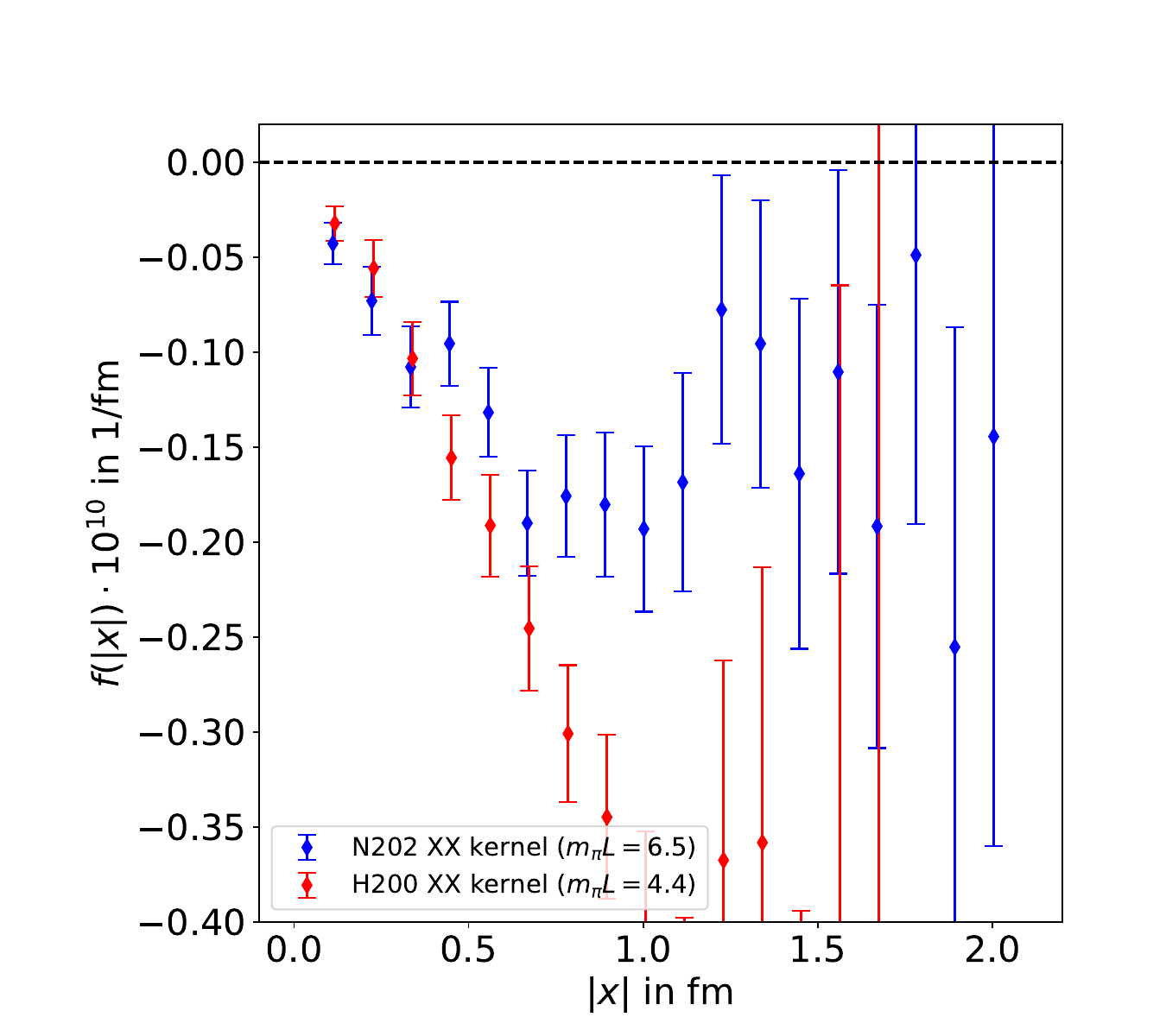}
    \caption{`XX' kernel $m_\pi=418(5)$ MeV}
\end{subfigure}
\begin{subfigure}{0.48\textwidth}
    \centering
    \includegraphics[width=0.99\textwidth]{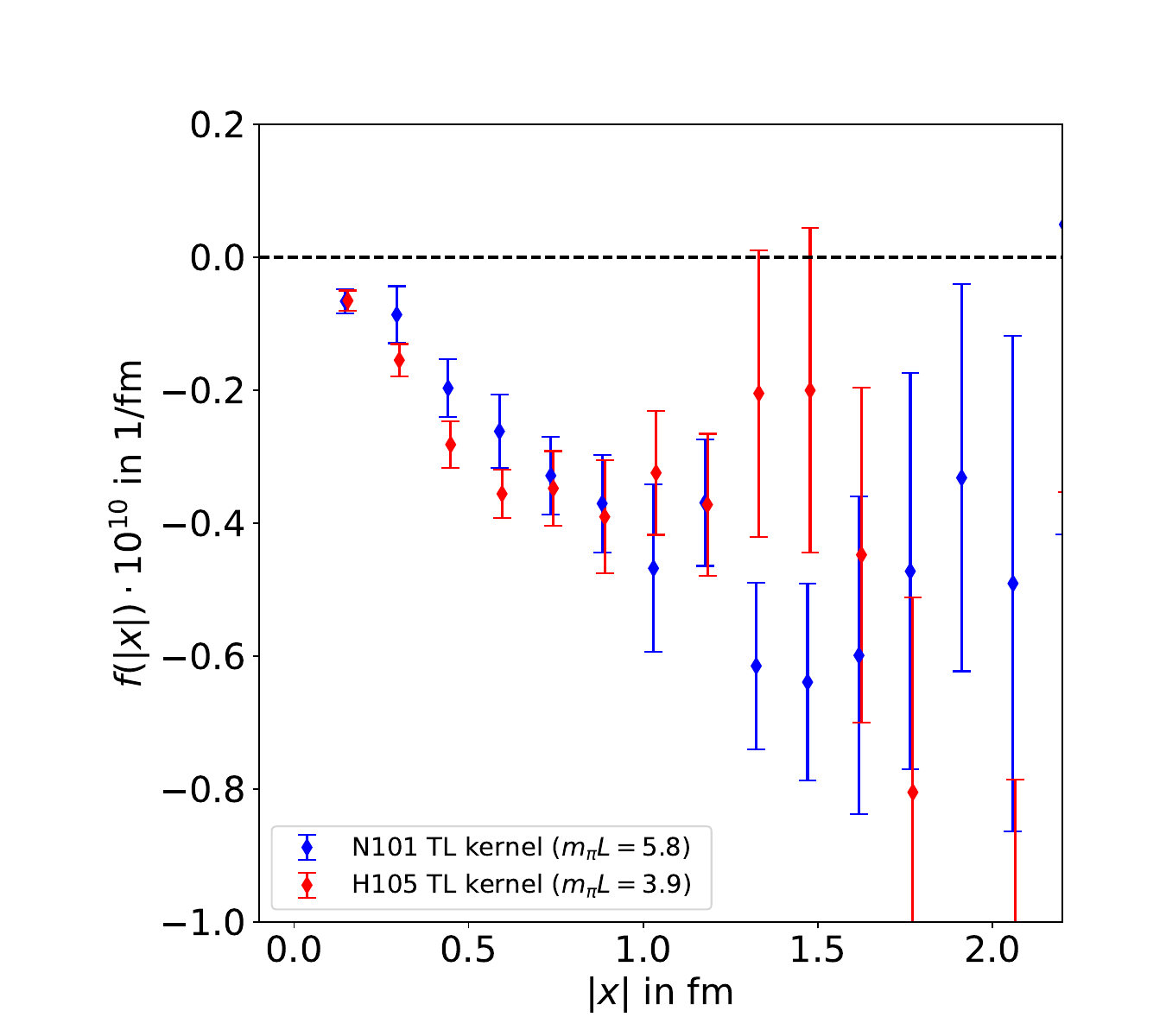}
    \caption{`TL' kernel $m_\pi=282(4)$ MeV}
\end{subfigure}
\begin{subfigure}{0.48\textwidth}
    \centering
    \includegraphics[width=0.99\textwidth]{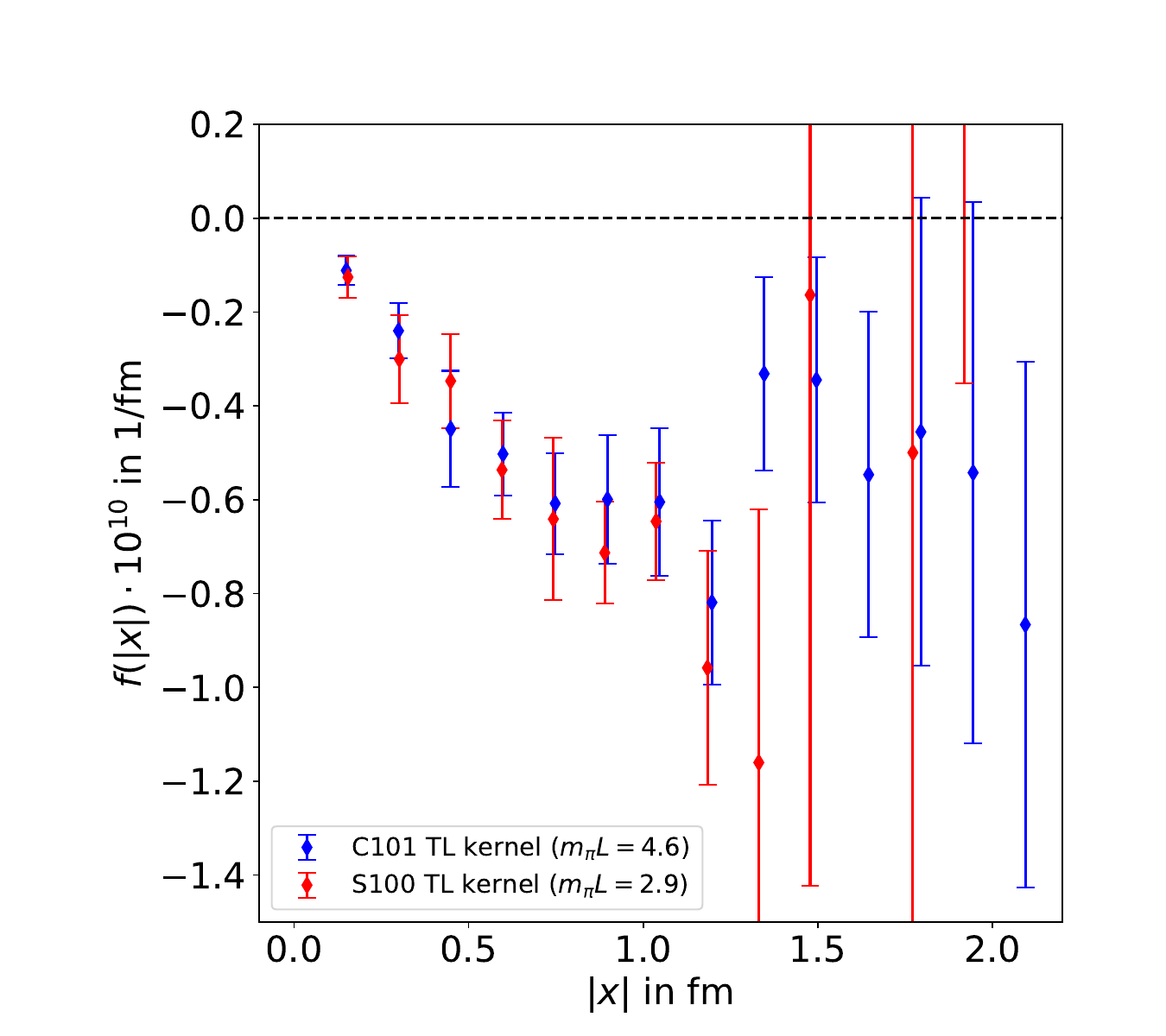}
    \caption{`TL' kernel $m_\pi= 222(3)$ MeV}
\end{subfigure}
\caption{Top: Comparison of the integrand of Eq.~\eqref{eq:2+2_master_a} between the ensembles H200 ($m_\pi L =4.4 $) and N202 ($m_\pi L=6.5$) for the `TL' kernel (left) and the `XX' kernel (right), defined in Eqs.~\eqref{eq:tl_kernel} and \eqref{eq:xx_kernel}.
Bottom: Comparison of the integrands for the `TL' kernel between ensembles N101 ($m_\pi L=5.8$), H105 ($m_\pi L=3.9$) (left) and between, C101 ($m_\pi L=4.6$) and S100 ($m_\pi L=2.9$) (right). }
\label{fig:nlohvp_fv}
\end{figure}

As discussed in section~\ref{sec:disco_def}, the e.m.\ kernel is not uniquely defined and this degree of freedom allows us to modify the profile of the integrand to get Eq.~\eqref{eq:2+2_master_a}.
In this work, our strategy is to choose between the `TL'-kernel and the `XX'-kernel defined in Eq.~\eqref{eq:tl_kernel} and Eq.~\eqref{eq:xx_kernel} the one with more suppressed finite-volume effects.
To this end, we compare the integrand of the $(2+2)a$ contribution Eq.~\eqref{eq:2+2_master_a} between the ensembles N202 and H200. 
Both ensembles are generated at the SU$(3)$ flavour symmetric point with the same value of coupling $\beta$, thus the same lattice spacing.
However, H200 has a significantly smaller $m_\pi L =4.4 $ compared to $m_\pi L=6.5$ for N202.  
Figure~\ref{fig:nlohvp_fv} shows this comparison for the `TL' and `XX' kernel. 
We observe that the integrands for the larger N202 computed from both kernels agree reasonably well. 
For the `TL' kernel the results computed on the smaller ensemble H200 are in good agreement with the results calculated on N202 up to $|x|\approx 1$ fm, which is about half the size of the spatial dimension of H200. 
Effects due to the finite-size become more significant for $|x|>1$ fm. 
For the `XX' kernel, however, the finite-size effects become significant at $|x|>0.75$ fm.
In addition to the ensembles at the SU$(3)$ flavour symmetric point, we perform a similar study on two ensembles at $m_\pi= 282(4)$ MeV, N101 ($m_\pi L = 5.8$) and H105 ($m_\pi L = 3.9$), and on two ensembles at $m_\pi = 222(3)$ MeV, C101 ($m_\pi L = 4.7$) and S100 ($m_\pi L = 2.9$). 
For these smaller pion masses, the behaviour for the `TL' kernel is consistent with the result at the SU$(3)$ flavour symmetric point: for $|x|<L/2$, there is no visible difference in the integrand computed with different box sizes within the statistical uncertainty. 
Moreover, the results with the `TL' kernel have smaller statistical error for each of the data points at small $|x|$. 
We therefore choose to work with the `TL' kernel in the rest of this work.

\section{Phenomenological model}
\label{sec:phenomenological_model}
While our formalism Eq.~\eqref{eq:ccs_nlohvp} offers a way to determine $\ahvpnlo$ from first principles, lattice calculations are limited by deteriorating signal quality at large distances. 
However, in this region, the dominant contributions to the observable are expected to be the light degrees of freedom as in the case of the HLbL calculation \cite{Chao:2021tvp}. In this section, we attempt to study those contributions -- namely the contributions from exchanging a single $\pi^0/\eta/\eta^\prime$ meson and the charged pion and kaon loops -- to $\ahvpnlo $ according to Eq.~\eqref{eq:ccs_nlohvp} using phenomenological models.
The study of the latter presented in this section serves two purposes for our lattice calculation of the $(2+2)a$ diagram: first, to reconstruct the tail of the integrand in Eq.~\eqref{eq:2+2_master_a}; and second, to determine the pion-mass dependence of the dominant $a_\mu^{(2+2)a-ll}$ contribution, thereby constraining the interpolation to the physical point.

In section \ref{sect:hvpnlo_modelparam}, we investigate the pseudoscalar meson contributions based on our earlier work in Ref. \cite{Biloshytskyi:2022ets}. We give the parameters that correspond to the ensembles on our chiral trajectory for the $\pi^0$, $\eta$ and $\eta'$ contribution.
In section \ref{sec:piloop-cont}, we compute the contribution from the charged pion loop in scalar QED, including vector-meson-dominance (VMD) parametrization for the form factors, followed by a discussion of the counterterms for this contribution that are obtained for different renormalization prescriptions in section \ref{sec:pure_sQED}. Lastly in section \ref{sec:model_precition_full_em}, we give a prediction for the total e.m. corrections to the HVP 
based on the phenomenological models studied in this section.

An important observation of this model study is that in the range $120$ MeV $<m_\pi< 450$ MeV, the contribution from the charged pion loop using VMD form factor can be well described by a simple power-law in the pion mass, $a_\mu^{\pi^+\pi^-} \propto m_\pi^{-3}$. Starting from the $SU(3)$ flavour symmetric point and going down to the physical point one sees a huge enhancement of the charged-pion-loop contribution compared to the pseudoscalar-meson-exchange contribution.
This is can be seen later explicitly in figure \ref{fig:tpta_pheno_full_model}.

\subsection{Pseudoscalar meson exchange contributions}
\label{sect:hvpnlo_modelparam}
A calculation of the $\pi^0$-exchange contribution to $\ahvp$, entering at O$(\alpha^3)$, was already carried out in Ref.~\cite{Biloshytskyi:2022ets}. 
Using a vector-meson-dominance (VMD) parametrization for the pion transition form factor
\be\label{eq:vmdtff}
\mathcal{F}_{\pi^0\gamma^*\gamma^*}(-p^2,-k^2) = \frac{m_V^4 F_\pi}{(p^2+m_V^2) (k^2+m_V^2)}.
\ee
Using physical parameters $m_V=775.26$ MeV and $F_\pi=0.274$ GeV$^{-1}$, one obtains the value $a_\mu^{\pi^0}=0.37(1) \times 10^{-10}$.
The O($\alpha^3$)-contributions of $\eta$ and $\eta^\prime$ to $\ahvp$ 
can be calculated based on the same techniques applied previously to the $\pi^0$ case. 
To give numerical estimates of these contributions, we will again make use of the VMD parametrization Eq.~\eqref{eq:vmdtff} for the $\eta,\eta^\prime\to \gamma^*\gamma^*$ transition form factors, with appropriate VMD masses $m_{V,\eta/\eta^\prime}$ and normalization constants $F_{\eta/\eta^\prime}$ to be determined.

We shall first give phenomenological estimates for the $\eta$ and $\eta^\prime$ contributions.
The normalization factors can be extracted from the measured $\eta,\eta^\prime\to \gamma \gamma$ decay widths~\cite{PDG}
\ba 
\label{eq:F_P_from_decay}
\Gamma (P\to \gamma \gamma) = \frac{\pi \alpha^2m_P^3}{4}|F_{P}|^2\,,\quad P=\eta,\eta^\prime\,
\ea 
from which we get $F_\eta = 0.275(5)$ GeV${}^{-1}$ and $F_{\eta^\prime} = 0.343(5)$ GeV${}^{-1}$.
On the other hand, the VMD masses $m_V$  for the $\eta$ and $\eta'$ mesons have been determined at the physical point by fitting the CLEO data~\cite{Nyffeler:2016gnb}
\ba 
m^{\text{phys}}_{V,\eta}= (774 \pm 29) \text{ MeV}, \qquad  m^{\text{phys}}_{V,\eta'}= (859\pm 28) \text{ MeV}\,.
\ea 
Altogether, we find
\ba 
\label{eq:apme_eta}
a_\mu^{\eta} &=& 0.20(1) \times 10^{-10}\,,\\
\label{eq:apme_eta_p}
a_\mu^{\eta'} &=& 0.23(1) \times 10^{-10}\,.
\ea
Note that the total contribution from the $\eta$ and $\eta^\prime$ obtained in this way is comparable to the contribution from $\pi^0$, despite the latter being much lighter.

For our purposes, we would like to have predictions on the $\pi^0$, $\eta$ and $\eta^\prime$ exchange contributions to $a_\mu^{\textrm{hvp1}\gamma^*}$ to compare with the lattice results for each ensemble included in the analysis.
To this end, we need to determine $(m_{P}, m_{V,P}, F_{P} )$ for $P=\pi^0,\eta,\eta^\prime$ corresponding to each ensemble.
For this work, we take 
the masses of the three pseudoscalar mesons from the determinations in Refs.~\cite{Strassberger:2021tsu,RQCD:2022xux,Bali:2021qem}.
Our $F_{\pi}$ and $m_{V,\pi}$ are obtained from a fit to the data from Ref.~\cite{gerardin:2019rua}, where the fit is restricted to the single-virtual case~\cite{Chao:2020kwq}. 
However, $F_{\eta}$, $F_{\eta^\prime}$, $m_{V,\eta}$ and $m_{V,\eta^\prime}$  are obtained in a more heuristic way by interpolating their values between the physical and the SU$(3)_{\rm f}$-symmetric point, corresponding to the ensemble N202 in table \ref{table:model_param}, linearly in the variable $\xi = m_K^2-m_\pi^2$. At the physical point we use Eqs.~\eqref{eq:apme_eta} and ~\eqref{eq:apme_eta_p} and $F_{\eta}$ and  $F_{\eta^\prime}$ obtained from Eq.~\eqref{eq:F_P_from_decay}.
At the SU$(3)_{\rm f}$-symmetric point, where the $\pi^0$ and $\eta$ are mass-degenerate we use 
\ba
F^{\textrm{SU}(3)}_{\eta} = \frac{F^{\textrm{SU}(3)}_{\pi}}{\sqrt{3}} , \qquad F^{\textrm{SU}(3)}_{\eta'} = \frac{2\sqrt{2}F^{\textrm{SU}(3)}_{\pi}}{\sqrt{3}}.
\ea 
While in the case of the $\eta$, this relation is a direct consequence of SU$(3)$-flavour symmetry, for the $\eta'$ it is an approximation justified at large $N_c$ that amounts to neglecting the disconnected diagram in the calculation of 
$F^{\textrm{SU}(3)}_{\eta'} $
We furthermore use $m^{\textrm{SU}(3)}_{V,\pi} = m^{\textrm{SU}(3)}_{V,\eta}$ due to the mass-degeneracy of $\pi^0$ and $\eta$.
For $m_{V,\eta'}$, whose value we do not know at the SU$(3)$ symmetric point, we assume the same slope in $\xi$ as for $m_{V,\eta}$. 
The model parameters calculated using the described procedure for the relevant ensembles are summarized in table~\ref{table:model_param}. Results for the pseudoscalar meson contributions along the chiral trajectory are summarized in table \ref{table:pme_results}.

\begin{table}[]
\small
\centering
\caption{The parameters, which define the chiral trajectory on the lattice ensembles from table \ref{table:nlohvp_ensemble}. The pion masses are determined in Ref.~\cite{Strassberger:2021tsu,RQCD:2022xux}. The masses for $\eta$ and $\eta'$ are obtained from Ref.~\cite{Bali:2021qem}. The other parameters are calculated from the procedure described in section~\ref{sect:hvpnlo_modelparam}. We use the same model parameters for all ensembles at roughly the same pion mass. For the ensemble E250, with a pion mass just below the physical point, we use the physical values for all model parameters other than the pion mass.
}
\label{table:model_param}
\begin{tabular}{|c|c|c|c|c|c|c|}
\hline
Id                      & N202     & N203     & N451     & C101    & E300        & physical              \\ \hline
$m_\pi$ {[}MeV{]}       & 418(5) & 349(4)   & 291(4)   & 222(3) & 177(2)      & 134.9768(5)        \\
$m_{V,\pi}$ {[}MeV{]}   & 952(15)       & 916(12)  & 869(15)  & 812(5)        & 785(4)      & 775.26(25)   \\
$F_\pi$ {[}1/GeV{]}     & 0.227(5)      & 0.238(5) & 0.249(5) & 0.264(4)      & 0.269(3)    & 0.274(3)    \\
\hline
$m_\eta$ {[}MeV{]}      & 418(5) & 487(6)   & 531(5)   & 523(9)        & 547.862(18) & 547.862(18)  \\
$m_{V,\eta}$ {[}MeV{]}  & 952(15)       & 894(15)  & 852(18)  & 807(24)       & 784(28)     & 774(29)        \\
$F_\eta$ {[}1/GeV{]}    & 0.131(2)      & 0.178(4) & 0.212(4) & 0.248(5)      & 0.260(6)     & 0.275(5)      \\
\hline
$m_{\eta'}$ {[}MeV{]}   & 1028(136)     & 876(159) & 941(113) & 980(76)       & 957.78 (6)  & 957.78 (6)    \\
$m_{V,\eta'}$ {[}MeV{]} & 1036(28)      & 978(21)  & 936(20)  & 891(24)       & 869(27)     & 859(28)        \\
$F_{\eta'}${[}1/GeV{]}  & 0.371(8)      & 0.362(6) & 0.355(5) & 0.348(5)      & 0.345(5)    & 0.343(5)      \\ \hline
\end{tabular}
\end{table}

\subsection{Continuum calculation of the charged pion loop}\label{sec:piloop-cont}
In scalar QED (sQED),
the charged-pseudoscalar-meson-loop contribution to the e.m.\ correction to the HVP is divergent when the UV regulator of the internal photon is absent.
We perform a calculation in momentum-space using the light-by-light scattering amplitude calculated in scalar QED in the continuum, based on the work in Ref.~\cite{Biloshytskyi:2022ets}. A similar calculation was also carried out in Ref.~\cite{Bijnens:2019ejw}.
In this approach, the charged pion loop can be regularized with a double Pauli-Villars regulated photon propagator. 
For large momenta, this regularization leads to the same asymptotic behaviour as if one introduces a VMD form factor for the  $\pi \pi \gamma $ interaction.

We employ the formalism developed in \cite{Biloshytskyi:2022ets} to derive the two-loop vacuum polarization $\Pi_\mathrm{4pt}$ in sQED using the Cottingham-like formula. This approach, which has been validated for providing a comprehensive set of two-loop corrections to vacuum polarization in QED, is now extended to the scalar case. Specifically, we apply the master formula for $\Pi_\mathrm{4pt}$ in the dispersive representation,
\begin{align}
\Pi_{{\rm 4pt}}(Q^2,\Lambda)
=& \frac{1}{3(2\pi)^3 Q^2}\int\limits_0^\infty d K^2 \,
K^2\,\left[\frac{1}{K^2}\right]_{\Lambda}\,
\Bigg[ \frac{\pi}{4}
\mathcal{M}(\bar\nu ,K^2,\,Q^2)\nn
&+ \int\limits_{\nu_\mathrm{thr}}^\infty d \nu  \left(\frac{2}{\nu+\sqrt{X}} - \frac{\nu}{\nu^{ 2} -\bar{\nu}^2 }\right) \sqrt{X}\,  \sigma(\nu,K^2,\,Q^2) 
\Bigg].
\label{eq:DRform}
\end{align}
This formula is written in terms of the dispersive part given by the integral of the unpolarized $\gamma^*\gamma^*$-fusion cross section, $\sigma = 4\sigma_{TT}-2\sigma_{LT}-2\sigma_{TL}+\sigma_{LL}$, which starts from the first inelastic threshold $\nu_\mathrm{thr}$, and the subtraction function -- the forward doubly-virtual LbL amplitude $\mathcal{M}$ evaluated at the subtraction point $\nu=\bar\nu$ (cf. appendix~\ref{app:SQEDamps} for exact expressions). Here  $K^2=-k^2$ and $Q^2=-q^2$ are virtualities of the scattered photons with momenta $k$ and $q$, correspondingly, denoting $\nu = k\cdot q$ and $X = \nu^2-Q^2K^2$.
This form is particularly well-suited for practical calculations, yielding stable numerical results. We perform the calculations at the subtraction point $\bar\nu = KQ$, since the corresponding subtraction function $\mathcal{M}(\bar\nu=KQ,K^2,Q^2)$ has a simple analytical expression (cf. appendix~\ref{app:SQEDamps}).

The  $O(\alpha^3)$ contribution to the muon $(g-2)$ that stems from the Cottingham formula without the counterterm reads
\begin{equation}
    a_\mu
    ^{\pi^+ \pi^-}
    = -\frac{\alpha}{2\pi}\Pi_\mathrm{4pt}(0,\Lambda)+\frac{\alpha}{\pi}\int_0^\infty dQ^2\mathcal{K}(Q^2)\Pi_\mathrm{4pt}(Q^2,\Lambda).
    \label{eq:AMMCottingham}
\end{equation}
The subtraction term $\Pi_\mathrm{4pt}(0,\Lambda)$ is given in Eq.~\eqref{eq:Pi0subtractionS} in appendix~\ref{app:CottinghamVerification}.
The continuum QED kernel $\mathcal{K}$ is given by
\begin{equation}
\mathcal{K}(Q^2) = \frac{1}{2m_\mu^2}\frac{(v-1)^3}{2v(v+1)},\quad v=\sqrt{1+\frac{4m_\mu^2}{Q^2}}.
\end{equation}

In order to regularize the pion loop behavior in our continuum calculation, we apply two different regularization schemes.
The first scheme -- the double Pauli-Villars regularization (dPV) of the internal photon propagator -- directly corresponds to the regularization proposed in Ref.~\cite{Biloshytskyi:2022ets} for a calculation of the one-photon irreducible QED effects to the HVP contribution in lattice QCD:
\begin{equation}
    \left[\frac{1}{K^2}\right]^\mathrm{dPV}_\Lambda = \frac{\zeta^2\Lambda^4}{K^2(K^2+\Lambda^2)(K^2+\zeta^2\Lambda^2)}, \quad \zeta = \frac{1}{\sqrt{2}}. \label{eq:doubly_pv}
\end{equation}
The second scheme -- the vector-meson-dominance regularization (VMD) --  incorporates the monopole VMD form factors,
\begin{equation}
    \mathcal{F}_{\pi^{\pm},\mathrm{VMD}}(Q^2,m_V) = \frac{1}{Q^2/m_V^2+1},
    \label{eq:VMDff}
\end{equation}
at each electromagnetic vertex of the pion loop.  This modification adjusts  Eq.~\eqref{eq:DRform} as follows:
\begin{equation}
    \left[\frac{1}{K^2}\right]^\mathrm{VMD}_{\Lambda\equiv m_V} = \frac{1}{K^2} \mathcal{F}_{\pi^{\pm},\mathrm{VMD}}^2(Q^2,m_V)\mathcal{F}_{\pi^{\pm},\mathrm{VMD}}^2(K^2,m_V),
    \label{eq:VMDreg}
\end{equation}
Notably, in contrast to the dPV regularization, the VMD regularization specifically impacts the $Q^2$-dependent component of the vacuum polarization.

Both dPV and VMD regularizations ensure the finiteness of the Cottingham formula.
The results obtained from evaluating  Eq.~\eqref{eq:AMMCottingham} under these regularization schemes, using parameters from the relevant lattice ensembles, are summarized in table~\ref{table:PionLoop}.

We note that the results for the charged pion loop using a VMD form factor, see table \ref{table:PionLoop}, can be well described by a simple power law $\propto m_\pi^{-3} $ in the region $120$ MeV $<m_\pi< 450$ MeV. While this is certainly not the expected asymptotic behavior in the chiral limit, 
empirically we found it to hold to a good approximation for pion masses in this intermediate region.

\begin{table}[h]
\centering
\caption{The pion-loop contribution calculated via the Cottingham formula in the continuum, evaluated in VMD and dPV regularization schemes. For the VMD regularization, the vector meson masses for each ensemble were taken from table~\ref{table:model_param}. We also provide the results evaluated at the physical mass of the neutral and charged pion using the physical $\rho^0$-meson mass in the VMD regularization. Both rows for dPV are used to cross check the lattice calculation of the charged pion described in section \ref{sec:chargedpion_lat}.}
\label{table:PionLoop}
\begin{tabular}{|c|c|c|c|c|c|c|c|c|}
\hline
$a_\mu^{\pi^+ \pi^-}$ $\times 10^{10}$ & N202 & N203 & N451 & C101 & E300 & E250& $m_{\pi^0}$&$m_{\pi^\pm}$\\ \hline
VMD & $-$0.22 & $-0.37$ & $-$0.60 & $-$1.25 & $-$2.39 & $-$5.85& $-$5.46&$-$4.92\\ \hline
dPV, $\Lambda=16 m_\mu$ & $-$0.15  & $-$0.33 & $-$0.68 & $-$1.94  & $-$4.57  & $-$13.47& $-$12.42& $-$10.99\\
dPV, $\Lambda=24 m_\mu$ & $-$0.36  & $-$0.78 & $-$1.57 & $-$4.44  & $-$10.38  & $-$30.46& $-$28.10&$-$24.88\\
\hline
\end{tabular}
\end{table}

\subsection{Renormalization of the charged pion loop}
\label{sec:pure_sQED}
It is interesting to compare the results from table \ref{table:PionLoop} with the renormalized, pure-sQED result for the electromagnetic corrections to $\ahvp$ in the on-mass-shell scheme (OMS). In this scheme, the entire effect of the photons on the pion pole mass is removed by a suitable mass counterterm.
Evaluated at the physical pion mass, corresponding to either the neutral or the charged pion, the sQED result reads
\begin{subequations}
\begin{align}
        a_{\mu,\mathrm{OMS}}^{\pi^+\pi^-}(m_{\pi^\pm}^\mathrm{}) = 0.700\times 10^{-10},\label{eq:sQEDrenPlus}\\
        a_{\mu,\mathrm{OMS}}^{\pi^+\pi^-}(m_{\pi^0}^\mathrm{}) = 0.742\times 10^{-10}\label{eq:sQEDrenNeutral}.
\end{align}
\label{eq:sQEDren}
\end{subequations}
We note that these values are small and positive owing to the contribution of the pion-mass counterterm.
It has become customary for lattice practitioners to expand around the isoQCD world with the pion mass assuming the value of the physical \emph{neutral} pion mass~\cite{FlavourLatticeAveragingGroupFLAG:2024oxs}.
After including the effects of photons, the charged pion mass should assume its physical value.
In order to account for this choice of the `FLAG' renormalization scheme, one can compute the (finite) difference of the counterterms $\Delta \text{CT}$ evaluated in the on-shell scheme and the FLAG scheme using Eq.~\eqref{eq:QEDct}

\ba 
\nonumber
    \Delta \text{CT}(m_{\pi^0}) &=& \text{CT}_{\mathrm{OMS}}^{}(m_{\pi^0})-\text{CT}_{\text{FLAG}}^{}(m_{\pi^0}) \\
    &=& 2 m_{\pi^0}
    \Delta m_\pi \left[\frac{\alpha}{\pi}\int_0^\infty dQ^2 \mathcal{K}(Q^2) \frac{\partial}{\partial 
    \nonumber
    m_{\pi}^2}\overline{\Pi}_{e^2}(Q^2,m_{\pi})\right]_{ m_\pi=m_{\pi^0}} \\
    &=& -4.49\times 10^{-10},
    \label{eq:AddCT}
\ea 
where $\Delta m_\pi\equiv m_{\pi^\pm}-m_{\pi^0}$ and $\overline{\Pi}_{e^2}$ is the one-loop sQED vacuum polarization subtracted at $Q^2=0$.
Again, this additional term (as compared to the customary on-shell scheme used in QED) describes the effect of the charged pion mass shifting from 134.97 to 139.57\,MeV. Thus, the final, pure sQED result corresponding to the FLAG renormalization scheme used by lattice practitioners reads
\begin{equation}
    a_{\mu,{\rm FLAG}}^{\pi^+\pi^-} = a_{\mu,\mathrm{OMS}}^{\pi^+\pi^-}(m_{\pi^0})+  \Delta \text{CT}^{}(m_{\pi^0}) = -3.75 \times 10^{-10}.
\end{equation}

In contrast to the above case, using sQED with VMD form factors associated with the electromagnetic vertices, the counterterm in the on-shell renormalization scheme is finite and reads 
 \begin{align}
\text{CT}_{\mathrm{OMS}}^{}(m_{\pi^0},m_V) &= \Sigma_2(m_{\pi^0},m_V)\Big[\frac{\alpha}{\pi}\int_0^\infty dQ^2 \mathcal{K}(Q^2)\mathcal{F}_{\pi^{\pm},\mathrm{VMD}}^2(Q^2,m_V) \nonumber \\
& \hspace{5.5cm} \times \frac{\partial}{\partial m_{\pi}^2}\overline{\Pi}_{e^2}(Q^2,m_{\pi})\Big]_{ m_\pi=m_{\pi^0}}\nn
 &= 3.74\times 10^{-10}.
  \label{eq:amu_OSS_VMD}
 \end{align}
Here, $m_{\pi^0}$ and $m_V=m_\rho=775.26$ MeV \cite{PDG}
are the physical masses of the neutral pion and the $\rho$ meson, respectively. The (finite) self-energy of the charged pion, $\Sigma_2$, is obtained applying the Cottingham formula to the tree-level Compton scattering amplitude in sQED (cf. Eq.~\eqref{eq:SE2} in appendix \ref{app:CottinghamVerification}), $\mathcal{M}_{e^2}$, supplemented with the VMD form factors:
\begin{equation}
    \Sigma_2(m_{\pi},m_V) = \int\frac{d^4k}{(2\pi)^4}\frac{1}{k^2}\mathcal{F}_{\pi^\pm,\mathrm{VMD}}^2(-k^2,m_V)\mathcal{M}_{e^2}(
    p\cdot k,k^2),
\end{equation}
where $p$ and $k$ denote the (on-shell) pion and virtual photon four-momenta, respectively.
The shift between the on-shell scheme and the FLAG renormalization scheme, similar to Eq.~\eqref{eq:AddCT}, computed with VMD form factor is given by
\begin{align}
\Delta \text{CT}^{}(m_{\pi^0},m_V) &= 2m_{\pi^0}\Delta m_\pi \Big[\frac{\alpha}{\pi}\int_0^\infty dQ^2 \mathcal{K}(Q^2)\mathcal{F}_{\pi^{\pm},\mathrm{VMD}}^2(Q^2,m_V) \nonumber \\
&\hspace{5cm} \times  \frac{\partial}{\partial m_{\pi}^2}\overline{\Pi}_{e^2}(Q^2,m_{\pi})\Big]_{ m_\pi=m_{\pi^0}} \nn
&= -3.94\times 10^{-10}.
\label{eq:amu_Delta_FLAG_VMD}
\end{align}
One can see that $\text{CT}_{\mathrm{OMS}}^{}(m_{\pi^0},m_V)+\Delta \text{CT}^{}(m_{\pi^0},m_V) \approx 0$. The corresponding mass shift, calculated using the Cottingham formula for the elastic Compton scattering with the VMD form factors attached, is
\begin{equation}
  m_{\pi^\pm}-m_{\pi^0} = \frac{1}{2m_{\pi^0}} \Sigma_2(m_{\pi^0},m_\rho)\approx  4.36\times 10^{-3}\,\mathrm{GeV}.
\end{equation}
Thus, the results for the different renormalization schemes are
\begin{align}
    a^{\pi^+\pi^-}_{\mu,\mathrm{OMS}}(m_{\pi^0},m_\rho) &= -1.71\times10^{-10},\\
    \label{eq:asQED_flag}
    a^{\pi^+\pi^-}_{\mu,\mathrm{FLAG}}(m_{\pi^0},m_\rho)&= -5.65 \times 10^{-10}.
\end{align}
The remarkable result is that the FLAG renormalization procedure approximately corresponds to just taking the non-renormalized result provided in table~\ref{table:PionLoop}, see the table entry for $(\text{VMD},m_{\pi^0})$ of $-5.46\times 10^{-10}$.

\subsection{Model prediction for the total electromagnetic corrections to the HVP}
\label{sec:model_precition_full_em}
Based on the models developed in the previous sections, we aim to provide an estimate for $\ahvpnlo$ that one can expect in the lattice practitioner's renormalization scheme (see section \ref{sec:pure_sQED}). To this end, we take into account the result for the charged pion loop computed with a VMD form factor in the FLAG renormalization scheme Eq.~\eqref{eq:asQED_flag}. We add the $\pi^0$ contribution $a_\mu^{\pi^0}=0.37(1) \times 10^{-10}$, which was already computed in Ref. \cite{Biloshytskyi:2022ets} with a VMD form factor\footnote{When one uses the $z$-expansion from Ref. \cite{gerardin:2019vio} for the parametrization of the pion transition form factor, one obtains $a_\mu^{\pi^0}=(0.18\pm0.08) \times 10^{-10}$. The conclusion that the contribution of the charged pion loop is by far the dominant one is unaffected.} and the contributions from $\eta$, Eq.~\eqref{eq:apme_eta}, and $\eta'$ , Eq.~\eqref{eq:apme_eta_p}. Lastly, one can approximate the contribution of the charged kaon loop by the same computation as given in section~\ref{sec:piloop-cont} assuming a VMD form factor for the kaon \cite{Aoyama:2020ynm, Stamen:2022uqh}
\ba 
\label{eq:kaon_FF_VMD}
\mathcal{F}_{K^\pm,\text{VMD}}(Q^2) = 1 - \frac{Q^2}{2}\Big( \frac{1}{m_\rho^2+Q^2}+\frac{1}{3} \frac{1}{m_\omega^2+Q^2} + \frac{2}{3} \frac{1}{m_\phi^2+Q^2}  \Big).
\ea 
Here, we use the physical meson masses from the PDG \cite{PDG},  $m_\rho = 775.26$ MeV, $m_\omega = 782.66$ MeV, $m_\phi = 1019.461$ MeV. 
As in the case of the pion, the bare mass of the charged kaon, which enters the loop, is chosen at the reference point of isospin-symmetric QCD, $(m_K)^{\text{isoQCD}} = 494.6$ MeV \cite{FlavourLatticeAveragingGroupFLAG:2024oxs}.
We fix the counterterm in the FLAG renormalization scheme by requiring that the electromagnetic corrections shift the charged kaon mass to its physical value, $m_{K^\pm} = 493.7$ MeV \cite{PDG}.  
Using now Eqs.~\eqref{eq:amu_OSS_VMD} and \eqref{eq:amu_Delta_FLAG_VMD} with the VMD form factor \eqref{eq:kaon_FF_VMD} and the kaon mass shift, $m_{K^\pm}-(m_K)^{\text{isoQCD}} = -0.9\,\mathrm{MeV}$, we obtain an estimate for the contribution of the charged kaon loop in the FLAG renormalization scheme,
\ba 
a^{K^+ K^-}_{\mu,\mathrm{FLAG}} = -0.056\times 10^{-10}.
\ea 
Our combined estimate for the electromagnetic corrections to the HVP contribution from the phenomenological model yields
\ba 
\nonumber
\Big( a^{\text{hvp}1\gamma^*}_{\mu}\Big)^{\text{pheno}} &=& a_\mu^{\pi^0} + a_\mu^{\eta} + a_\mu^{\eta'} + a^{\pi^+\pi^-}_{\mu,\mathrm{FLAG}} + a^{K^+ K^-}_{\mu,\mathrm{FLAG}}  \\
 &=& - 4.91(2.46) \times 10^{-10}.
 \label{eq:amu1g_pheno}
\ea 
Note that this estimate contains a large systematic error due to our choices of models. This error can be naively addressed by comparing the prediction for the integrand of the $(2+2)a$ contribution, given in Eq. \ref{eq:full_model}, using the model parameters from table \ref{table:model_param} and the integrand obtained from a fit of the charged pion loop to the lattice data, see figure \ref{fig:D450_integrand}. At the physical point, both methods differ by at most $50\%$ of the prediction. Since the charged pion loop dominates the phenomenological contributions to Eq.  \eqref{eq:amu1g_pheno}, we have  assigned a $50\%$ uncertainty to the result for $\Big( a^{\text{hvp}1\gamma^*}_{\mu}\Big)^{\text{pheno}} $.

We note that our result Eq.~\eqref{eq:amu1g_pheno} is significantly larger in magnitude and of opposite sign than the phenomenological estimate of Hoferichter et al.~\cite{Hoferichter:2023sli}. The main qualitative difference lies in the treatment of the $P=(\pi^0,\eta,\eta')$ intermediate states.
While our approach is akin to that used in the treatment of the hadronic light-by-light contribution to $(g-2)_\mu$~\cite{Jegerlehner:2009ry,Nyffeler:2016gnb}, these contributions are evaluated via the measured $e^+e^-\to P\gamma$ cross sections in~\cite{Hoferichter:2023sli} and result in much larger positive contributions, by more than an order of magnitude in the case of the pion.
The respective merit of these two approaches will likely become clearer in the future, with more lattice calculations becoming available. For now, we naively note that the 
lattice result presented below for the $(2+2)a$ diagram does not appear to leave room for a ten to twelve times enhanced $\pi^0$ pole contribution as compared to our estimate -- see figures~\ref{fig:tpta_pheno_full_model} and \ref{fig:chiral_fit}.

Effects that have not been taken into account in the estimate Eq.~\eqref{eq:amu1g_pheno} are for instance rescattering effects of pions. In particular, one may expect some enhancement in the $s$ wave, by analogy with the dispersive analysis of the HLbL contribution to the muon $(g-2)$~\cite{Colangelo:2017fiz,Danilkin:2018qfn}. Other resonances coupling to two photons could be included as well, such as the axial-vector $(a_1,f_1)$ and the tensor mesons $(a_2,f_2)$. Ultimately, the asymptotic behaviour predicted for large $\Lambda$ by the operator-product expansion~\cite{Biloshytskyi:2022ets} is driven by the partonic degrees of freedom, so that a matching procedure is necessary between the low- and high-energy regimes.
\section{Coordinate-space integrand from the model contributions}
\label{sec:model_prediction_for_integrand}
In section \ref{sec:phenomenological_model}, we have established a model for the electromagnetic corrections to the HVP, Eq.~\eqref{eq:ccs_nlohvp}. It is furthermore interesting to compare the profile of the $|x|$-integrand as given in Eq.~\eqref{eq:2+2_master_a} from the model prediction to the lattice data. This is particularly useful for the purpose of approximating the tail of the $(2+2)a$ contribution, where lattice data is affected by noise.
To this end, we perform a continuum calculation of the coordinate-space integrand of the $\pi^0$ contribution in section \ref{sect:nlohvp_pi0} using a VMD form factor \eqref{eq:vmdtff}. The integrand for the $\eta$ and $\eta'$ contribution is obtained from the same functional form using the parameters from table \ref{table:model_param}.
To obtain the $|x|$-integrand for the charged pion loop, we use a different strategy than in the case of the $\pi^0$ exchange.
We perform a calculation of the charged pion loop in coordinate-space, see section~\ref{sec:chargedpion_lat}, by implementing sQED in lattice regularization. In this approach, we are able to obtain the integrand for a fixed distance between the vertex $x$ and the origin with the integrations over $y$ and $z$ performed. The lattice parameters are chosen to be `near' the continuum and infinite-volume limit. The resulting integrand can then be directly compared to the integrand of the lattice-QCD calculation.

\subsection{Integrand for the $\pi^0$-exchange contribution}
\label{sect:nlohvp_pi0}
In order to compare the profile of the $|x|$-integrand as given in Eq.~\eqref{eq:2+2_master_a} from lattice data to the prediction from the $\pi^0$-exchange, we define the integrand $f^{\pi^0}(|x|)$ as 
\ba
\label{eq:pi0_integranddef}
a_\mu^{\pi^0} \equiv \int_0^\infty d|x| f^{\pi^0}(|x|).
\ea
We start from the coordinate-space vector correlator for the $\pi^0$ exchange,
\ba 
\label{eq:pi0veccor}
\langle j_\sigma(z) j_\mu(y) j_\nu(x) j_\lambda(0) \rangle^{\pi^0} = \frac{1}{(2\pi)^{12}} \int_{q,k,p} e^{i(p\cdot z + q\cdot x + k\cdot y)} \Pi_{\sigma\mu\nu\lambda}(p,q,k),
\ea 
where $\int_{p}:= \int_{\mathbb{R}^4} d^4p$. Substituting Eq.~\eqref{eq:pi0veccor} for the four-point vector-correlator in Eq.\ \eqref{eq:ccs_nlohvp}, we obtain the following expression for the integrand
\ba
f^{\pi^0}(|x|) &=& -\frac{e^2}{2} 2\pi^2 |x|^3 \frac{1}{(2\pi)^{12}} \int_{z,x} H_{\sigma\lambda}(z) \delta_{\mu \nu} \Big[G_0(x-y)\Big]_\Lambda \int_{q,k,p} e^{i(p\cdot z + q\cdot x + k\cdot y)} \Pi_{\sigma\mu\nu\lambda}(p,q,k)
\nonumber
\\
&=& -\frac{e^2}{2} 2\pi^2 |x|^3 \frac{1}{(2\pi)^{12}} \int_k e^{ik\cdot x} \Big[\frac{1}{k^2}\Big]_\Lambda \int_p  \tilde H_{\sigma\lambda}(p) \int_q e^{iq\cdot x}\,\Pi_{\sigma\mu\mu\lambda}(p,q,k),
\label{eq:pme_integrand_momspace}
\ea
where the Fourier transform of the CCS kernel is
\be
\tilde H_{\sigma\lambda}(p) = \int_z H_{\sigma\lambda}(z)\,e^{-ip\cdot z}\,.
\ee
Using Eqs. (20-24) from Ref.~\cite{Meyer:2017hjv}, we find
\be
\label{eq:Hp}
\tilde H_{\sigma\lambda}(p) = \frac{\delta_{\sigma\lambda}(2\pi)^4}{3\pi^2} \Big[\Big(3+p^2 \frac{d}{dp^2}\Big) \Big(\frac{g(p^2)}{|p|^4}\Big) -2\alpha^2\frac{\delta^{(1)}(|p|)}{|p|^5}\Big]
+ p_\lambda p_\sigma (\dots)\,,
\ee
with 
\be
g(p^2) = 2\alpha^2 \frac{m_\mu^4}{p^6}y(|p|)^4,\qquad y(|p|) = \frac{2|p|}{|p|+\sqrt{4m_\mu^2+p^2}},
\ee
where the term proportional to $p_\lambda p_\sigma$ does not contribute to $f^{\pi^0}(|x|)$, owing to the transversality property
$p_\sigma \Pi_{\sigma\mu\mu\lambda}(p,q,k)=0$. 
We use the momentum-space polarization tensor $\Pi_{\sigma\mu\nu\lambda}(p,q,k)$ given in Ref.\ \cite{Biloshytskyi:2022ets} (which is based on the earlier Ref.\ \cite{Knecht:2001qf}), 
\ba
\label{eq:pol_tensor_knecht_maintext}
\Pi_{\sigma\mu\mu\lambda}(p,q,k) &=&
\epsilon_{\sigma\mu\alpha\beta} \epsilon_{\mu\lambda\gamma\delta} \,p_\alpha
\Big(\frac{{\cal F}(-p^2,-k^2)\;{\cal F}(-q^2,-(p+k+q)^2)}{(p+k)^2+m_\pi^2}\; k_\beta\, q_\gamma(p+k)_\delta
\nonumber\\ && + \frac{{\cal F}(-p^2,-q^2) \,{\cal F}(-k^2,-(p+k+q)^2)}{(p+q)^2 + m_\pi^2}\; q_\beta\,k_\gamma (p+q)_\delta\Big).
\ea
We assume a VMD parametrization of the pion transition form factor in the spacelike region, Eq.~\eqref{eq:vmdtff},
which is characterized by the mass of the vector meson $m_V$ and the factor $F_\pi$, which determines the normalization of the form factor in the real-photon limit, $F_\pi:=\mathcal{F}_{\pi^0\gamma^*\gamma^*}(0,0)$.
Using the method of Gegenbauer polynomials for evaluating the angular integrals, we can simplify Eq.~\eqref{eq:pme_integrand_momspace} to a series of two-dimensional integrals, see Eq.~\eqref{eq:pi0_integrand_final}. The details of this calculation are provided in appendix~\ref{sect:pi0_integrand}.

\subsection{Lattice calculation of the charged pion loop integrand in scalar QED}
\label{sec:chargedpion_lat}
To compare the charged-pion-loop contribution to the lattice integrand data, Eq.~\eqref{eq:2+2_master_a}, we shall write the former as a one-dimensional integral in the same coordinate-space parametrization 
\ba 
\label{eq:integrand_pipi}
a_\mu^{\pi^+\pi^-}\equiv\int_0^\infty d|x| f^{\pi^+\pi^-}(|x|)\,.
\ea 
A direct semi-analytic evaluation of $f^{\pi^+\pi^-}$ is more challenging than computing the integrated contribution $a_\mu^{\pi^+\pi^-}$.
To get the size of the charged-pion-loop contribution in the continuum and infinite-volume, we instead compute the former in discrete and finite space-time and then take the continuum and infinite-volume limit.
As we are primarily interested in the long-distance region, scalar QED with point-like $\gamma\pi\pi$ and $\gamma\gamma\pi\pi$ interaction vertices should give a realistic description.
On the lattice, the scalar-QED action which describes the charged pions reads
\begin{equation}\label{eq:s-sqed-lat}
S = a^4\sum_x \left\{
\nabla_\mu\phi(x)\nabla_\mu\phi^*(x) + m_\pi^2\phi(x)\phi^*(x)
\right\}\,,
\end{equation}
where the covariant derivative of the scalar field $\phi(x)$ associated with a background U(1)-electromagnetic field $A_\mu(x)$ is given by
\begin{equation}
\nabla_\mu\phi(x)=\frac{1}{a}\left(
e^{-iaeA_\mu(x)}\phi(x+\hat{\mu})-\phi(x)
\right)\,.
\end{equation}
A remarkable difference between the lattice action Eq.~\eqref{eq:s-sqed-lat} and its continuum counterpart is that, at O($\alpha^2$), there are additional contact terms arising when three or four scalar fields are evaluated at the same point. 
These contact terms are crucial as they guarantee the U(1)-Ward identity on the four-point vector-current correlation function to hold in the sense of distribution. 
We have explicitly verified this to machine precision in our numerical implementation of the four-point vector-current correlation functions. 
We have noticed that typically, an $m_\pi L \gtrsim 8-9$ is needed to be in the asymptotic infinite-volume limit in our setup, which we conclude by comparing the shape of the integrand $f^{\pi^+\pi^-}(|x|)$ in the large-$|x|$ region as we vary $m_\pi L$.
As for the continuum limit, we have verified that, at a simulated large volume and fixed lattice spacing $a$ and pion mass, the total $a_\mu^{\pi^+ \pi^-}$ from numerically integrating $f^{\pi^+ \pi^-}(|x|)$ scaled linearly with $a^2$, with a continuum-extrapolated result agreeing to the percent level with that obtained from the continuum, momentum-space approach described in section~\ref{sec:piloop-cont}.
As these extrapolated results are also in percent-level agreement with the integrated values from our finest simulated lattice for each given $m_\pi$, we treat the results from those finest boxes as if they corresponded to the integrand $f^{\pi^+ \pi^-}(|x|)$ calculated in the continuum and infinite volume.

\begin{figure}
\begin{subfigure}{0.48\textwidth}
    \centering
    \includegraphics[width=1.1\textwidth]{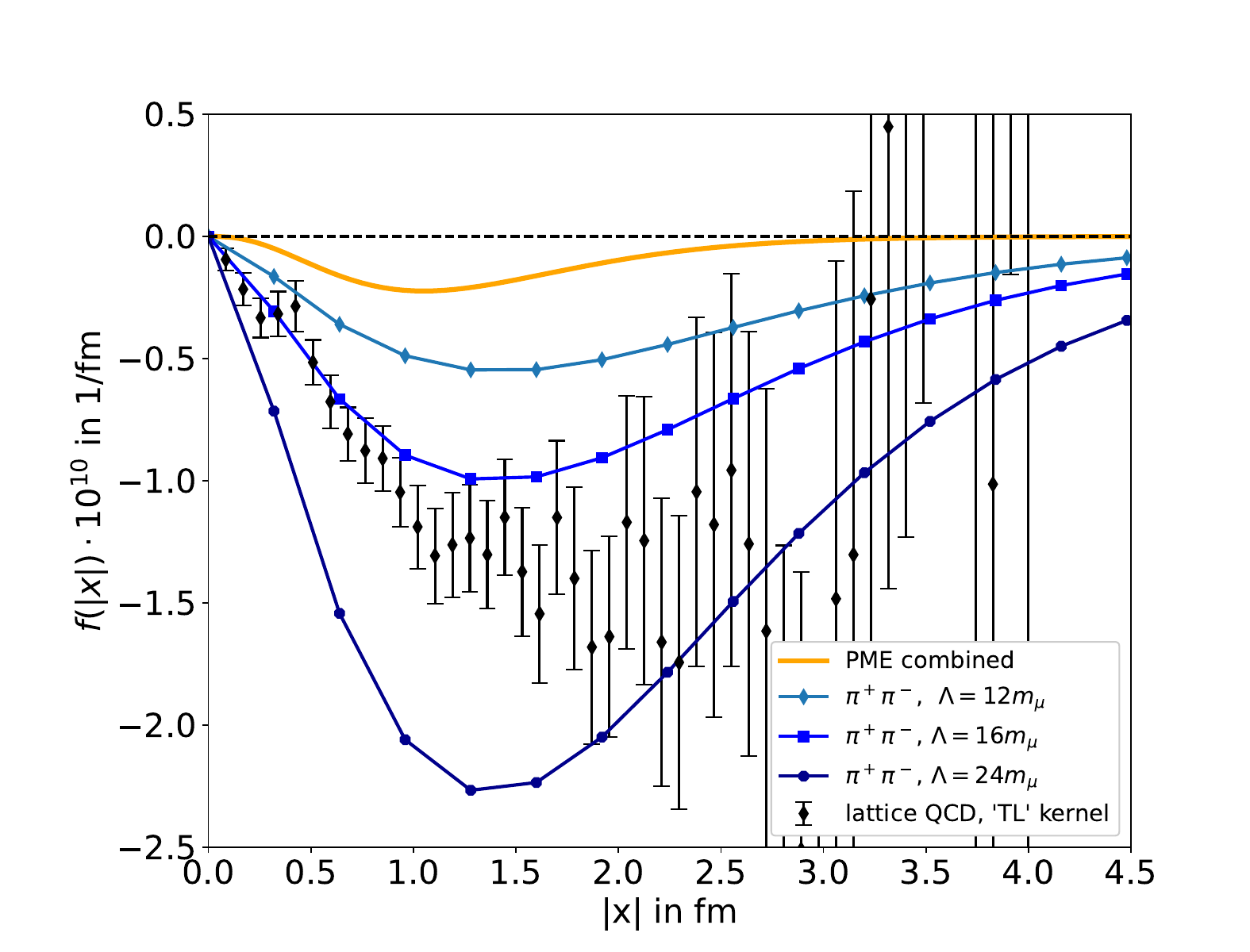}
\end{subfigure}
\begin{subfigure}{0.48\textwidth}
    \centering
    \includegraphics[width=1.1\textwidth]{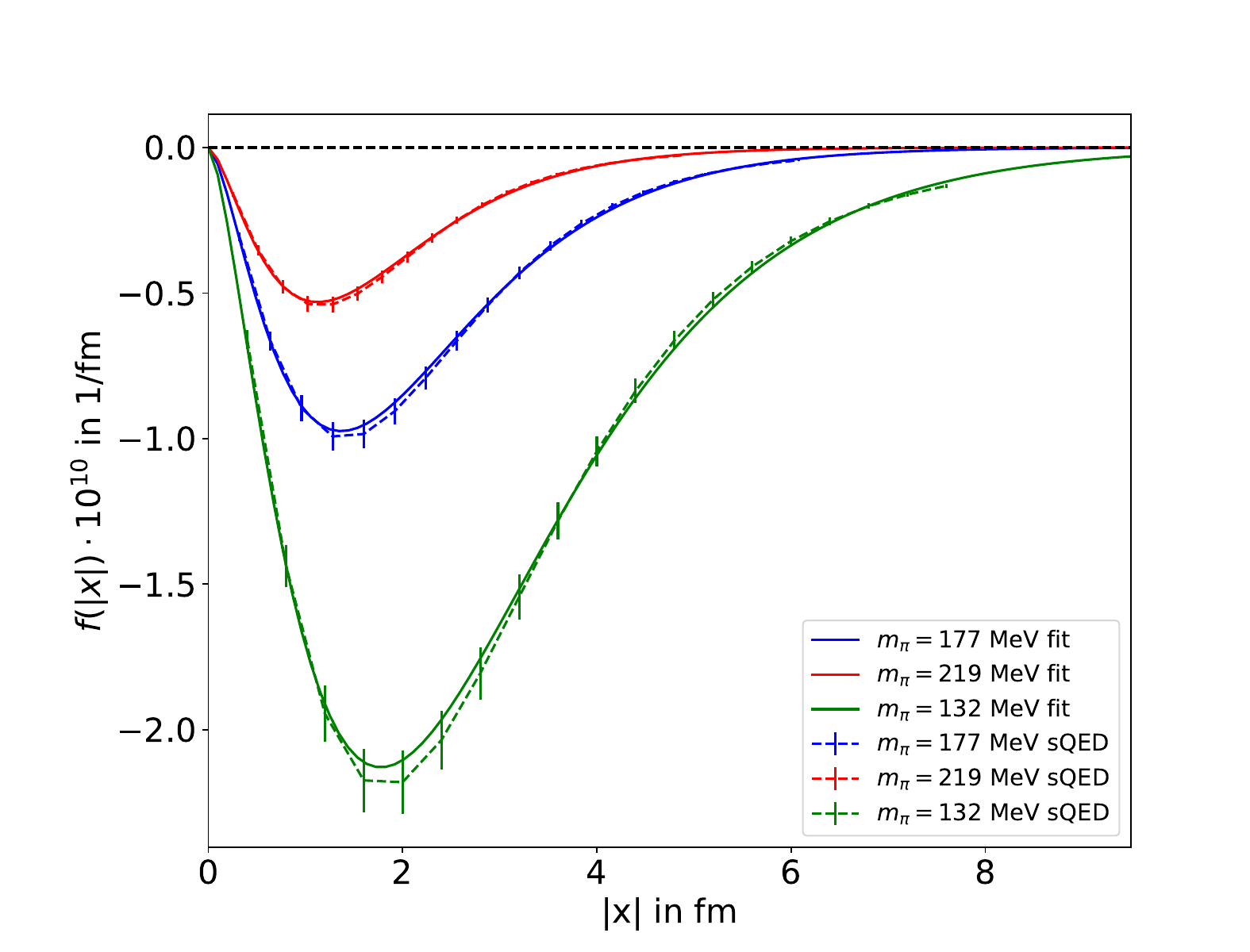}
\end{subfigure}
	\caption{Left: Comparison between the lattice data and the different phenomenological contributions for the integrand of the $(2+2)a$ contribution Eq.~\eqref{eq:2+2_master_a}. The integrand for the $\pi^+ \pi^-$ contribution is calculated in scalar QED with several values of the Pauli-Villars regulator $\Lambda$ on the lattice according to section~\ref{sec:chargedpion_lat}. Right: The integrand from sQED for the $\pi^+ \pi^- $ contribution, Eq.~\eqref{eq:integrand_pipi}, for $\Lambda=16 m_\mu$ for different values of the pion mass. The dashed lines correspond to the calculated values. The solid lines are fits using the ansatz Eq.~\eqref{eq:chargedpion_fitfunction}.}
\label{fig:nlohvp_sQED}
\end{figure}
For several values $\Lambda=\{10,12,16,24\}m_\mu$ for the Pauli-Villars regulator in the photon propagator \eqref{eq:doubly_pv}, the form of the integrand can be very well described by the following fit ansatz
\ba 
\label{eq:chargedpion_fitfunction}
f^{\pi^+\pi^-}(|x|) \simeq A |x|^n \exp(-{\textstyle\frac{4}{3}} m_\pi|x|)
\ea 
where $A$ and $n$ are fitted parameters. Here, $n=1.65(5)$ is fitted globally for each pion mass along the chiral trajectory and $\Lambda=\{10,12,16,24\}m_\mu$. In this calculation, the prefactor $A$ grows proportionally to $\Lambda^2$ for the range that we are interested in. This is precisely the expected asymptotic behaviour, as the scalar QED result is quadratically divergent as $\Lambda$  goes to infinity. Here, we note that the exponential function $\exp(-{\textstyle\frac{4}{3}} m_\pi|x|)$ in Eq.~\eqref{eq:chargedpion_fitfunction} differs from the naively expected $\exp(-2m_\pi |x|)$ given by the asymptotic behaviour of the massive propagator. In figure \ref{fig:nlohvp_sQED}, on the left we show a comparison between the charged pion loop integrand for different values of $\Lambda$ and the integrand calculated with Eq.~\eqref{eq:2+2_master_a} on the ensemble E300. On the right of this figure, we show different fits of Eq. \eqref{eq:chargedpion_fitfunction} to the integrand computed in scalar QED for different values of the pion mass.
\section{Analysis of the contribution $a^{(2+2)a-ll}_\mu$}\label{sec:analysis}
Optimizing the trade-off between statistical and systematic uncertainties from the interplay between phenomenology and lattice data has become a common strategy to improve the overall error.
For lattice calculations based on coordinate-space QED kernels, this strategy turns out to be effective. 
For instance, in the HLbL contribution to $a_\mu$, the tail of the lattice integrand data has been modeled successfully via the exchanges of single pseudoscalar mesons~\cite{Chao:2020kwq}.
As another example, when computing the intermediate window observable of $a_\mu^{\rm hvp}$ with a coordinate-space kernel~\cite{Chao:2022ycy}, the large-distance behaviour is well captured by two-pion intermediate states.

In section \ref{sect:hvpnlo_matching}, we show how to match the contributions from the phenomenological model to the QCD Wick contraction labeled by $(2+2)a$, see figure \ref{fig:FiniteDiags}. Eq.~\eqref{eq:full_model} summarizes the result, where the contributions of $\pi^0$, $\eta$, $\eta'$ and the charged pion loop each enter with a multiplicative matching coefficient in the prediction for the light-light component of the $(2+2)a$ contribution.
In section \ref{sect:nlohvp_tail},
we present a procedure to treat the noisy tail of the integrand of Eq.~\eqref{eq:2+2_master_a} computed in lattice QCD, making use of the model studies carried out in section \ref{sec:model_prediction_for_integrand}.
In section \ref{sect:nlohvp_extrapolation}, we extrapolate the results computed on different ensembles to the physical point.

\subsection{Matching coefficients for the $(2+2)a$ topology}
\label{sect:hvpnlo_matching}
\begin{figure}
     \centering
     \includegraphics[width=0.75\textwidth]{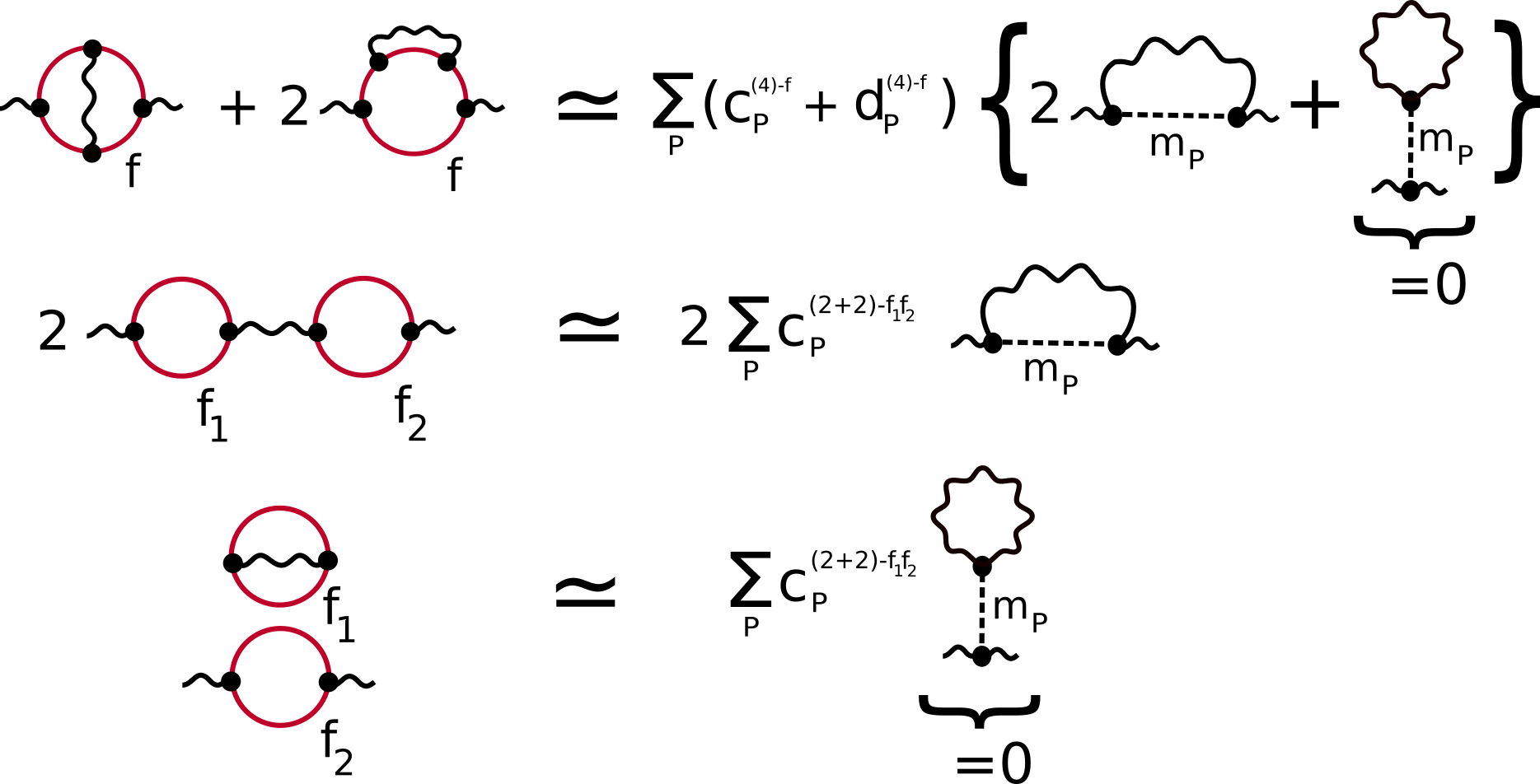}
     \caption{Diagram matching between the QCD Wick contractions on the left and the pseudoscalar meson exchange contributions on the right, where $\rm{f,f_1,f_2}$ denote the quark flavour in the loop and the sum on the right-hand side runs the contributions from the pseudoscalar mesons $P=\pi^0, \eta, \eta^{\prime}$. In the first row the sum of the diagrams of the $(4)$ topology is depicted. In the second row the matching for the $(2+2)a$ contribution is given, where $f_1$,$f_2$ can either be a light or strange quark. The $(2+2)b$ diagram receives no contribution from the pseudoscalar meson exchange (PME), since the corresponding diagram vanishes. The coefficients $c_P$ and $d_P$ are given in Ref. \cite{Chao:2021tvp}.}
     \label{fig:matching}
 \end{figure}

In the previous sections \ref{sec:phenomenological_model} and \ref{sec:model_prediction_for_integrand}, we have considered the phenomenological contributions to the total electromagnetic correction to the leading order HVP contribution \eqref{eq:ccs_nlohvp}. 
In fact, it is possible to match the contributions from $\pi^0$, $\eta$, $\eta'$ and charged pion and kaon loop to the individual QCD Wick contractions in figure~\ref{fig:FiniteDiags} \cite{Chao:2021tvp, Bijnens:2016hgx}. 
The diagram matching for the $(4)$ and $(2+2)$ topology is depicted in figure~\ref{fig:matching}. From this we see that the pseudoscalar exchange does not contribute to the $(2+2)b$ diagram, because the self-contracted photon loop results in a Kronecker delta multiplying the Wess-Zumino-Witten interaction term, which vanishes due to the anti-symmetry of the term. The pseudoscalar meson exchange (PME) diagram that matches the $(2+2)a$ topology is precisely the one computed from the polarization tensor Eq.~\eqref{eq:pol_tensor_knecht_maintext}.
Each flavor combination of the two quark loops takes a portion of the total $\pi^0$, $\eta$ and $\eta^\prime$ contributions. We will refer to their respective weights as the \textit{matching coefficients}.

In the case of $N_{\rm{f}}=2+1$ quark flavours, we have to distinguish between the light-light (ll), light-strange (ls), strange-light (sl) and strange-strange (ss) components of the $(2+2)a$ diagram. Only the (ll) component receives a contribution from the $\pi^0$. 
To get its matching coefficient, we distribute the total contribution $a_\mu^{\pi^0}$ among the different involved diagrams according to the mapping condition depicted in figure~\ref{fig:matching} at the level of the QCD-Wick contraction, weighted by the proper electric charge factors. Concretely, 
 \ba 
a_\mu^{(2+2)a-ll,\pi^0}=2\frac{25}{81}c^{ll}_\pi \Big[2 c^{ll}_\pi\frac{25}{81} + 2\frac{17}{81}\left(c^{(4)-l}_\pi+d^{(4)-l}_\pi\right) \Big]^{-1} a_\mu^{\pi^0}=-\frac{25}{9} a_\mu^{\pi^0}
 \ea 
 where the factor 2 arises from the fact that there are two possible contractions with the photon propagator which give rise to the same diagram. The coefficients $c_P$ are given in table 10 of Ref.~\cite{Chao:2021tvp}. For simplicity, we do not write the '$(2+2)$' explicitly in the superscript of the coefficients $c^{f_1 f_2}_P:=c^{(2+2)-f_1 f_2}_P $.

The case of the $\eta$ and $\eta'$ is more complicated. Neither contributes to the light quark component of the fully-connected (4) topology. Instead, their contributions are distributed among the (ll), (ls), (sl) and (ss) components of the $(2+2)a$ diagram 
 \ba 
 \label{eq:matching_eta_ll}
a_\mu^{(2+2)a-ll,(\eta,\eta')}&=& 2\frac{25}{81}c^{ll}_{(\eta,\eta')} \Big[2 c^{ll}_{(\eta,\eta')}\frac{25}{81} + 2\frac{5}{81}c^{ls}_{(\eta,\eta')}+2\frac{1}{81}c^{ss}_{(\eta,\eta')}  \Big]^{-1}  a_\mu^{{(\eta,\eta')}},\\
 \label{eq:matching_eta_ls}
a_\mu^{(2+2)a-ls,{(\eta,\eta')}}&=& 2\frac{5}{81}\frac{c^{ls}_{(\eta,\eta')}}{2} \Big[2 c^{ll}_{(\eta,\eta')}\frac{25}{81} + 2\frac{5}{81}c^{ls}_{(\eta,\eta')}+2\frac{1}{81}c^{ss}_{(\eta,\eta')} \Big]^{-1}  a_\mu^{{(\eta,\eta')}}\\
\nonumber
&=&a_\mu^{(2+2)a-sl,{(\eta,\eta')}},\\
  \label{eq:matching_eta_ss}
a_\mu^{(2+2)a-ss,{(\eta,\eta')}}&=&2\frac{1}{81}c^{ss}_{(\eta,\eta')} \Big[2 c^{ll}_{(\eta,\eta')}\frac{25}{81} + 2\frac{5}{81}c^{ls}_{(\eta,\eta')}+2\frac{1}{81}c^{ss}_{(\eta,\eta')}  \Big]^{-1}  a_\mu^{{(\eta,\eta')}},
 \ea 
where the coefficients for the (ls) and the (sl) components are identical.

At first, we will focus on the SU$(3)$ flavour symmetric point, where the $\eta$ is a true octet state and the $\eta'$ is a true singlet state. For this case we can obtain the coefficients for $\eta$  from table X of Ref. \cite{Chao:2021tvp}. We have $c^{ll}_\eta=\frac{1}{3}$, $c^{ls}_\eta=-\frac{4}{3}$ and $c^{ss}_\eta=\frac{4}{3}$ and thus the matching coefficients are
\ba 
\nonumber
a_\mu^{(2+2)a-ll,\eta} &=& \frac{25}{9} a_\mu^{\eta}, \quad a_\mu^{(2+2)a-ls,\eta} = -\frac{1}{2} \frac{20}{9} a_\mu^{\eta}, \quad a_\mu^{(2+2)a-ss,\eta} = \frac{4}{9} a_\mu^{\eta} \quad (\small \textrm{SU$(3)_{\rm f}$ sym. point}).\\
& &
\ea 
The singlet state splits according to the charge factors, leading to
\ba 
\nonumber
a_\mu^{(2+2)a-ll,\eta'} &=& \frac{25}{36} a_\mu^{\eta'}, \hspace{0.25cm} a_\mu^{(2+2)a-ls,\eta'} = \frac{1}{2} \frac{10}{36} a_\mu^{\eta'}, \hspace{0.25cm}  a_\mu^{(2+2)a-ss,\eta'} = \frac{1}{36} a_\mu^{\eta'} \hspace{0.25cm}  ( \small \textrm{SU$(3)_{\rm f}$ sym. point}).\\
& &
\ea 
The (ll) and (ls) components of the  $(2+2)a$ contribution at the SU$(3)_{\rm f}$ symmetric point in terms of the PME are then given by 
\ba 
a_\mu^{(2+2)a-ll,\textrm{PME}} &=& -\frac{25}{9} a_\mu^{\pi^0}+\frac{25}{9} a_\mu^{\eta}+\frac{25}{36} a_\mu^{\eta'}  \qquad (\textrm{SU$(3)_{\rm f}$ sym. point}),\\
\label{eq:matching_ls_SU3}
a_\mu^{(2+2)a-ls,\textrm{PME}} &=& \frac{1}{2} \Big(-\frac{20}{9} a_\mu^{\eta}+\frac{5}{18} a_\mu^{\eta'}\Big) \hspace{1.35cm} (\textrm{SU$(3)_{\rm f}$ sym. point}).
\ea 
In the (ss) channel, an unphysical $s\overline{s}$-meson is present in Ref.~\cite{Chao:2021tvp}, which does not contribute to the total result for $\ahvpnlo$.

Away from the SU$(3)_{\rm f}$ symmetric point, the matching condition in \cite{Chao:2021tvp} requires the knowledge of the $\eta-\eta^\prime$ mixing angle $\theta$ (cf. Eqs.(A.29-A.36) of Ref. \cite{Chao:2021tvp}).
 In large $N_c$ chiral perturbation theory (L$N_c\chi$PT), the mixing angle at leading order, denoted by $\theta^{[0]}$, is fully determined by the physical meson masses \cite{Bickert:2016fgy}
\ba 
\label{eq:lo_mixingangle}
\sin(2\theta^{[0]}) = -\frac{4\sqrt{2}(m_K^2-m_\pi^2)(-4m_K^2+3m_{\eta'}^2+m_\pi^2)}{3\Big[-8m_K^2(m_{\eta'}^2+m_\pi^2)+8m_K^4+3m_{\eta'}^2+2m_\pi^2m_{\eta'}^2+3m_\pi^4 \Big]}.
\ea  
A computation of $\theta^{[0]}$ with physical meson masses at leading order in $\chi$PT using Eq.~\eqref{eq:lo_mixingangle} yields a value of $\theta^{[0],\textrm{phys}}=-19.7^\circ$, whereas at next-to-leading order one finds  $\theta^{[1],\textrm{phys}} = -11.1(6)^\circ$ \cite{Bickert:2016fgy}.
Due to this discrepancy we choose to estimate the mixing angle on each ensemble by linearly interpolating $\sin(\theta^{[1]})$ between $\theta^{\textrm{SU}(3)}=0^\circ$ and the NLO result for the mixing angle $\theta^{[1],\rm{phys}} = -11.1(6)^\circ$ with respect to the parameter $\xi = m_K^2-m_\pi^2$.  
In the following, we use the matching coefficients calculated with the interpolated mixing angles $\theta^{[1]}$, given in table \ref{table:mixing_angle_coefficients}.
If one instead uses the mixing angles calculated with Eq.~\eqref{eq:lo_mixingangle} for the calculation of the matching factors, this does not change the prediction for the $\eta$ and $\eta^\prime$ contribution for $a_\mu^{(2+2)a-ll}$ significantly.

\begin{figure}
    \centering
    \includegraphics[width=0.7\textwidth]{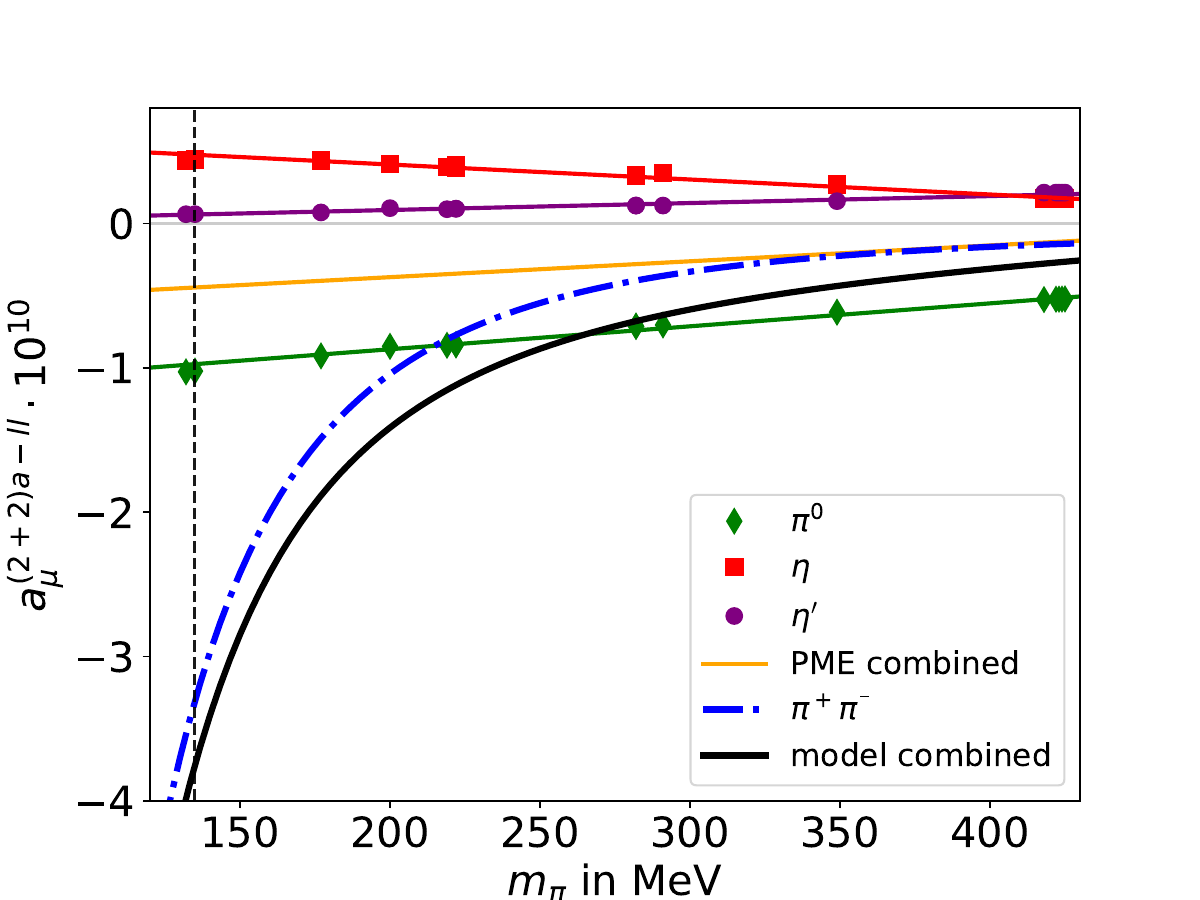}
    \caption{Plot of the pion-mass dependence of different model contributions in Eq.~\eqref{eq:full_model}, including the matching coefficients for the individual contributions. The combination of $\pi^0$, $\eta$ and $\eta'$ is depicted in yellow as `PME combined'. 
    The black solid line gives the combined result for all contributions. The grey dashed vertical line lies at the physical mass of the neutral pion.}
    \label{fig:tpta_pheno_full_model}
\end{figure}

To obtain the matching coefficient for the charged pion loop, we use table 8 and 9 from Ref. \cite{Chao:2021tvp}. Here, all different Wick contractions contribute equally to the (ll) component of the $(2+2)$ topology with equal weight. With the quark charge factor $25/81$ and a factor $2$ for the two possible contractions with the photon propagator for the $(2+2)a$ contribution, we obtain
\ba 
a_\mu^{(2+2)a-ll,\pi^+\pi^-} = \frac{50}{81} a_\mu^{\pi^+\pi^-} 
\ea 
Our full model prediction for the light-quark component of the $(2+2)a$ contribution is given by 
\ba 
\nonumber
a_\mu^{(2+2)a-ll,\textrm{model}} &=& -\frac{25}{9} a_\mu^{\pi^0}(m_\pi,m_{V,\pi},F_\pi)
+\hat{c}^{(ll)}_\eta(m_\pi,m_K,m_\eta,m_{\eta'},\theta) a_\mu^{\eta}(m_\eta,m_{V,\eta},F_\eta)\\
\nonumber
&&+\hat{c}^{(ll)}_{\eta'}(m_\pi,m_K,m_\eta,m_{\eta'},\theta) a_\mu^{\eta'}(m_{\eta'},m_{V,{\eta'}},F_{\eta'})+\frac{50}{81} a_\mu^{\pi^+\pi^-}(m_\pi,m_{V,\pi}),
\\
\label{eq:full_model}
\ea 
where the parameters for the model are given in table \ref{table:model_param}. The matching coefficients $\hat{c}^{(ll)}_{(\eta,\eta')}$ for $\eta$ and $\eta'$ are obtained from Eq.~\eqref{eq:matching_eta_ll} and given in table \ref{table:mixing_angle_coefficients}.
We plot the dependence of Eq.~\eqref{eq:full_model} on $m_\pi$ along our chiral trajectory in figure~\ref{fig:tpta_pheno_full_model}.
Here, we observe that the combined result for pseudoscalar meson exchange contributions is almost linear with a small slope along the chiral trajectory. However, the charged pion loop increases with $m_\pi^{-3}$ when approaching the physical point. We expect this behaviour to show up in the results computed in lattice QCD as well. This makes the use of ensembles near the physical pion mass crucial in order to perform a more robust chiral extrapolation. Lastly, it is worth mentioning that our model does not take into account pion rescattering effects and relies on the VMD parametrization of the form factors. Therefore, it is not expected that this model agrees precisely with the result computed in lattice QCD, but rather gives an impression on the dominant part of the contribution. A comparison to the results computed in lattice QCD is shown in figure~\ref{fig:chiral_fit}.

\subsection{Approximation of the tail}
\label{sect:nlohvp_tail}
\begin{figure}
\begin{subfigure}{0.49\textwidth}
    \includegraphics[width=1.1\linewidth]{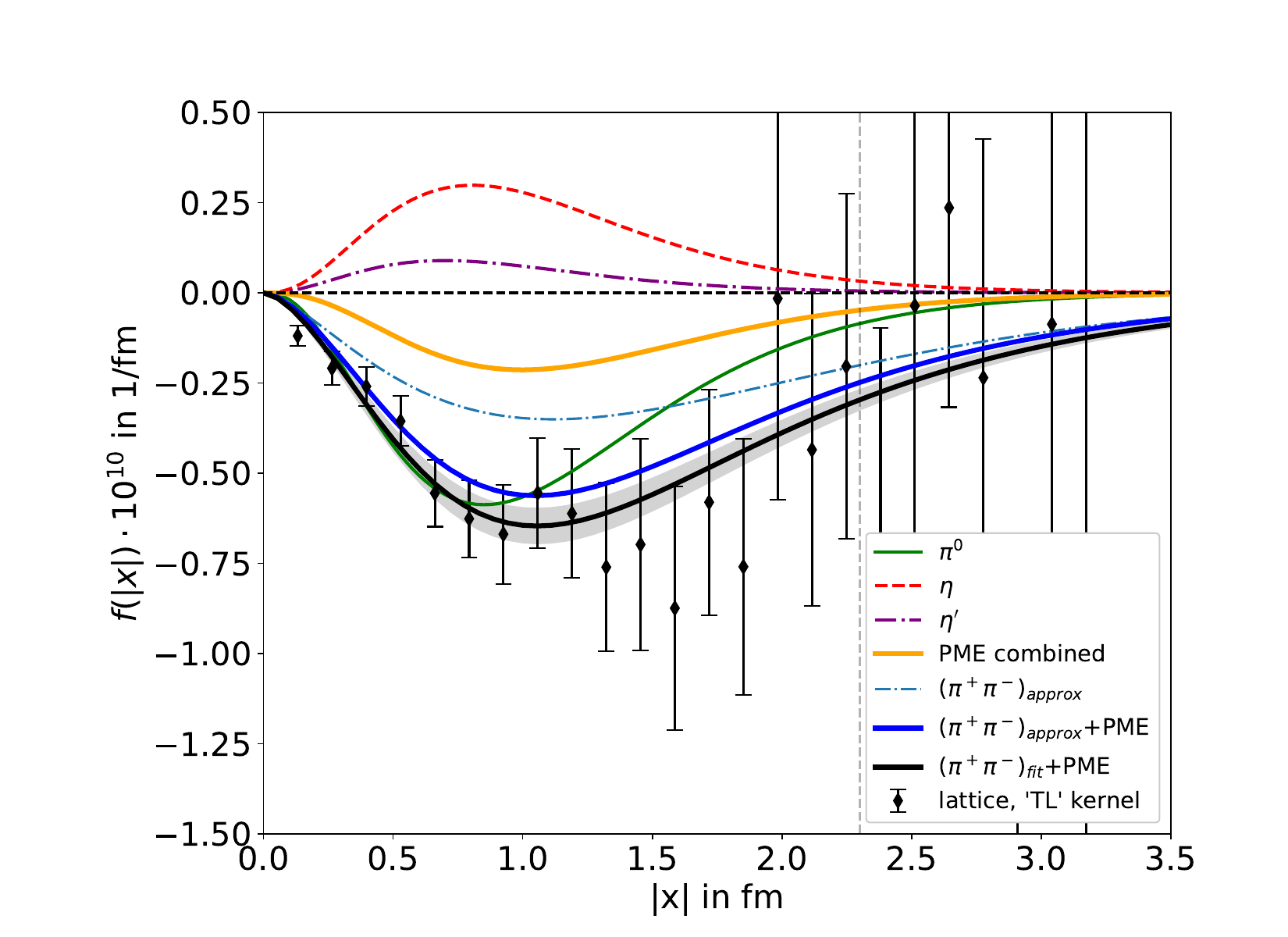}
\end{subfigure}
\begin{subfigure}{0.49\textwidth}
    \includegraphics[width=1.1\linewidth]{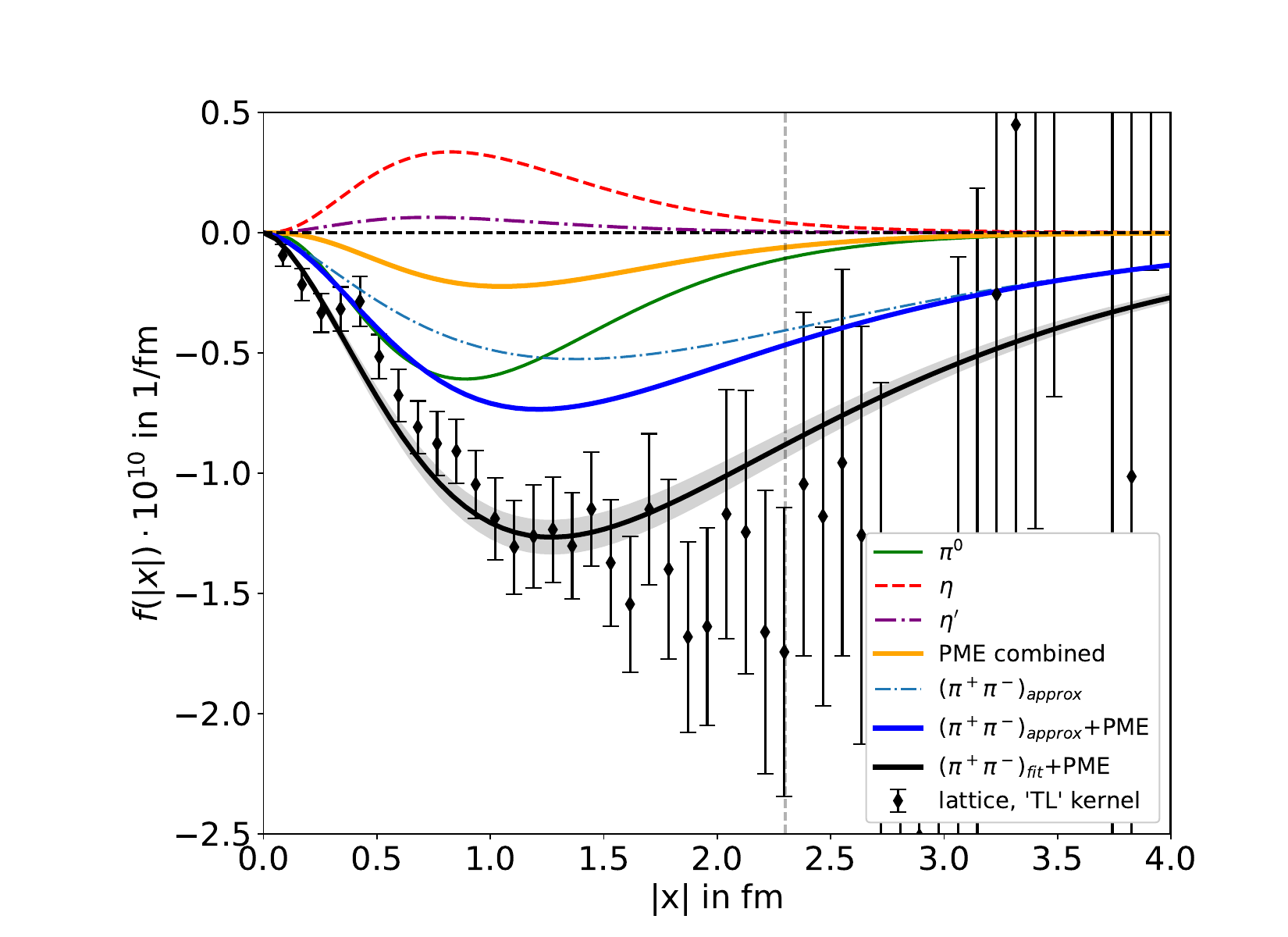}
\end{subfigure}
    \centering
    \caption{Integrands of Eq.~\eqref{eq:2+2_master_a} (black data points) calculated on the ensemble D450 (left) and E300 (right). We also show the prediction for the integrand from the pseudoscalar meson exchange Eq.~\eqref{eq:pme_integrand} including the matching coefficients for the $(2+2)a$ diagram. The combined result from the pseudoscalar mesons Eq.~\eqref{eq:pme_integrand} is shown in yellow. The solid black line corresponds to Eq.~\eqref{eq:integrand_model} where the prefactor $A$ of the charged pion contribution is obtained from a fit to the lattice data.
    The dashed blue line is obtained from Eq. \eqref{eq:chargedpion_fitfunction}, where the prefactor $A$ is fixed using the results for $a_\mu^{\pi^+\pi^-}$ computed with a VMD form factor (see table~\ref{table:PionLoop}). 
    The solid blue line is obtained by adding the yellow and dashed blue lines. The vertical dashed line gives the value of $|x|_{\textrm{cut}}$ from which we use the fit to obtain the contribution from the tail.} 
    \label{fig:D450_integrand}
\end{figure}

From figure~\ref{fig:D450_integrand} for the integrand in Eq.~\eqref{eq:2+2_master_a} computed on the lattice, we see that the large-$|x|$ tail is very noisy. In order to reduce the statistical uncertainty due to the large fluctuations at long distances, we suggest the following strategy: for each ensemble, we integrate the lattice data for Eq.~\eqref{eq:2+2_master_a} up to a value of $|x|=|x|_{\textrm{cut}}$. For $|x|>|x|_{\textrm{cut}}$, we estimate the tail of the integrand using an ansatz motivated by the models considered in section~\ref{sec:phenomenological_model}. This ansatz takes into account the contributions of the $\pi^0$, $\eta$, $\eta'$ and the charged pion loop. For the pseudoscalar meson exchange we use the VMD form factors according to sections~\ref{sect:nlohvp_pi0} and \ref{sect:hvpnlo_modelparam}. Analogous to Eq.~\eqref{eq:full_model}, we obtain for the integrand with the appropriate matching coefficients
\ba 
\nonumber
f^{\textrm{PME}}(|x|) &=& -\frac{25}{9} f^{\pi^0}(|x|,m_\pi,m_{V,\pi},F_\pi)
+\hat{c}^{(ll)}_\eta(m_\pi,m_K,m_\eta,m_{\eta'},\theta) f^{\eta}(|x|,m_\eta,m_{V,\eta},F_\eta)\\
&&+\hat{c}^{(ll)}_{\eta'}(m_\pi,m_K,m_\eta,m_{\eta'},\theta) f^{\eta'}(|x|,m_{\eta'},m_{V,{\eta'}},F_{\eta'})\,.
\label{eq:pme_integrand}
\ea 
After subtracting the PME contributions, we expect the $(2+2)a$ integrand to be dominated by the charged-pion loop. As discussed in section~\ref{sec:piloop-cont}, a simple scalar QED model fails to give a finite result after the UV regulator on the internal photon is lifted. In the same time, it has been observed that, with a double-Pauli-Villars regulator at a cutoff scale $\Lambda$ around 1 to 2 GeV, the charged-pion loop integrand can be well approximate by a simple ansatz Eq.~\eqref{eq:chargedpion_fitfunction}. Due to the similar asymptotic behavior between the double-Pauli-Villars regulator and the more realistic VMD parametrization of the $\pi\pi\gamma$ form factor which leads to a finite contribution to the $(2+2)a$ diagram, we expect Eq.~\eqref{eq:chargedpion_fitfunction} to provide a reasonable description of the integrand data after subtracting the PME. As a result, we fit our lattice integrand data to the following ansatz ensemble by ensemble:
\ba 
\label{eq:integrand_model}
f^{\textrm{model}}(|x|) = f^{\textrm{PME}}(|x|) + f^{\pi^+\pi^-}(|x|),
\ea 
where the exponent in Eq.~\eqref{eq:chargedpion_fitfunction} is fixed to $n=1.65(5)$ and $A$ is the only fit parameter.
To obtain the tail contribution, we choose to fit the parameter $A$ to the lattice data on each ensemble individually. 
One could also fix the parameter $A$ by using the results for $a_\mu^{\pi^+ \pi^-}$ computed with a VMD form factor in section \ref{sec:pure_sQED}, given in table \ref{table:PionLoop}. For $m_\pi>200$ MeV both procedures give consistent results. However, for $m_\pi<200$ MeV, we see that the latter method results in a smaller value of $A$.
The lattice data on D450 and E300, as well as the model predictions are depicted in figure~\ref{fig:D450_integrand}.
We choose $|x|_{\rm cut} = L/2$ for each ensemble. The lattice data and their tail correction according to the prescription described so far are summarized in table~\ref{tab:ensemble_results}.
We assign 50\% of the integrated tail as systematic error.

Finally, we recall that the choice for the 'TL' kernel in section~\ref{sect:nlohvp_fv} is motivated by the observed milder finite-volume effects for $|x| < L/2$. The difference between the central values of the integrand of C101 and S100, with $m_\pi L = 4.6$ and $2.9$ respectively, is below 15\%. We thus assign a conservative 15\%-systematic error for the finite-volume effects for the lattice data below $|x|_{\rm cut}$.
Since all ensembles that are taken into account for the final result have $m_\pi L > 4$, this estimate is expected to give an upper bound for the error from the finite-size effect.

\subsection{Extrapolation to the physical point}
\label{sect:nlohvp_extrapolation}
In order to obtain a result at the physical point, we use the dimensionless parameter 
\ba 
\tilde{y} = \frac{m_\pi^2}{16\pi^2 (f_\pi^{\text{phys}})^2}
\label{eq:sy_def}
\ea 
computed with the pion mass on each ensemble and the physical value for the pion decay constant $f_\pi^{\text{phys}}=92.32$ MeV.
To get the result at the physical point, we interpolate the result for the $a_\mu^{(2+2)a-ll}$ contribution in $\sy$ in order to match the mass of the physical neutral pion $m_{\pi^0}= 134.9768 \textrm{ MeV}$ and extrapolate to the continuum $a\to 0$ and infinite volume $L \to \infty$.
We fit the lattice data to several ans\"atze of type
\ba 
\nonumber
a^{(2+2)a-ll}_\mu(\sy,a,m_\pi L) &=& A_0 + A_1 (\sy)^{-3} + A_2  (\sy)^{-2}+ A_3  (\sy)^{-1} + A_4 (\sy)\\
\nonumber
& & +A_5  (\sy)^2+A_6  (\sy)^3  +A_7 \sy \log(\sy) +B_0 a +B_1 a^2\\
& & + B_2 a^3  + B_3 a\log(a)  +B_4 a/ \log(a)+ C_0 \exp(-m_\pi L),
\label{eq:chiral_fit_form}
\ea 
where $A_i$, $B_i$ and $C_0$ are real coefficients.
This is again inspired by the phenomenological model of the $(2+2)a$ contribution in terms of the pseudoscalar meson exchange and the charged pion loop, given in Eq.~\eqref{eq:full_model}.
We expect a steep enhancement when decreasing the pion mass to approach its physical value as dictated by the charged pion loop, which scales heuristically as $a_\mu^{\pi^+\pi^-}\propto m_\pi^{-3}$ for $m_\pi^{\textrm{phys}}\leq m_\pi <m_\pi^{\textrm{SU}(3)}$  -- see figure~\ref{fig:tpta_pheno_full_model}.

\begin{figure}
    \centering
    \includegraphics[width=0.8\linewidth]{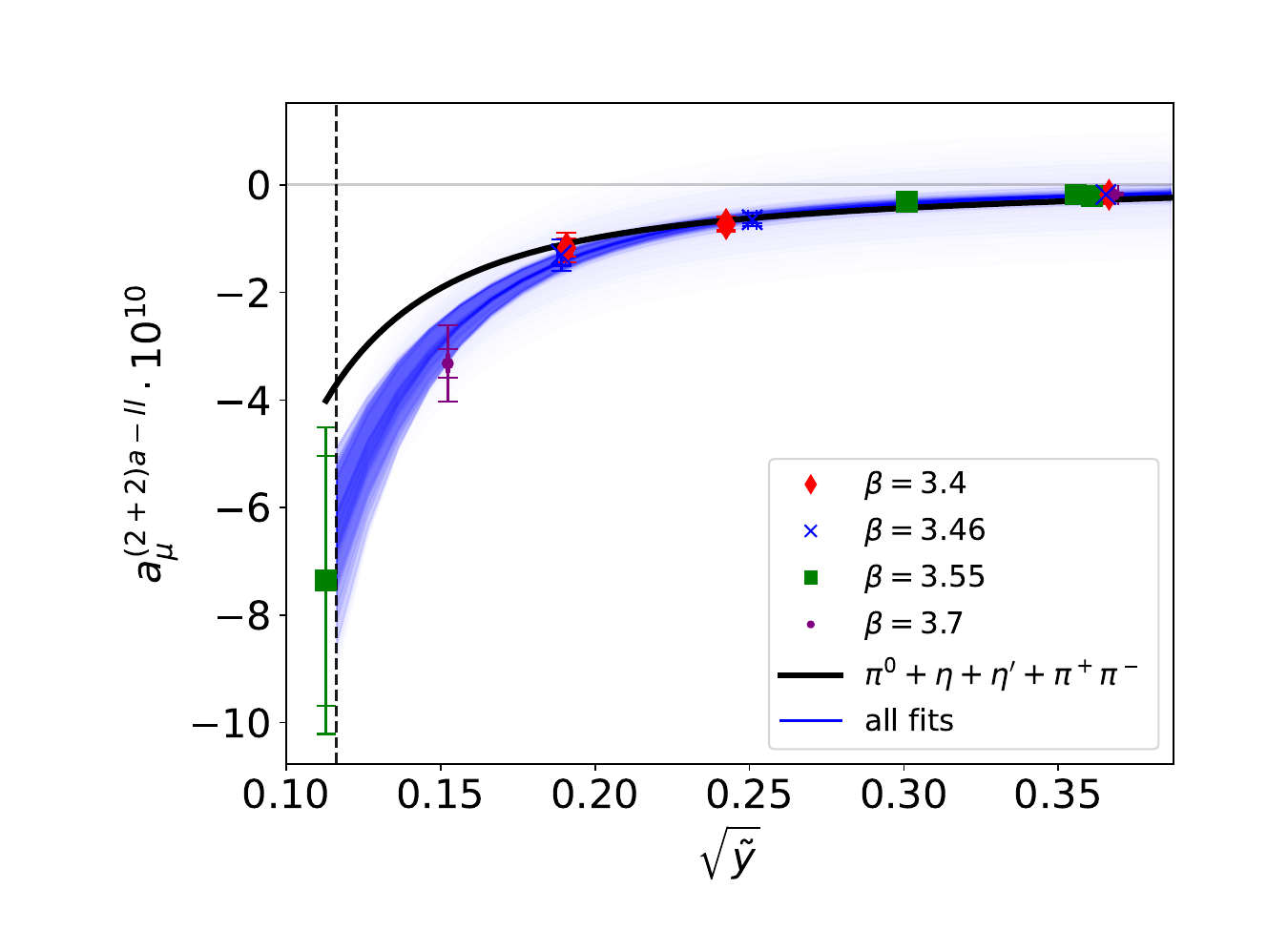}
    \caption{Results computed on different lattice ensembles and collection of all fitted curves Eq.~\eqref{eq:chiral_fit_form}. The opacity of the fits is proportional to the associated AIC weight. We also plot the prediction from the model Eq.~\eqref{eq:full_model} (black solid line). The inner error bar of the lattice data points gives only the statistical uncertainty, the outer error bar includes the uncertainty from approximating the tail of the integrand and the finite-size error. The vertical dashed line corresponds to the physical mass of the neutral pion.}
    \label{fig:chiral_fit}
\end{figure}

We do not fit all parameters in Eq.~\eqref{eq:chiral_fit_form} simultaneously. Instead, we perform multiple fits, where we drop several of the terms. The terms multiplying $A_0$ and $A_1$ are always included, but we keep at maximum two other terms in all possible combinations. We apply the above fitting scheme also to subsets of the data with 1.) those with $m_\pi>400$ MeV dropped and 2.) those with $m_\pi < 150$ MeV dropped. 
We use the Akaike Information criterion (AIC) \cite{Akaike1998} to compute the weight for each fit
\ba 
w_i = \frac{1}{N}\exp\Big[-\frac{1}{2}( \chi^2 + 2k -2n )\Big]\,,
\ea 
where $k$ is the number of fit parameters and $n$ is the number of data points that are included. The normalization factor $N$ is chosen such that the sum of all weights adds up to unity.
We found that the fit with only the two terms $A_0 + A_1 (\sy)^{-3} $ has by far the largest AIC weight. For all fits the pion-mass dependence  is governed by the $(\sy)^{-3}$ term numerically, as seen in figure~\ref{fig:chiral_fit}.
Additionally, we observe that there is almost no visible shifts due to the lattice spacing $a$ and $m_\pi L$ within the quoted statistical uncertainties.
The central value of our final result is obtained by averaging the results from the individual fits with their assigned AIC weights~\cite{Ce:2022kxy},
\ba 
\label{eq:nlohvp_model_av}
\overline O = \sum_i w_i O_i\,.
\ea 
In order to disentangle the statistical errors from the systematic uncertainties associated with our tail reconstruction and finite-size effects, we first compute Eq.~\eqref{eq:nlohvp_model_av} and the AIC weights without assigning further systematic errors for the last two for each jackknife sample.
The statistical error is then determined from the variance $(\sigma^2)_{\rm stat}$ of the AIC-weighted result obtained in this way.
Similarly, we infer the error associated to the tail reconstruction (resp. finite-volume effects) from the variance of Eq.~\eqref{eq:nlohvp_model_av} by considering the finite-size effects (resp. tail reconstruction) as exact with the statistical fluctuation neglected ($(\sigma^2)_{\textrm{tail/finite-size}}$).
The uncertainty associated with the chiral extrapolation is calculated by the variance of the weighted fit results
\ba 
(\sigma^2)_{\textrm{extr}} = \sum_i w_i (O_i- \overline O)^2.
\ea 
Treating these different sources of uncertainty as uncorrelated, we have, for the total variance,
\ba 
(\sigma^2)_{\text{total}} = (\sigma^2)_{\textrm{stat}}+(\sigma^2)_{\textrm{tail}}+(\sigma^2)_{\textrm{finite-size}}+(\sigma^2)_{\textrm{extr}} .
\ea 
Our final result is given by
\ba 
\nonumber
 a_\mu^{(2+2)a-ll} = \Big(-6.42\pm(0.89)_{\textrm{stat}} \pm(0.81)_{\textrm{tail}}\pm(0.51)_{\textrm{finite-size}}\pm(0.30)_{\textrm{extr}} \Big)\times 10^{-10}.\\
 \label{eq:tpta_result}
\ea  
We postpone the discussion of this result to section \ref{sect:discu}.
 As a remark, using the value of $f_\pi$ on each ensemble instead of the physical value $f_\pi^{\rm phys}$ modifies the dependence on the lattice spacing in the proxy variable $\sy$ \eqref{eq:sy_def} . These effects are enhanced by the $(\sy)^{-3}$ behaviour when approaching the physical point, resulting in a larger uncertainty from the chiral extrapolation due to the introduced cutoff effects, while giving a compatible result.
\section{Additional contributions to $a_\mu^{{\rm hvp}1\gamma^*}$}\label{sec:add-contrib}
In this section, we present exploratory calculations of two additional contributions to $a_\mu^{{\rm hvp}1\gamma^*}$. Rather than performing a full analysis, we provide further evidence for the viability of our coordinate-space formalism. In section \ref{sec:strange},
we calculate the contributions from the cases where one or both of the quark loops are comprised of strange quarks.
This study is only carried out on one ensemble at a pion mass of $m_\pi=291(4)$ MeV.
Secondly, we show a calculation of the $(2+2)b $ contribution using Eq.~\eqref{eq:2+2_master_b}. Since this diagram is UV-divergent, it is necessary to specify a full renormalization scheme for QED+QCD before taking the limit $\Lambda \to \infty$, which we do not attempt to do in this work. Here, we want to demonstrate that with our coordinate-space method, the statistical fluctuation is of comparable in relative size to the $(2+2)a$ case.

\subsection{Light-strange and strange-strange components}
\label{sec:strange}
\begin{figure}
\centering
\begin{subfigure}{0.49\textwidth}
    \centering
    \includegraphics[width=0.99\textwidth]{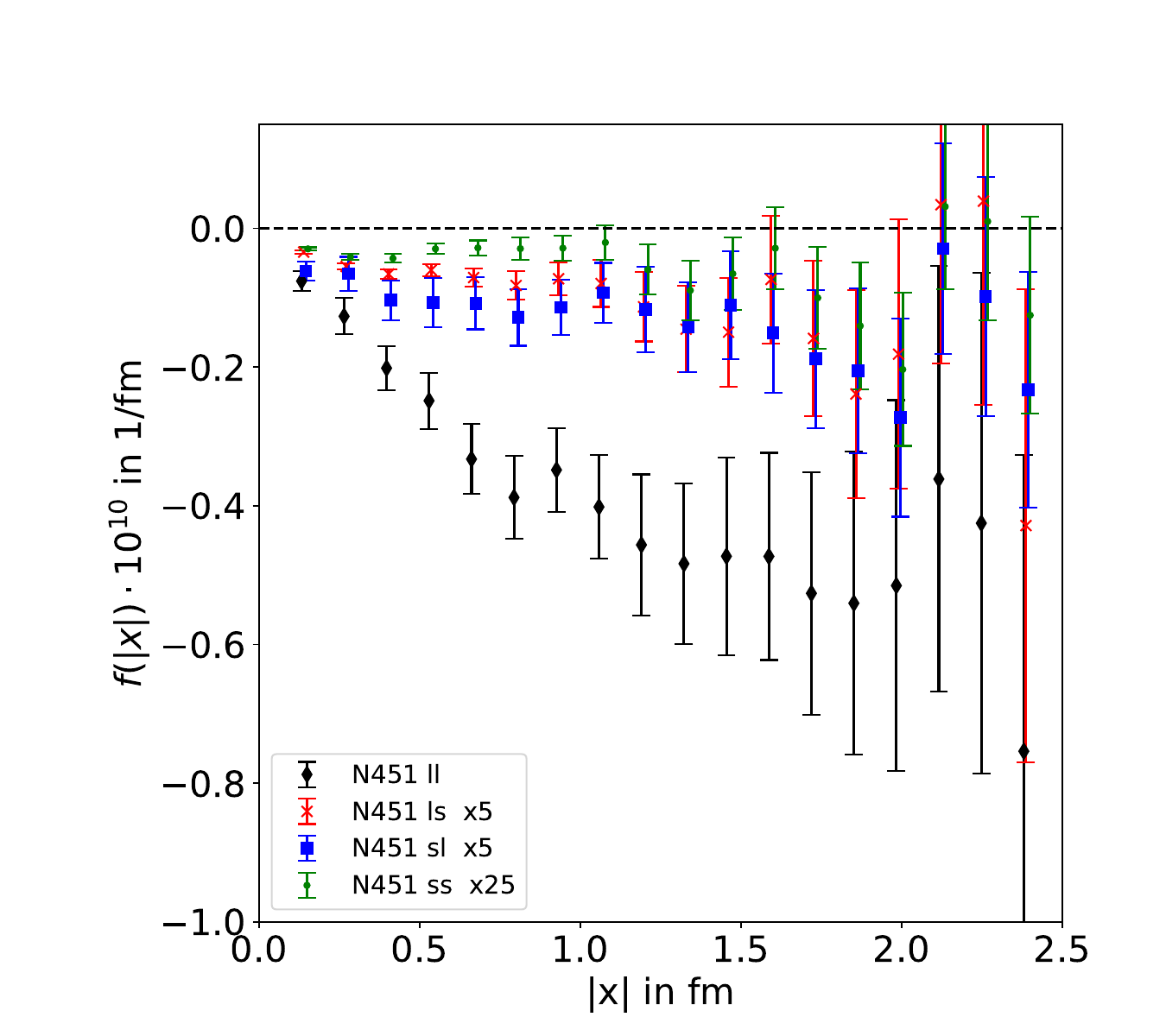}
    \caption{}
    \label{fig:N451_all}
\end{subfigure}
\begin{subfigure}{0.49\textwidth}
    \centering
    \includegraphics[width=0.99\textwidth]{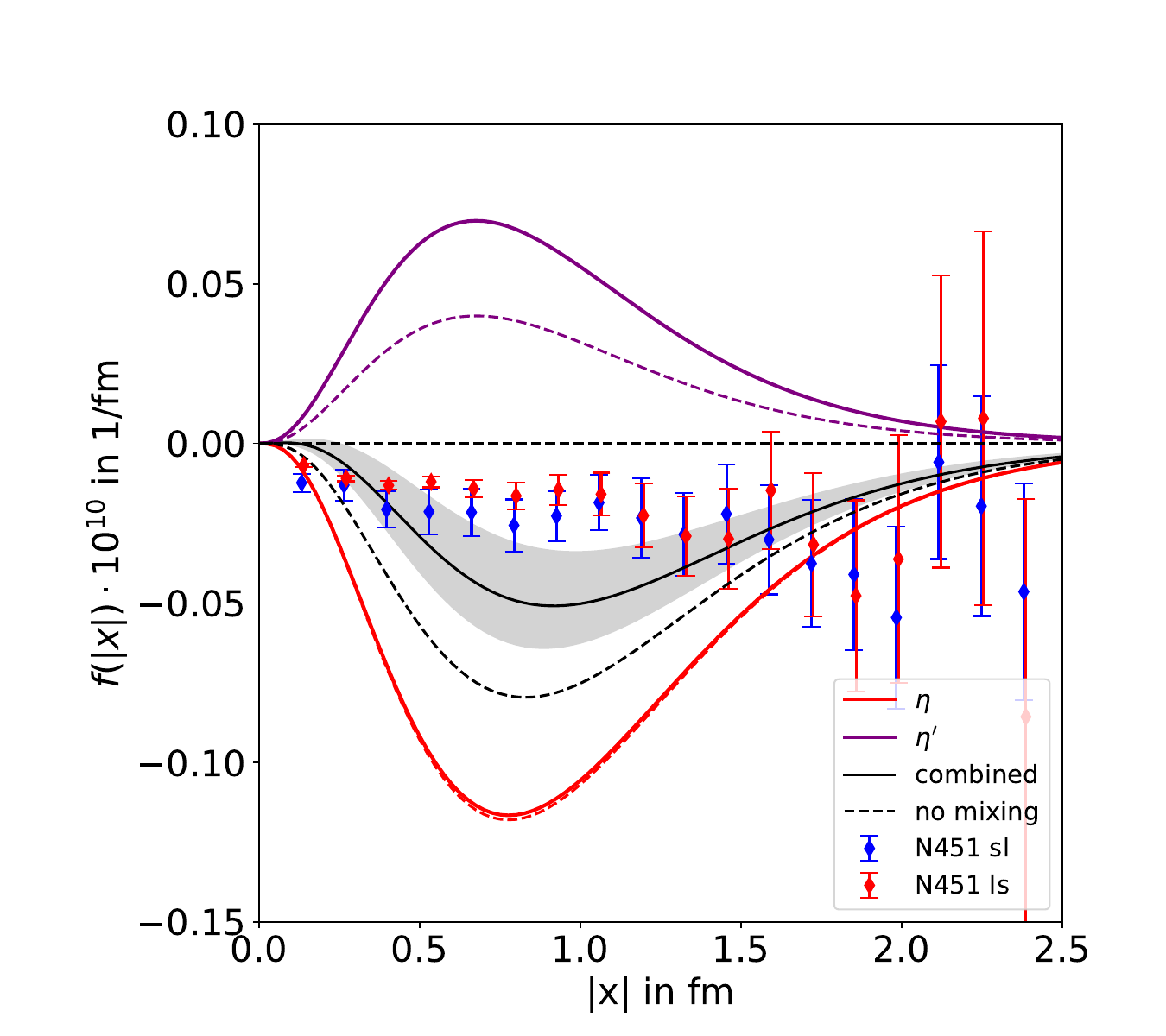}
    \caption{}
    \label{fig:ls_comparison}
\end{subfigure}
    \caption{(a) Comparison between the $(ll)$, $(ls)$, $(sl)$ and $(ss)$ components of Eq.~\eqref{eq:2+2_master_a} on the ensemble N451. All the components are scaled to have the same charge factor as the $(ll)$ component. (b) Comparison between the result computed in lattice QCD and the model predictions for the $(ls)$ and $(sl)$ contribution, using the matching coefficients  Eq.~\eqref{eq:matching_eta_ls} for the $(ls)$ and $(sl)$ components with mixing angle $\theta=6.47^\circ$. The dashed lines correspond the the case $\theta=0$.}
\end{figure}

\begin{table}[h]
\centering
\caption{Results on the ensemble N451 for contribution to $a_\mu^{(2+2)a}$ \eqref{eq:2+2_master_a}, where the lattice results and the prediction of the $\eta+\eta'$ contribution for the $(ls)$ component are integrated up to $|x|=1.8$ fm, corresponding to $L/2$ on that ensemble. The results for all components contain the relevant charge factors $C^{(ff')}$.}
\label{table:N451_results}
\begin{tabular}{|c|c|c|c|c|c|}
\hline
& $(ll)$& $(ls)$& $(sl)$& $(ss)$&  $\eta+\eta'$ $(ls)$\\ \hline
$a_\mu^{(2+2)a} \times 10^{10}$& -0.565(72)& -0.028(6)& -0.037(8)& -0.003(1)&  -0.054(18)\\
\hline
\end{tabular}
\end{table}
Here, we want to look at the light-strange $(ls)$, strange-light $(sl)$ and strange-strange $(ss)$ components of the $(2+2)a$ contribution Eq.~\eqref{eq:2+2_master_a}. We have computed these on  the ensemble N451, on which the pion mass and kaon masses are $m_\pi=291(4)$ MeV and $m_K=468(5)$ MeV, see figure~\ref{fig:N451_all}. At the SU$(3)_f$ symmetric point, the only difference between the $(ll)$, $(ls)$, $(sl)$ and $(ss)$ components is that they contribute with different charge factors $C^{(ls)}=C^{(sl)}=\frac{5}{81}$ and $C^{(ss)}=\frac{1}{81}$. But, following the chiral trajectory to the physical point, the correlation functions calculated with strange quarks become much smaller in size compared to the $(ll)$ case as the difference $m_K^2-m_\pi^2$ increases. In figure~\ref{fig:N451_all}, we see that, if one scales all the components to have the same charge factor as the $(ll)$ component, the latter is significantly larger on the ensemble N451. We can also confirm that the $(ls)$ and $(sl)$ components agree within error. However, the $(ls)$ component, where the integral $I^{(2)}$ Eq.~\eqref{eq:I2} is calculated with the light-quark two-point function and the integral $I^{(3)}$ Eq.~\eqref{eq:I3} is calculated with the strange-quark two-point function, has smaller statistical uncertainties in the small-$|x|$ region.

We also examine the prediction of the pseudoscalar meson exchange for the $(ls)$ and $(sl)$ components. Among the processes considered in section~\ref{sec:phenomenological_model}, these components receive only contributions from the $\eta$, $\eta'$ and the charged kaon loop.
In figure~\ref{fig:ls_comparison}, we display the integrand of the prediction for the $\eta$ and $\eta'$ contribution with appropriate matching factors Eq.~\eqref{eq:matching_eta_ls} calculated with a mixing angle $\theta=-6.37^\circ$, given in table \ref{table:mixing_angle_coefficients}. We see a strong cancellation between both contributions. Because of this cancellation and the fact that the matching factors for the $(ls)$ and $(sl)$ components are very sensitive to the variations on $m_\eta$ and $m_K$, we see a relatively large uncertainty from the model prediction\footnote{Since for the $(ll)$ component the $\pi^0$ is by far the dominant contribution, the uncertainty of the model prediction due to the matching coefficients is largely reduces for this component compared to the $(ls)$ and $(sl)$ components.}.
However, we can observe that the model prediction falls in the bulk of the result computed on the lattice.
We also see that if one assumes the matching factors to be the same as on the SU$(3)_{\rm f}$ symmetric point Eq.~\eqref{eq:matching_ls_SU3} the agreement between the model and the lattice data is worse.

The charged kaon also contributes to the $(ls)$ and $(sl)$ components, which in principle can be treated similarly to the charged pion loop.
But given the large statistical uncertainty of the lattice data and the mediocre prediction from the pseudoscalar meson exchange, repeating the fitting procedure that we applied for the charged pion for the $(ll)$ component in section \ref{sect:nlohvp_tail} is not feasible for the case of the $(ls)$ and $(sl)$ components.

Looking at table \ref{table:N451_results}, we see that the results for the combined contribution from the $(ls)$, $(sl)$ and $(ss)$ components amount to about $60\%$ of the size of the quoted uncertainty of
the $(ll)$ contribution. Since the $(ls)$ and $(sl)$ components are dominated by the $\eta$ and $\eta'$ exchange and the charged kaon loop contribution, which do not increase significantly when approaching the physical point, we expect the $(ls)$ and $(sl)$ components at the physical point to be negligible compared to the uncertainty of our result for $a_\mu^{(2+2)a-ll}$, given in Eq.~\eqref{eq:tpta_result}. As the strange-quark mass gets heavier on our chiral trajectory, when approaching the physical point, we also expect the $(ss)$ component to be negligible compared to the $(ll)$ component.

\subsection{Computing the $(2+2)b$ contribution with a double Pauli-Villars regulator} \label{sec:tptb}
\begin{figure}
    \includegraphics[width=0.8\linewidth]{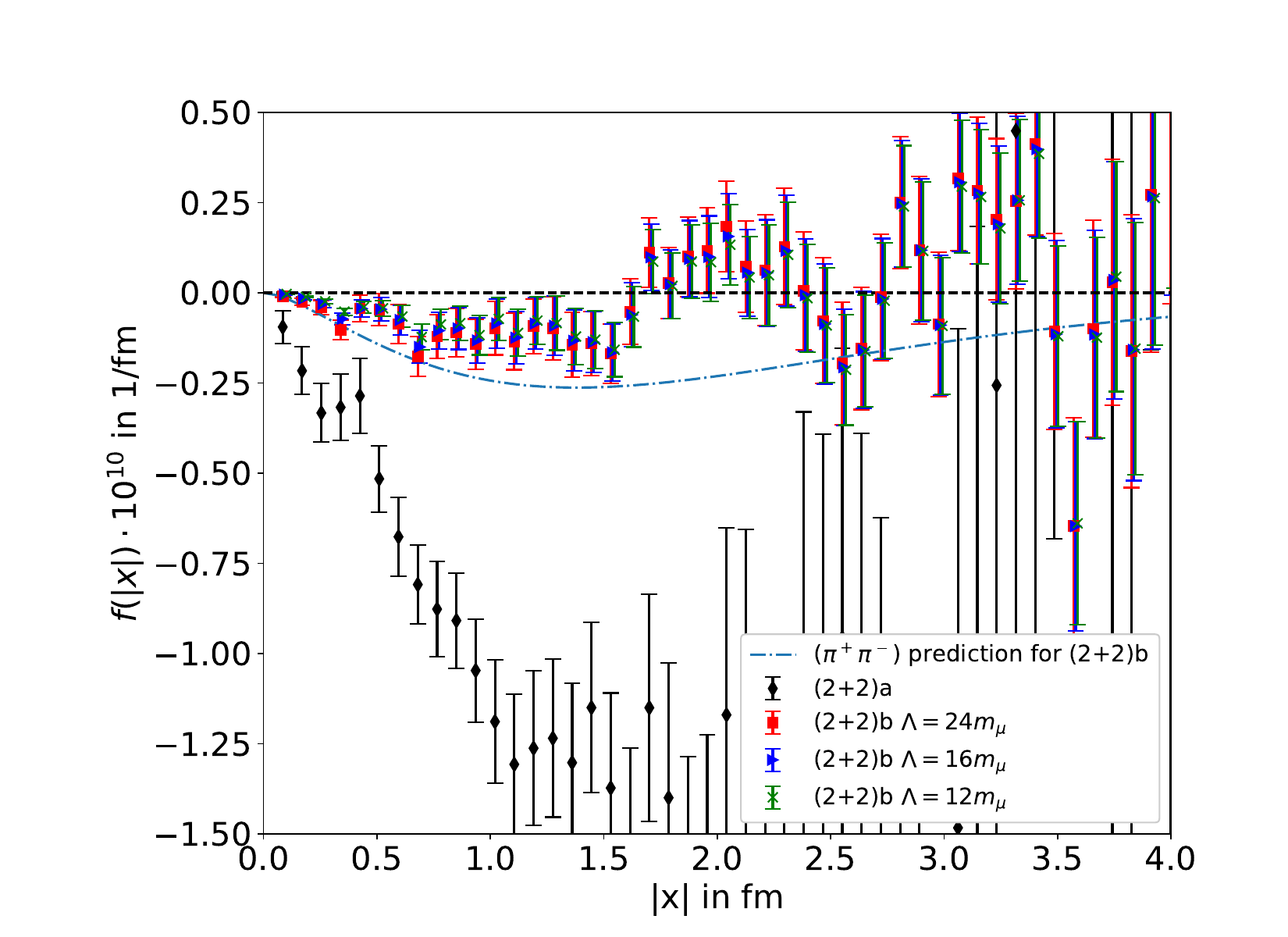}
    \centering
    \caption{Comparison of the integrand of the $(2+2)a$ contribution (black dots) Eq.~\eqref{eq:2+2_master_a} and the $(2+2)b$ contribution Eq.~\eqref{eq:2+2_master_b} on the ensemble E300 for three values of the photon mass $\Lambda=\{12, 16, 24\} m_\mu $. We also plot the prediction for the $(2+2)b$ contribution from the charged pion loop  (dashed blue line).} 
    \label{fig:tptb}
\end{figure}

As mentioned previously in section~\ref{sec:disco_def}, the $(2+2)b$ diagram can be computed using Eq.~\eqref{eq:2+2_master_b} with similar lattice techniques as in the $(2+2)a$ case.
The major difference compared to the latter is that this contribution is UV-divergent by itself when the UV-regulator on the internal photon is removed.
To arrive at a finite, physical quantity, one needs to renormalize the contribution from the $(2+2)b$ diagram together with other UV-divergent ones. 
Since we do not attempt to perform a full calculation of all diagrams in this work, we can only obtain a finite result for the $(2+2)b$ contribution if we keep the cutoff scale $\Lambda$ of the regulator finite. For the current study, we will use the double Pauli-Villars regulator Eq.\ \eqref{eq:I1}.
Using the same lattice setup as described in section \ref{sect:nlohvp_setup}, we perform a calculation of this diagram on the ensemble E300 for $\Lambda =\{12,16,24\}m_\mu$. The integrands for the $(2+2)b$ diagram are depicted in figure~\ref{fig:tptb}. 
For a comparison, we also give the integrand of the $(2+2)a$ contribution with $\Lambda \to \infty$. Despite the fact that the charge factor of the $(2+2)b$ contribution is only a factor of two smaller than that of the $(2+2)a$ contribution, we find a much smaller amplitude for the $(2+2)b$ contribution. We also observe that the dependence on the value of $\Lambda$ is  weak in the explored range of $\Lambda$. 
For a naive phenomenological estimate we also display a prediction from the charged pion loop, which contributes with $25/81$ to the $(2+2)b$ topology. The integrand for the latter is obtained by using the ansatz Eq.~\eqref{eq:chargedpion_fitfunction}, where prefactor $A$ is fixed, such that the integrated result matches $a_\mu^{\pi^+ \pi^-}$ given in table \ref{table:PionLoop}
calculated with a VMD form factor according to the description in section~\ref{sec:piloop-cont}. 

\section{Discussion of the results \label{sect:discu}}
In this work, we have investigated some aspects of the leading electromagnetic corrections to the O($\alpha^2$) hadronic vacuum polarization contribution to the anomalous magnetic moment of the muon using the coordinate-space-based lattice QCD formalism proposed in Ref.~\cite{Biloshytskyi:2022ets}.
In particular, we have identified a class of quark-contraction diagrams that remain finite when the UV-regulator on the internal photon is removed. 
These quantities can thus by themselves serve as benchmark quantities that can be compared among different lattice collaborations.
For the total isospin-breaking corrections to the HVP however, a full renormalization scheme for QCD+QED needs to be specified, which necessitates the computation of additional counterterms for the diagrams that are not finite in the  sense described above.

We have calculated one of those UV-finite contributions with two two-point quark-loops, which we refer to as $(2+2)a$ (see figure~\ref{fig:FiniteDiags}).
For the dominant component, where only light quarks are present in the loops, we obtain
\begin{equation}
a_\mu^{(2+2)a-ll} = \left(-6.4\pm1.3 \right)\times 10^{-10},
 \label{eq:tpta_result}
\end{equation}
which amounts to $-0.89\%$ of the leading order HVP contribution calculated by the Mainz group~\cite{Djukanovic:2024cmq}.
Our result is in good agreement with the result of the RBC/UKQCD collaboration\footnote{The complete calculation of the BMW collaboration~\cite{Borsanyi:2020mff} for $\ahvpnlo$ contains in particular the diagram (2+2)a, however no result is quoted for this diagram on its own.} on one gauge ensemble, where they obtained $a_\mu^{(2+2)a-ll}= (-6.9\pm2.9)\times 10^{-10}$~\cite{RBC:2018dos}.
We point out that the method used in this work can furthermore be applied to a calculation of the other UV-finite QED corrections, depicted in boldface in figure~\ref{fig:FiniteDiags}.

In order to better understand the pion-mass dependence of $\ahvpnlo$, we have calculated the O($\alpha^3$)-contributions to $\ahvp$ of the $\pi^0/\eta/\eta^\prime$-exchanges and charged-pion and kaon loops based on phenomenological models in section~\ref{sec:model_precition_full_em}.
For comparison, their contributions to the $(2+2)a$ QCD-Wick-contraction diagram have been worked out in section~\ref{sect:hvpnlo_matching}.
Numerically, the charged-pion loop becomes most dominant when the pion mass is close to its physical value. 
In the range of pion masses included in our final analysis, the charged-pion-loop contribution is found to be approximately proportional to $m_\pi^{-3}$.
This behavior is also observed in our lattice data, making it essential to perform lattice calculations at close-to-physical pion masses in order to have a more robust chiral extrapolation.
A qualitative explanation for the sign and size of the $(2+2)a$ diagram is that it contains the effect of the electromagnetic self-energy of the pion, \emph{de facto} raising the pion mass-squared by a finite but sizeable amount, thereby leading to a reduction of $a_\mu^{\rm hvp}$, i.e.\ the diagram contributes negatively.
For orientation, we have also used phenomenological models to obtain a prediction for the total leading electromagnetic corrections to the HVP, starting from the isospin-symmetric world defined by FLAG~\cite{FlavourLatticeAveragingGroupFLAG:2024oxs} and fixing the counterterms to account for the mass shifts of the charged pion and kaon from their physical values. The overall correction is negative. The interested reader is invited to consult section~\ref{sec:model_precition_full_em}.

Although our main result is  for the $(2+2)a$ contribution, one can use the model described in section \ref{sec:phenomenological_model} to estimate the contributions for other classes of diagrams. In the recent $a_\mu^{\rm hvp}$ calculation of the Mainz group \cite{Djukanovic:2024cmq}, some aspects of this model have been used to account for those QED corrections that the collaboration has not yet computed on the lattice.
One observes a cancellation between the disconnected and connected contributions, reducing the absolute size of the total correction. 
However, in terms of uncertainties, this correction presently makes a sizeable contribution to the error budget of $a_\mu^{\rm hvp}$.
Targeting a precision of $0.2\%$ for $\ahvp$, it is crucial to gain better control over these effects in the near future.

\acknowledgments
This work was supported in part by the European Research Council (ERC) under the
  European Union's Horizon 2020 research and innovation programme
  through grant agreement 771971-SIMDAMA, as well as by  
 the Deutsche
  Forschungsgemeinschaft (DFG) through the Cluster of Excellence \emph{Precision Physics, Fundamental Interactions, and Structure of Matter} (PRISMA+ EXC 2118/1) within the German Excellence Strategy (Project ID 39083149). 
  V.\ P.\ and V.\ B.'s work was supported by the DFG
within the Research Unit FOR\;5327
``Photon-photon interactions in the Standard Model and
beyond -- exploiting the discovery potential from MESA
to the LHC'' (grant 458854507).
E.-H.C.'s work was supported by the US Department of Energy grant No.~DE-SC0011941.
Calculations for this project were partly performed on the HPC clusters ``Mogon II'' and ``Mogon NHR'' at JGU Mainz. 

Our simulation programs use the deflated SAP+GCR solver from the \texttt{openQCD} package~\cite{Luscher:2012av}, as well as the \texttt{QDP++} library
\cite{Edwards:2004sx}.
The measurement codes were developed based on the C++ library \texttt{wit}, a coding effort led by Renwick J.~Hudspith. 
The lattice charged-pion loop program was developed based on \texttt{Grid}~\cite{grid}.
E.-H.C. thanks Renwick J.~Hudspith for useful discussions on the numerical work related to the latter program.

We are grateful to our colleagues in the CLS initiative for sharing ensembles. We would also like to acknowledge Nils Hermansson-Truedsson for providing a code that allows for an independent cross-check of the scalar-QED two-loop vacuum polarization in the continuum.

\appendix
\section{One-loop LbL amplitudes, subtraction functions and cross sections in scalar QED}\label{app:SQEDamps}
The one-loop helicity-averaged forward LbL amplitude in sQED with unit mass of the charged scalar particle in the loop is given by:
\begin{align}
	&\mathcal{M}(\nu,K^2,Q^2) = 2\alpha^2\bigg(-24+\bigg\{
	2\sqrt{Q^2+4}\log\left[\frac{1}{2}Q(Q+\sqrt{Q^2-4})\right]\nonumber\\
	&\times \bigg(K^2Q^4(4+7K^2)(4+K^2+Q^2)^2\nonumber\\
	&-4\nu^2Q^2\left[(4+K^2)(4+15K^2)+Q^2(4+17K^2+7K^4)\right]+96\nu^4
	\bigg)\nonumber \\
    & \bigg/\bigg(Q\big[(KQ)^4(4+K^2+Q^2)^2\nonumber
	-4(KQ)^2\left(K^2(Q^2+2)+2(Q^2+4)\right)\nu^2+16\nu^4\big]\bigg) \nonumber\\
    & +\left\{K\leftrightarrow Q\right\}\bigg\}\nonumber\\
	&-\bigg\{\frac{2\sqrt{1+\frac{4}{K^2+Q^2+2\nu}}\log\left[\frac{1}{2}\left(\sqrt{(K^2+2\nu+Q^2)(K^2+2\nu+Q^2+4)}+K^2+2\nu+Q^2+2\right)\right]}{K^2Q^2(K^2+Q^2+2\nu+4)-4\nu^2}\nonumber\\
	&\times\big(
	(16+4(K^2+Q^2)+5K^2Q^2)(\nu+2(K^2+Q^2))+8((K^2+Q^2)K^2Q^2-\nu^2)
	\big)\nonumber\\
	&+\bigg(\frac{\left(K^2+Q^2+2\nu+6\right)(K^2+Q^2)+\frac{1}{2}K^2Q^2+8(\nu+1)}{\nu} \nonumber \\
    &\times C_0(-K^2,-Q^2,-K^2-2\nu-Q^2;1,1,1) \bigg)+\left\{\nu\to -\nu\right\}\bigg\}
	\bigg)
\end{align}
Here $C_0$ is the triangle scalar loop integral defined in a standard way, following {\tt LoopTools} \cite{Hahn:1998yk} and {\tt Package-X} \cite{Patel:2015tea,Patel:2016fam} conventions.
The expressions for the subtraction function at the subtraction points $\bar\nu=0$ and $\bar\nu = KQ$ are, respectively:
\begin{align}
	&\mathcal{M}(0 ,K^2,\,Q^2)= 4\alpha^2 \Bigg\{-12+4(K^2+Q^2+4)C_0(-K^2,-Q^2,-K^2-Q^2;1,1,1)\nonumber\\
	&+\frac{\sqrt{K^2+4}\left[-2K^4+2K^2(Q^2-2)+5Q^2(Q^2+4)\right]}{KQ^2(K^2+Q^2+4)}
	\log\left[\frac{1}{2}K\left(K+\sqrt{K^2+4}\right)+1\right]\nonumber\\
	&+\frac{\sqrt{Q^2+4}\left[-2Q^4+2Q^2(K^2-2)+5K^2(K^2+4)\right]}{QK^2(K^2+Q^2+4)}
	\log\left[\frac{1}{2}Q\left(Q+\sqrt{Q^2+4}\right)+1\right]\nonumber\\
	&+2\sqrt{\frac{K^2+Q^2+4}{K^2+Q^2}}\frac{(K^4+Q^4)}{K^2Q^2} \log\left[\frac{1}{2}\left(K^2+Q^2 \sqrt{(K^2+Q^2)(K^2+Q^2+4)}\right)+1\right]
	\Bigg\}
\end{align}

\begin{align}
	&\mathcal{M}(KQ ,K^2,\,Q^2)= 4\alpha^2 \Bigg\{-12+\frac{\left(K^2+Q^2-2\right) \sqrt{(K+Q)^2+4}}{K^2 Q^2 (K+Q)}\nonumber\\
	&\times\left[(K^2+Q^2)\left((K+Q)^2+4\right)+3\left(8+2(K+Q)^2+K^2Q^2\right)\right]\nonumber\\
	&\times\log \left[\frac{1}{2} (K+Q)
	\left(\sqrt{(K+Q)^2+4}+K+Q\right)+1\right]\nonumber\\
	&+\frac{\left(K^2+Q^2-2\right) \sqrt{(K-Q)^2+4}}{K^2 Q^2 | K-Q| }\nonumber\\
	&\times\left[(K^2+Q^2)\left((K-Q)^2+4\right)+3\left(8+2(K-Q)^2+K^2Q^2\right)\right]\nonumber\\
	&\times\log \left[\frac{1}{2} | K-Q|  \left(| K-Q|
	+\sqrt{(K-Q)^2+4}\right)+1\right]\nonumber\\
	&+\frac{ \sqrt{K^2+4} \left[2(K^2+4)^2+4(2K^2-1)Q^2-9Q^4\right] }{K Q^2  \left(K^2-Q^2\right)}\nonumber\\
	&\times\log \left[\frac{1}{2} K
	\left(\sqrt{K^2+4}+K\right)+1\right]\nonumber\\
	&+\frac{\sqrt{Q^2+4} \left[2(Q^2+4)^2+4(2Q^2-1)K^2-9K^4\right] }{K^2 Q 
		\left(Q^2-K^2\right)}\nonumber\\
	&\times\log \left[\frac{1}{2} Q \left(\sqrt{Q^2+4}+Q\right)+1\right]
	\Bigg\}.
\end{align}

The cross sections that correspond to the
fusion of two polarized photons, either transverse ($T$) or longitudinal ($L$),  to two charged spin-$0$ particles, read:
\begin{subequations}
\begin{eqnarray}
	\sigma_{TT}(\nu,Q^2,K^2) &=& \frac{1}{2}\left(\sigma_\parallel+\sigma_\perp\right)\nonumber\\
	&=&\frac{\alpha^2\pi}{4}\frac{s^2\nu^3}{X^3} \bigg\{
	\sqrt{a}\left[2-a+\left(1-\frac{2X}{s\nu}\right)^2\right]\nonumber\\
	&&-(1-a)\left[3+a-\frac{4X}{s\nu}\right]L
	\bigg\}\theta(\nu-\nu_\mathrm{thr}),\\
	\sigma_{LT}(\nu,Q^2,K^2) &=& \sigma_{TL}(\nu,K^2,Q^2)\nonumber\\
	&=& \frac{\pi\alpha^2}{2}Q^2\frac{s\nu}{X^3}(\nu-K^2)^2\bigg\{-3\sqrt{a}+(3-a)L\bigg\}\theta(\nu-\nu_\mathrm{thr}),\\
	\sigma_{LL}(\nu,Q^2,K^2) &=& \pi\alpha^2Q^2K^2 \frac{s^2\nu}{X^3}\bigg\{
	\sqrt{a}\left[ 2+\frac{\left(1-\frac{X}{s\nu}\right)^2}{1-a}\right]\nonumber\\
	&&-\left[3+\frac{X}{s\nu}\right]\left[1-\frac{X}{s\nu}\right]L
	\bigg\}\theta(\nu-\nu_\mathrm{thr}).
\end{eqnarray}
\end{subequations}
These expressions are written following the conventions 
provided in Refs.~\cite{Pascalutsa:2012pr,Budnev:1975poe}:
\begin{equation}
	L \equiv \log\left(\frac{1+\sqrt{a}}{\sqrt{1-a}}\right),\quad 
	a \equiv \frac{X}{\nu^2}\left(1-\frac{4}{s}\right), \quad X = \nu^2-Q^2K^2,
\end{equation}
where $s=(k+q)^2=2\nu-K^2-Q^2$, $\nu=k\cdot q$; $Q^2=-q^2$ and $K^2=-k^2$ are the spacelike photon virtualities. The threshold energy in this case is given by $\nu_\mathrm{thr} = 2 + \nicefrac12 (K^2 +Q^2)$.
Under these conventions, the total unpolarized cross section, which is required for the Cottingham formula \eqref{eq:DRform}, is given by
\begin{eqnarray}
    \sigma &=& 4\sigma_{TT}-2\sigma_{LT}-2\sigma_{TL}+\sigma_{LL} = \frac{\pi\alpha^2}{X\nu}
	\Bigg\{\nu\sqrt{X\frac{4+K^2+Q^2-2\nu}{K^2+Q^2-2\nu}}\nonumber\\
	&&\times\frac{K^2[K^2(5Q^2+4)+8(2-\nu)]+Q^2[Q^2(5K^2+4)+8(2-\nu)]-2\nu[8(2+\nu)+5K^2Q^2])}{K^2Q^2(4-2\nu+K^2+Q^2)-4\nu^2}\nonumber\\
	&&+\left[2(K^2+Q^2)(K^2+Q^2+2(3-\nu))+K^2Q^2+16(1-\nu)\right]\nonumber\\
    &&\times\log
	\frac{1+\sqrt{\frac{X}{\nu^2}\frac{4+K^2+Q^2-2\nu}{K^2+Q^2-2\nu}}}{\sqrt{1-\frac{X}{\nu^2}\frac{4+K^2+Q^2-2\nu}{K^2+Q^2-2\nu}}}\Bigg\}\theta(\nu-\nu_\mathrm{thr}).
\end{eqnarray}

\section{Verification of the Cottingham-like formula in scalar QED}\label{app:CottinghamVerification}

\begin{figure}[h]
	\centering
	\includegraphics[width=\textwidth]{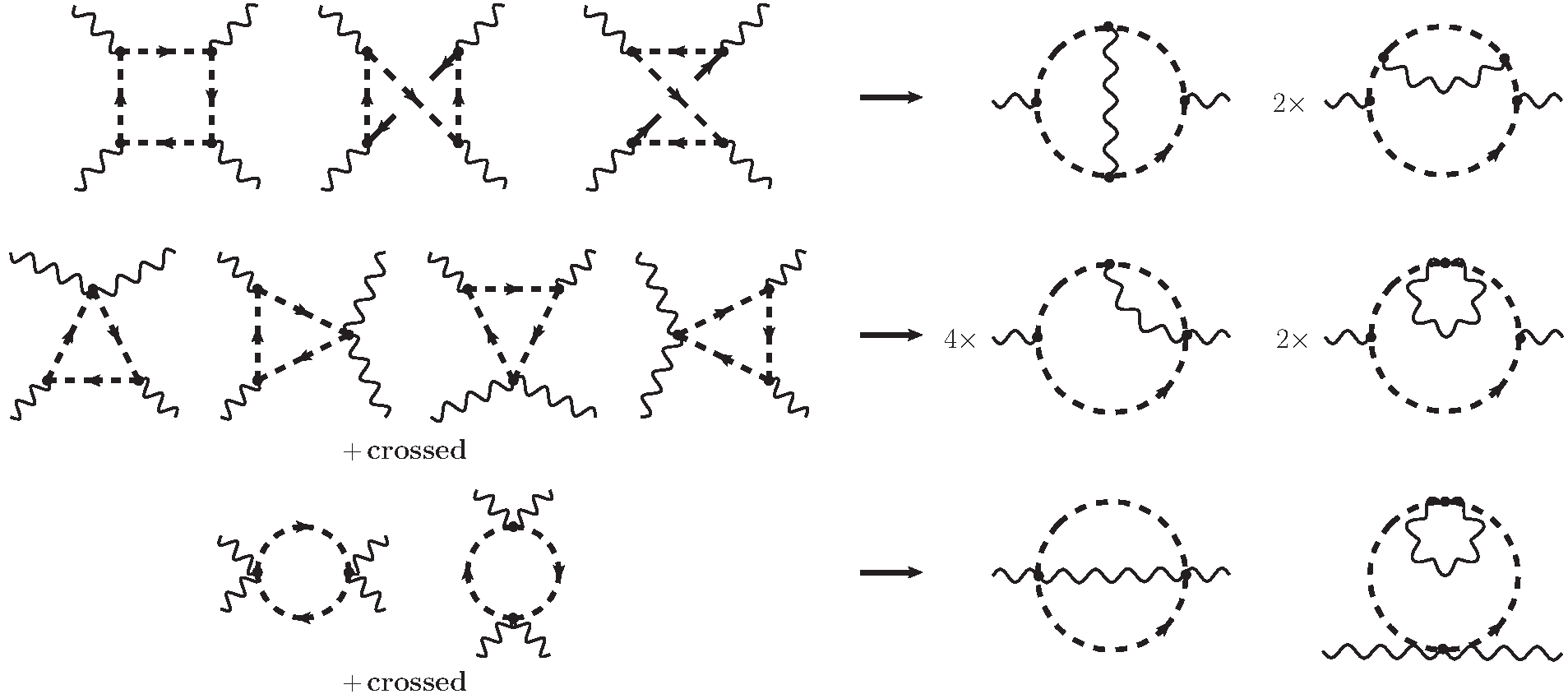}
	\caption{One-loop LbL scattering (left three diagrams) and the resulting two-loop vacuum polarization in sQED.}
	\label{fig:TwoLoopsQED}
\end{figure}
In this appendix, we perturbatively verify the Cottingham-like formula \eqref{eq:DRform} in sQED by considering the one-loop LbL amplitude, which is shown in figure~\ref{fig:TwoLoopsQED}.
Substituting the LbL amplitude into the Cottingham-like formula with a sharp cut-off in the internal photon momentum $K$, $K<\Lambda$, yields a complicated expression, which
we can expand into a series in the photon virtuality $Q^2$:
\begin{align}
	&\Pi_{{\rm 4pt}}(Q^2,\Lambda) 
	= \frac{\alpha^2}{\pi^2}
	\left[
	\frac{3}{16}
	+\frac{707}{25920} \frac{Q^2}{m_\pi^2}
	-\frac{589}{302400} \left(\frac{Q^2}{m_\pi^2}\right)^2
	+\frac{26113}{127008000} \left(\frac{Q^2}{m_\pi^2}\right)^3
	+\mathcal{O}(Q^8)
	\right]
	\nonumber\\ &\quad
	+\frac{\alpha^2}{\pi^2}\left(\frac{\Lambda^2}{m_\pi^2}+\log\frac{\Lambda^2}{m_\pi^2}\right)
	\left[-\frac{3}{16}
	-\frac{1}{160} \frac{Q^2}{m_\pi^2}
	+\frac{1}{1120} \left(\frac{Q^2}{m_\pi^2}\right)^2
	-\frac{1}{6720} \left(\frac{Q^2}{m_\pi^2}\right)^3
	+O(Q^8)
	\right] 
	\nonumber\\ &\quad
	+\frac{\alpha^2}{4\pi^2}\frac{\Lambda^2}{m_\pi^2}
	,
	\label{eq:CottinghamSQEDseries}
\end{align}
where $m_\pi$ stands for the charged pion mass. The value of the two-loop vacuum polarization at zero photon virtuality, $Q^2=0$, reads
\begin{align}
	\Pi_{{\rm 4pt}}(0,\Lambda) = \frac{\alpha^2}{4\pi^2}\,\int\limits_0^\infty d K^2 K^2\left[\frac{1}{K^2}\right]_\Lambda
	\Bigg\{&
	\frac{3 m_\pi^4 \log \left[\frac{1}{2} \frac{K^2}{m_\pi^2}  \left(\sqrt{1+\frac{4m_\pi^2}{K^2}}+1 \right)+1 \right]}{K ^3 \left(K ^2+4m_\pi^2 \right)^{3/2}}
	\nonumber \\&  
	+\frac{ K^4+m_\pi^2 K^2-6m_\pi^4}{4m_\pi^2 K ^2 \left(K ^2+4 m_\pi^2\right)}\Bigg\}.
	\label{eq:Pi0subtractionS}
\end{align}
Here, $\left[\nicefrac{1}{K^2}\right]_\Lambda$ represents an arbitrary regularization scheme. To maintain consistency with Eq.~\eqref{eq:CottinghamSQEDseries}, the same sharp cutoff regularization must be applied to the momentum $K$.

Due to gauge invariance, the structure of the counterterm in sQED is similar to that in QED \cite{Biloshytskyi:2022ets}. It can be expressed in terms of the one-loop pion self energy $\Sigma_2$ and the one-loop vacuum polarization $\overline\Pi_{e^2}$ as
\begin{equation}\label{eq:QEDct}
\overline{\Pi}_{\mathrm{ct}}(q^2,\Lambda) = -\Sigma_2(m_\pi^2)\frac{\partial}{\partial m_\pi^2}\overline{\Pi}_{e^2}(q^2).
\end{equation}
The one-loop vacuum polarization in sQED has the form
\begin{equation}
	\overline\Pi_{e^2}(q)=-\frac{\alpha}{2\pi}\int_0^1 dx x (2x-1) \log\left[\frac{m^2_\pi}{m^2_\pi-q^2x(1-x)}\right],
\end{equation}
so that the derivative with respect to the mass reads
\begin{align}
\frac{\partial}{\partial m_\pi^2}\overline\Pi_{e^2}(q) &= \frac{\alpha}{ \pi }\left(\frac{ \sqrt{4 m^2_\pi-q^2} \tan ^{-1}\left(\frac{q}{\sqrt{4m^2_\pi-q^2}}\right)}{q^3}+\frac{1}{12m^2_\pi}-\frac{1}{q^2}\right)\nonumber\\
&=\frac{\alpha}{\pi}\frac{1}{m_\pi^2}\left[-\frac{q^2/m_\pi^2}{120 \pi }-\frac{(q^2/m_\pi^2)^2}{840 \pi }-\frac{(q^2/m_\pi^2)^3}{5040 \pi }+\mathcal{O}\left(q^8\right)\right].
\end{align}

\begin{figure}[h!]
	\centering
	\includegraphics[width=\textwidth]{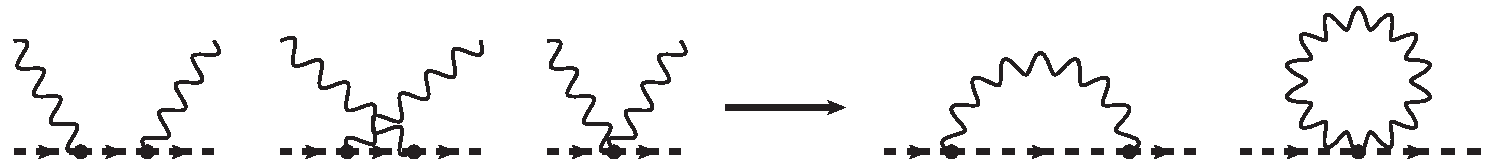}
	\caption{Self energy in sQED via the Cottingham formula.}
	\label{fig:ScalarQED_SE}
\end{figure}
The more gradual asymptotic behavior of the LbL amplitude in sQED, compared to QED, leads to a stronger dependence of the finite part of the UV-divergent pion self-energy on the regularization scheme. In particular, unlike in QED, the finite parts of the self-energy at one loop differ when using sharp cutoff, Pauli-Villars, or dimensional regularization. In order to retain the same regularization scheme in both the Cottingham-like formula and the counterterm calculations, we can also apply the Cottingham formula for derivation of the one-loop self-energy counterterm $\Sigma_2(m_\pi^2)$, as depicted in figure~\ref{fig:ScalarQED_SE}. For that purpose, we need to integrate the tree-level forward doubly-virtual Compton amplitude over the photon momentum. Denoting the pion four-momentum as $p$, $p^2=m_\pi^2$, and the photon four-momentum as $k$, $k^2=-K^2$, this amplitude contracted with $g_{\mu\nu}$ has the following form
\begin{align}
\mathcal{M}_{e^2}(\nu,k^2)\equiv& g_{\mu\nu}\mathcal{M}^{\mu\nu}_{e^2}(\nu,k^2)\nonumber\\ =& 4\pi\alpha\left(\frac{4m^2_\pi-K^2+2(s-m^2_\pi+K^2)}{s-m^2_\pi}+\frac{4m^2_\pi-K^2-2(s+K^2-m^2_\pi)}{-2K^2+m^2_\pi-s}-8\right),
\end{align}
where $s = (p+k)^2=m^2_\pi-K^2+2\nu$, $\nu=p\cdot k$. Introducing the variable $x=\nu/(i m_\pi K)$, one arrives at the following simple expression for the traced Compton amplitude
\begin{equation}
\mathcal{M}_{e^2}(x,K^2) = -8\pi\alpha\frac{4+8x^2+3K^2/m_\pi^2}{4x^2+K^2/m_\pi^2}.
\end{equation}
Substituting this amplitude into the Cottingham formula and integrating over the photon momentum, we obtain the on-shell pion self energy
\begin{align}
\Sigma_2(p^2=m^2_\pi) &=
\int\frac{d^4k}{(2\pi)^4}\frac{1}{k^2}\mathcal{M}_{e^2}(\nu,k^2)=\int_0^\Lambda KdK\int_0^1 dx \sqrt{1-x^2}\mathcal{M}_{e^2}(x,K^2) \nonumber\\
&= -\frac{\alpha}{\pi}m_\pi^2\left[\frac{9}{8}+\frac{3}{4}\left(\frac{\Lambda^2}{m_\pi^2}+\log\frac{\Lambda^2}{m_\pi^2}\right)\right].
\label{eq:SE2}
\end{align}
Here, the sharp cutoff regularization was applied in the same way as in Eq.~\eqref{eq:CottinghamSQEDseries}.
Finally, we obtain the following series expansion for the counterterm
\begin{equation}
\overline\Pi_\mathrm{ct}(Q,\Lambda) = \frac{\alpha^2}{\pi^2}\left[\frac{3}{2}+\left(\frac{\Lambda^2}{m_\pi^2}+\log\frac{\Lambda^2}{m_\pi^2}\right)\right]\left(\frac{Q^2/m_\pi^2}{160}-\frac{(Q^2/m_\pi^2)^2}{1120}+\frac{(Q^2/m_\pi^2)^3}{6720}+\mathcal{O}(Q^8)\right)
\end{equation}
This counterterm, when added to $\overline{\Pi}^\mathrm{sQED}_\mathrm{4pt}$, yields the on-shell renormalized two-loop vacuum polarization with the following series expansion for small $Q^2$:
\begin{equation}
	\Delta\overline{\Pi}(Q^2) = \frac{\alpha^2}{\pi^2}\frac{Q^2}{m_\pi^2}\Bigg(\frac{95}{2592}- \frac{71}{21600}\frac{Q^2}{m_\pi^2}+\frac{54463}{127008000}\left(\frac{Q^2}{m_\pi^2}\right)^2+\mathcal{O}(Q^6)\Bigg).
	\label{eq:VPseriesSQEDcorrect}
\end{equation}
It exactly corresponds to the known result, which can be found either via the dispersion relation
\begin{equation}
\Delta\overline{\Pi}(Q^2) = -\frac{Q^2}{\pi}\int_{4m^2}^{\infty}\;
\frac{d t}{t(t+Q^2)}\im \Pi(t)
\label{DispersiveAnswer}
\end{equation}
with the correct imaginary part given by \cite{SchwingerTextBook}\footnote{In order to get the correct result, one should derive it from Eqs. (5-4.71), (5-4.80) and (5-4.87) instead of using the final expression given by Eq.(5-4.132), which contains a typo.}
\begin{eqnarray}
\im \Pi(t) &=& -\frac{\alpha^2}{12\pi} \Bigg\{
\frac{3}{2}v(1+v^2)-2v^3\left(2\log v-3\log\frac{1+v}{2}\right)\nonumber\\
&&-\frac{1}{4}(1-v)\left[3+v\left(3+v(5v-7)\right)\right]\log\frac{1+v}{1-v}\nonumber\\
&&+v^2(1+v^2)\bigg[\frac{5}{12}\pi^2-\frac{1}{2}\log\frac{2v^4}{1+v}\log\frac{1+v}{1-v}\nonumber\\
&&+4\mathrm{Li}_2(v)+\mathrm{Li}_2\left(v^2\right)+\mathrm{Li}_2\left(\frac{2v}{v-1}\right)-\mathrm{Li}_2\left(\frac{2v}{v+1}\right)-\mathrm{Li}_2\left(\frac{v+1}{v-1}\right)\nonumber\\
&&+\frac{3}{2}\left(\mathrm{Li}_2\left(\frac{1+v}{2}\right)-\mathrm{Li}_2\left(\frac{1-v}{2}\right)\right)\bigg]
\Bigg\},
\end{eqnarray}
where $v=\sqrt{1-\nicefrac{4m_\pi^2}{t}}$,
or by performing exact two-loop calculation in dimensional regularization \cite{Bijnens:2019ejw,Nils}. Therefore, we conclude that the Cottingham-like formula accurately reproduces the complete set of two-loop diagrams for vacuum polarization in sQED, as illustrated in figure~\ref{fig:TwoLoopsQED}. Furthermore, when combined with the appropriate counterterm, this formula yields the correct finite expression for the renormalized two-loop vacuum polarization.
\section{Calculating the integrand for the pseudoscalar meson exchange}
\label{sect:pi0_integrand}
In this appendix, we calculate the $\pi^0$-exchange contribution to the integrand given by Eq.~\eqref{eq:pme_integrand_momspace}.
Due to gauge invariance, only the part of the Fourier-transformed CCS kernel  proportional to $\delta_{\sigma\lambda}$ (cf. Eq.~\eqref{eq:Hp}) contributes when contracted with the polarization tensor. Consequently, this contraction effectively picks out the trace of the polarization tensor:
\begin{equation}
    \tilde H_{\sigma\lambda}(p) \Pi_{\sigma\mu\mu\lambda}(p,q,k) = \tilde h(p)\Pi_{\sigma\mu\mu\sigma}(p,q,k),
\end{equation}
where $\tilde h(p)$ is the coefficient multiplying $\delta_{\lambda\sigma}$ in Eq.~(\ref{eq:Hp}).

The traced polarization tensor can be written as
\ba
\Pi_{\sigma\mu\mu\sigma}(p,q,k) &=& -2
\Big(\frac{{\cal F}(-p^2,-k^2)\;{\cal F}(-q^2,-(p+k+q)^2)}{(p+k)^2+m_\pi^2}\;
\nonumber\\ && ( (p\cdot q) (k\cdot(p+k)) - (p\cdot(p+k)) (k\cdot q))
\nonumber\\ && + \frac{{\cal F}(-p^2,-q^2) \,{\cal F}(-k^2,-(p+k+q)^2)}{(p+q)^2 + m_\pi^2}\;
\nonumber\\ && ( (p\cdot k) (q\cdot(p+q)) - (p\cdot(p+q)) (q\cdot k))
\Big)
\nonumber\\ &=& \Pi^{(1)}_{\sigma\mu\mu\sigma}(p,q,k)  + \Pi^{(1)}_{\sigma\mu\mu\sigma}(p,k,q),\label{eq:PiTraced}
\\ \Pi^{(1)}_{\sigma\mu\mu\sigma}(p,q,k)  &\equiv &
-2 \frac{{\cal F}(-p^2,-k^2)\;{\cal F}(-q^2,-(p+k+q)^2)}{(p+k)^2+m_\pi^2}\;
\nonumber\\ && ( (p\cdot q) (k\cdot(p+k)) - (p\cdot(p+k)) (k\cdot q)).
\ea
Clearly, in Eq.~\eqref{eq:PiTraced} the second term equals the first term upon interchaging $k$ and $q$. Therefore, we can rewrite Eq.~\eqref{eq:pme_integrand_momspace} in terms of the traced polarization tensor as follows:
\ba
\label{f12}
f^{\pi^0}(|x|) &=& -\frac{e^2}{2} 2\pi^2 |x|^3 \frac{1}{(2\pi)^{12}} \int_p  \tilde h(p^2)  \int_{k,q} e^{i(k+q)\cdot x} \Big[\frac{1}{k^2} + \frac{1}{q^2}\Big]_\Lambda  \Pi_{\sigma\mu\mu\sigma}^{(1)}(p,q,k).
\ea
Set
\be
{\cal F}(-p^2,-k^2) = \frac{m_V^4 F_\pi}{(p^2+m_V^2) (k^2+m_V^2)}
  \ee
to the VMD form.
Handling the first term inside the square bracket of Eq.\ (\ref{f12}), we successively obtain 
\ba
&& \Big[\frac{1}{k^2}\Big]_\Lambda\int_q e^{iq\cdot x}  \Pi_{\sigma\mu\mu\sigma}^{(1)}(p,q,k)
= \Big[\frac{1}{k^2}\Big]_\Lambda   \frac{(-2m_V^4 F_\pi){\cal F}(-p^2,-k^2)}{(p+k)^2+m_\pi^2}
\\ &&  \Big( (p\cdot q) (k\cdot(p+k)) - (p\cdot(p+k)) (k\cdot q)\Big)_{q=-i\nabla_x}
\underbrace{\int_q   \frac{e^{iq\cdot x}}{(q^2+m_V^2) ((p+q+k)^2+m_V^2)}}_{=L(m_V,m_V;p+k,x)}\,,
\nonumber
\ea
\ba
\nonumber
&& \int_k \frac{e^{ik\cdot z}}{k^2+\Lambda^2}\int_q e^{iq\cdot x}  \Pi_{\sigma\mu\mu\sigma}^{(1)}(p,q,k) \\
&& = \frac{(-2(m_V^4 F_\pi)^2)}{p^2+m_V^2} \Big( (p\cdot q) (k\cdot(p+k)) - (p\cdot(p+k)) (k\cdot q)\Big)_{\hspace{-0.2cm} \footnotesize 
\begin{array}{cc}
     & q=-i\nabla_x \\
     & k=-i\nabla_z
\end{array}}
\nonumber\\ 
&& \hspace{2cm} \int_k   \frac{e^{ik\cdot z} }{[k^2+\Lambda^2][(p+k)^2+m_\pi^2][k^2+m_V^2]}  L(m_V,m_V;p+k,x),
\ea
and 
\ba
&& \int_k   \frac{e^{ik\cdot z} }{[k^2+\Lambda^2][(p+k)^2+m_\pi^2][k^2+m_V^2]}  L(m_V,m_V;p+k,x)
\nonumber\\ && = \frac{1}{\Lambda^2-m_V^2}\int_k  \Big(\frac{1}{k^2+M_V^2}-\frac{1}{k^2+\Lambda^2}\Big)
\frac{e^{ik\cdot z} }{(p+k)^2+m_\pi^2}  L(m_V,m_V;p+k,x)
\nonumber\\ && = \frac{(8\pi^2)^{-1}}{\Lambda^2-m_V^2}\int_k  \Big(\frac{1}{k^2+M_V^2}-\frac{1}{k^2+\Lambda^2}\Big)
\frac{e^{ik\cdot z} }{(p+k)^2+m_\pi^2}
\nonumber \\ && \int_0^1 d\alpha \,e^{-i\alpha(p+k)\cdot x}\,K_0\Big(|x|\sqrt{\alpha(1-\alpha)(p+k)^2 + m_V^2}\Big)
\nonumber\\ && \stackrel{k\to k-p}{=} \frac{e^{-ip\cdot z}}{4(\Lambda^2-m_V^2)} \int_0^1 d\alpha \,
\int_0^\infty d|k|\,|k|^3   \frac{1 }{k^2+m_\pi^2}\,\,K_0\Big(|x|\sqrt{\alpha(1-\alpha)k^2 + m_V^2}\Big)
\nonumber \\ &&   \Big\<e^{ik\cdot (z-\alpha x)}\Big(\frac{1}{(k-p)^2+m_V^2}-\frac{1}{(k-p)^2+\Lambda^2}\Big) \Big\>_{\hat k}
\nonumber\\ && = \frac{e^{-ip\cdot z}}{4(\Lambda^2-m_V^2)} \int_0^1 d\alpha \,
\int_0^\infty d|k|\,|k|^3   \frac{1}{k^2+m_\pi^2}\,\,K_0\Big(|x|\sqrt{\alpha(1-\alpha)k^2 + m_V^2}\Big)
 \nonumber \\ && \frac{2}{|k||p|}\sum_{n=0}^\infty i^n \frac{J_{n+1}(|k||z-\alpha x|)}{|k||z-\alpha x|}
   \Big(Z_{|k|,|p|}^{(m_V)}{}^{n+1} -  Z_{|k|,|p|}^{(\Lambda)}{}^{n+1}\Big)   C_n^{(1)}(\hat p\cdot\widehat{z-\alpha x}).
\ea
Performing the angular integral over $\hat p$,
\ba
&& 2\pi^2\Big\<\int_k   \frac{e^{ik\cdot z} }{[k^2+\Lambda^2][(p+k)^2+m_\pi^2][k^2+m_V^2]}  L(m_V,m_V;p+k,x)\Big\>_{\hat p}
\\ && = \frac{2\pi^2}{2|p|(\Lambda^2-m_V^2)} \int_0^1 \frac{d\alpha}{|z-\alpha x|} \,
\int_0^\infty d|k|\,   \frac{|k|}{k^2+m_\pi^2}\,\,K_0\Big(|x|\sqrt{\alpha(1-\alpha)k^2 + m_V^2}\Big)
 \nonumber \\ && \sum_{n=0}^\infty i^n J_{n+1}(|k||z-\alpha x|)
 \Big(Z_{|k|,|p|}^{(m_V)}{}^{n+1} -  Z_{|k|,|p|}^{(\Lambda)}{}^{n+1}\Big)  \Big\< e^{-ip\cdot w} C_n^{(1)}(\hat p\cdot\widehat{z-\alpha x})\Big\>_{\hat p}\hspace{0.1cm} \Big|_{w=z}
\\ && = \frac{2\pi^2}{2|p|(\Lambda^2-m_V^2)} \int_0^1 \frac{d\alpha}{|z-\alpha x|} \,
\int_0^\infty d|k|\,   \frac{|k|}{k^2+m_\pi^2}\,\,K_0\Big(|x|\sqrt{\alpha(1-\alpha)k^2 + m_V^2}\Big)
 \nonumber \\ && \sum_{n=0}^\infty i^n J_{n+1}(|k||z-\alpha x|)
 \Big(Z_{|k|,|p|}^{(m_V)}{}^{n+1} -  Z_{|k|,|p|}^{(\Lambda)}{}^{n+1}\Big)
 2 (-i)^n\frac{J_{n+1}(|p||w|)}{|p||w|} C_n^{(1)}(\hat w\cdot \widehat{z-\alpha x}) \Big|_{w=z}
 \nonumber
\\ && = \frac{2\pi^2}{p^2|w|(\Lambda^2-m_V^2)} \int_0^1 \frac{d\alpha}{|z-\alpha x|} \,
\int_0^\infty d|k|\,   \frac{|k|}{k^2+m_\pi^2}\,\,K_0\Big(|x|\sqrt{\alpha(1-\alpha)k^2 + m_V^2}\Big)
 \nonumber \\ && \sum_{n=0}^\infty   \Big(Z_{|k|,|p|}^{(m_V)}{}^{n+1} -  Z_{|k|,|p|}^{(\Lambda)}{}^{n+1}\Big)
J_{n+1}(|k||z-\alpha x|) \,J_{n+1}(|p||w|)\, C_n^{(1)}(\hat w\cdot \widehat{z-\alpha x}) \Big|_{w=z},
 \nonumber
 \ea
 we obtain
 \ba
&&  f^{\pi^0}_{1}(|x|) = 
-\frac{e^2\pi^2|x|^3}{(2\pi)^8}\frac{-4\pi^2(m_V^4 F_\pi)^2}{\Lambda^2-m_V^2}
 \Big( (p\cdot q) (k\cdot(p+k)) - (p\cdot(p+k)) (k\cdot q)\Big)_{\hspace{-0.2cm} \footnotesize 
 \begin{array}{cc}
     & q=-i\nabla_x \\
     & k=-i\nabla_z \\
     & p=i\nabla_w
\end{array}}
\nonumber  \\ &&  \int_0^\infty d|p|\,\frac{|p|\,\tilde h(p^2)}{p^2+m_V^2}\int_0^1 \frac{d\alpha}{|z-\alpha x|} \,
\int_0^\infty d|k|\,   \frac{|k|}{k^2+m_\pi^2}\,\,K_0\Big(|x|\sqrt{\alpha(1-\alpha)k^2 + m_V^2}\Big)
 \\ && \frac{1}{|w|}\sum_{n=0}^\infty   \Big(Z_{|k|,|p|}^{(m_V)}{}^{n+1} -  Z_{|k|,|p|}^{(\Lambda)}{}^{n+1}\Big)
 J_{n+1}(|k||z-\alpha x|) \,J_{n+1}(|p||w|)\, C_n^{(1)}(\hat w\cdot \widehat{z-\alpha x}) \Big|_{w=z} \Big|_{z=x}.
 \nonumber
 \ea
 The notation $\nabla_z\nabla_w(.)\Big|_{w=z} \Big|_{z=x}$ indicates that one has to first evaluate the derivatives with respect to $w$ and set $w=z$, and afterwards evaluate the derivatives with respect to $z$ and then set $z=x$.
 The second term of Eq.~\eqref{f12} can be calculated analogously:
\ba
&&  \nonumber  f^{\pi^0}_{2}(|x|) = 
-\frac{e^2\pi^2|x|^3}{(2\pi)^8}\frac{-4\pi^2(m_V^4 F_\pi)^2}{\Lambda^2-m_V^2}
 \Big( (p\cdot q) (k\cdot(p+k)) - (p\cdot(p+k)) (k\cdot q)\Big)_{\hspace{-0.2cm} \footnotesize 
 \begin{array}{cc}
     & q=-i\nabla_x \\
     & k=-i\nabla_z \\
     & p=i\nabla_w
\end{array}}
\\ && \int_0^\infty d|p|\,\frac{|p|\,\tilde h(p^2)}{p^2+m_V^2}\int_0^1 \frac{d\alpha}{|z-\alpha x|} \,
\int_0^\infty d|k|\,   \frac{|k|}{k^2+m_\pi^2}\,\,
 \\ && 
\Big[K_0\Big(|x|\sqrt{\alpha(1-\alpha)k^2 + m_V^2}\Big)-\Big(K_0\Big(|x|\sqrt{\alpha(1-\alpha)k^2 + \alpha m_V^2+(1-\alpha) \Lambda^2}\Big) \Big]
 \nonumber \\ && \frac{1}{|w|}\sum_{n=0}^\infty   Z_{|k|,|p|}^{(m_V)}{}^{n+1}
 J_{n+1}(|k||z-\alpha x|) \,J_{n+1}(|p||w|)\, C_n^{(1)}(\hat w\cdot \widehat{z-\alpha x}) \Big|_{w=z} \Big|_{z=x}.
 \nonumber
 \ea
 The final result for the integrand is obtained as the sum 
 \ba
 \label{eq:pi0_integrand_final}
f^{\pi^0}(|x|) = f^{\pi^0}_{1}(|x|) + f^{\pi^0}_{2}(|x|) .
 \ea 

\section{Tables}
In this appendix, we provide several tables that might be useful for possible crosschecks. We display the results of the model calculation, described in section \ref{sec:phenomenological_model}, for the $\pi^0$ $\eta$ and $\eta'$ contribution along our chiral trajectory in table \ref{table:pme_results}. Additionally, we provide the matching coefficients (table \ref{table:mixing_angle_coefficients}) and the results for $a_\mu^{(2+2)a-ll}$ (table \ref{tab:ensemble_results}) computed on each ensemble listed in table \ref{table:nlohvp_ensemble}.

\begin{table}[h]
\centering
\caption{Results of the pseudoscalar meson contributions along the chiral trajectory defined by the parameters in table \ref{table:model_param}. All results are given in units of $10^{-10}$ and without any multiplicative matching coefficient.}
\label{table:pme_results}
\begin{tabular}{|c|c|c|c|c|c|c|}
\hline
Id                      & N202     & N203     & N451     & C101    & E300        & physical              \\ \hline
$m_\pi$ {[}MeV{]}       & 418(5) & 349(4)   & 291(4)   & 222(3) & 177(2)      & 134.9768(5)        \\
\hline
$a_\mu^{\pi^0}$&0.190(9)&0.221(10)&0.253(11)&0.302(9)&0.331(8)&0.368(8)\\
$a_\mu^{\eta}$&0.063(2)& 0.103(5)& 0.134(6)& 0.177(9)& 0.184(10)&0.204(10)\\
 $a_\mu^{\eta'}$&0.307(33)& 0.313(40)& 0.274(26)& 0.243(18)& 0.237(10)& 0.232(10)\\
\hline
\end{tabular}
\end{table}

\begin{table}[h]
\centering
\caption{The mixing angles and matching coefficients for the $\eta$ and $\eta'$ meson in Eq. \eqref{eq:full_model} along the chiral trajectory. The superscript $^{[0]}$ denotes leading order in $\chi$PT and indicates that the angle is evaluated with Eq. \eqref{eq:lo_mixingangle}, while  $^{[1]}$ denotes next-to-leading order and is interpolated between the SU(3) flavour symmetric point and the physical result from Ref. \cite{Bickert:2016fgy}. The coefficients $\hat{c}$ are computed using the procedure described in section \ref{sect:hvpnlo_matching} and the NLO result for the mixing angle $\theta^{[1]}$ on each ensemble.}
\label{table:mixing_angle_coefficients}
\begin{tabular}{|c|c|c|c|c|c|c|}
\hline
Id                      & N202     & N203     & N451     & C101    & E300        & E250/physical              \\ \hline
$m_\pi$ {[}MeV{]}       & 418(5) & 349(4)   & 291(4)   & 222(3) & 177(2)      & 132(2)/134.9768(5)        \\
\hline
$\theta^{[0]}$  [$^\circ$] &0 &-7.2(3.7) &-10.8(3.6) &-14.7(3.2) &-18.5(6) &-19.7 \\
$\theta^{[1]}$  [$^\circ$] &0 & -3.6(3)& -6.3(4)& -9.1(5)& -10.4(6)&-11.1(6)\\
\hline
 $\hat{c}^{(ll)}_\eta$&25/9 & 2.27(1.49)& 2.43(69)& 2.14(12)& 2.24(6)& 2.19(3)\\
  $\hat{c}^{(ll)}_{\eta'}$&25/36 & 0.51(20)& 0.47(14)& 0.42(11)& 0.30(4)& 0.28(3)\\
 $\hat{c}^{(ls)}_\eta$&-20/9 & -1.15(58)& -1.21(21)& -0.67(9)& -0.75(5)& -0.71(2)\\
$\hat{c}^{(ls)}_{\eta'}$&10/36 & 0.22(9)& 0.23(6)& 0.25(4)& 0.29(1)& 0.30(1)\\
\hline
\end{tabular}
\end{table}

\begin{table}[h]
    \centering
    \small
    \caption{Results for each ensemble obtained with the method discussed in section \ref{sect:nlohvp_tail} to approximate the tail. Results are also provided for the contribution below $|x|=L/2$ from the integrand calculated on the lattice and the contribution of the tail computed with the phenomenological model. All values are given in units of $10^{-10}$.}
    \label{tab:ensemble_results}
    \begin{tabular}{|c|c|c|c|}
    \hline
       Id  & $a_\mu^{(2+2)a-ll}$ [combined] & $a_\mu^{(2+2)a-ll}$ [$<L/2$ (lattice)] & $a_\mu^{(2+2)a-ll}$ [$>L/2$ (model)]  \\
       \hline
H200 & 
    $-0.18 (0.01)_\textrm{stat}  (0.01)_\textrm{fs} (0.06)_\textrm{tail}  $ &
    $-0.06 (0.01)_\textrm{stat} (0.01)_\textrm{fs}  $ &
    $-0.12  (0.06)_\textrm{tail}  $  \\
    
N202 & 
    $-0.21 (0.01)_\textrm{stat}  (0.02)_\textrm{fs} (0.03)_\textrm{tail}  $ &
    $-0.15 (0.01)_\textrm{stat} (0.02)_\textrm{fs}  $ &
    $-0.06  (0.03)_\textrm{tail}  $  \\
    
N203 & 
    $-0.31 (0.03)_\textrm{stat}  (0.03)_\textrm{fs} (0.04)_\textrm{tail}  $ &
    $-0.23 (0.03)_\textrm{stat} (0.03)_\textrm{fs}  $ &
    $-0.08  (0.04)_\textrm{tail}  $  \\
    
N451 & 
    $-0.65 (0.06)_\textrm{stat}  (0.07)_\textrm{fs} (0.07)_\textrm{tail}  $ &
    $-0.50 (0.06)_\textrm{stat} (0.07)_\textrm{fs}  $ &
    $-0.15  (0.07)_\textrm{tail}  $  \\
    
C101 & 
    $-1.17 (0.17)_\textrm{stat}  (0.12)_\textrm{fs} (0.18)_\textrm{tail}  $ &
    $-0.80 (0.17)_\textrm{stat} (0.12)_\textrm{fs}  $ &
    $-0.36  (0.18)_\textrm{tail}  $  \\
    
D450 & 
    $-1.30 (0.20)_\textrm{stat}  (0.15)_\textrm{fs} (0.15)_\textrm{tail}  $ &
    $-1.00 (0.20)_\textrm{stat} (0.15)_\textrm{fs}  $ &
    $-0.30  (0.15)_\textrm{tail}  $  \\
    
E300 & 
    $-3.35 (0.26)_\textrm{stat}  (0.33)_\textrm{fs} (0.59)_\textrm{tail}  $ &
    $-2.18 (0.26)_\textrm{stat} (0.33)_\textrm{fs}  $ &
    $-1.17  (0.59)_\textrm{tail}  $  \\
    
E250 & 
    $-7.36 (2.32)_\textrm{stat}  (0.65)_\textrm{fs} (1.53)_\textrm{tail}  $ &
    $-4.31 (2.32)_\textrm{stat} (0.65)_\textrm{fs}  $ &
    $-3.05  (1.53)_\textrm{tail}  $  \\
    
H101 & 
    $-0.18 (0.01)_\textrm{stat}  (0.02)_\textrm{fs} (0.04)_\textrm{tail}  $ &
    $-0.10 (0.01)_\textrm{stat} (0.02)_\textrm{fs}  $ &
    $-0.08  (0.04)_\textrm{tail}  $  \\
    
N300 & 
    $-0.18 (0.01)_\textrm{stat}  (0.02)_\textrm{fs} (0.04)_\textrm{tail}  $ &
    $-0.10 (0.01)_\textrm{stat} (0.02)_\textrm{fs}  $ &
    $-0.08  (0.04)_\textrm{tail}  $  \\
    
B450 & 
    $-0.18 (0.01)_\textrm{stat}  (0.02)_\textrm{fs} (0.04)_\textrm{tail}  $ &
    $-0.10 (0.01)_\textrm{stat} (0.02)_\textrm{fs}  $ &
    $-0.08  (0.04)_\textrm{tail}  $  \\
    
N101 & 
    $-0.73 (0.10)_\textrm{stat}  (0.10)_\textrm{fs} (0.04)_\textrm{tail}  $ &
    $-0.64 (0.10)_\textrm{stat} (0.10)_\textrm{fs}  $ &
    $-0.09  (0.04)_\textrm{tail}  $  \\
    
H105 & 
    $-0.57 (0.03)_\textrm{stat}  (0.04)_\textrm{fs} (0.15)_\textrm{tail}  $ &
    $-0.26 (0.03)_\textrm{stat} (0.04)_\textrm{fs}  $ &
    $-0.31  (0.15)_\textrm{tail}  $  \\
    
S100 & 
    $-1.33 (0.08)_\textrm{stat}  (0.07)_\textrm{fs} (0.44)_\textrm{tail}  $ &
    $-0.44 (0.08)_\textrm{stat} (0.07)_\textrm{fs}  $ &
    $-0.88  (0.44)_\textrm{tail}  $  \\
       \hline
    \end{tabular}
\end{table}

\newpage
\bibliographystyle{JHEP}
\bibliography{biblist}
\end{document}